# An epitaxial graphene platform for zero-energy edge state nanoelectronics

*Vladimir S. Prudkovskiy[1, 2, 3], Yiran Hu[2], Yue Hu[2], Kaimin Zhang[1], Peixuan Ji[1], Grant Nunn[2], Jian Zhao[1], Chenqian Shi[1], Antonio Tejeda[4, 5], David Wander[3], Alessandro De Cecco[3], Clemens B. Winkelmann[3], Yuxuan Jiang[6], Tianhao Zhao[2], Katsunori Wakabayashi[7, 8], Zhigang Jiang[2], Lei Ma[1, 9, †], Claire Berger[3,10], Walt A. de Heer[1,2 *]*

[1] Tianjin International Center for Nanoparticles and Nanosystems, (TICNN) Tianjin University, Nankai District, China, 30007
[2] School of Physics, Georgia Institute of Technology, Atlanta, Georgia 30332, United States
[3] Institut Néel, Univ. Grenoble Alpes, CNRS, Grenoble INP, 38000 Grenoble, France
[4] Laboratoire de Physique des Solides, CNRS, Univ. Paris-Sud, 91405 Orsay, France
[5] Synchrotron SOLEIL, L'Orme des Merisiers, Saint-Aubin, 91192 Gif sur Yvette, France
[6] National High Magnetic Field Laboratory, Tallahassee, Florida 32310, United States
[7] School of Science and Technology, Kwansei Gakuin University, Gakuen 2-1, Sanda 669-1337, Japan
[8] Center for Spintronics Research Network (CSRN), Osaka University, Toyonaka 560-8531, Japan
[9] State key Laboratory of Precision Measurement Technology and Instruments, Tianjin University, Nankai District, China, 300072
[10] Laboratoire de Recherche International 2958 Georgia Tech-CNRS, 57070 Metz, France

**Corresponding authors**
* e-mail: walter.deheer@physics.gatech.edu
† e-mail maleixinjiang@tju.edu.cn

## Abstract

Graphene's original promise to succeed silicon faltered due to pervasive edge disorder in lithographically patterned deposited graphene and the lack of a new electronics paradigm. Here we demonstrate that the annealed edges in conventionally patterned graphene epitaxially grown on a silicon carbide substrate (epigraphene) are stabilized by the substrate and support a protected edge state. The edge state has a mean free path that is greater than 50 microns, 5000 times greater than the bulk states and involves a theoretically unexpected Majorana-like zero-energy non-degenerate quasiparticle that does not produce a Hall voltage. In seamless integrated structures, the edge state forms a zero-energy one-dimensional ballistic network with essentially dissipationless nodes at ribbon-ribbon junctions. Seamless device structures offer a variety of switching possibilities including quantum coherent devices at low temperatures. This makes epigraphene a technologically viable graphene nanoelectronics platform that has the potential to succeed silicon nanoelectronics.

## Introduction

A viable nanoelectronics platform can be defined as a material that can be processed using conventional nanoelectronics technology, as required to produce high density, high performance commercial nanoelectronics. Currently, silicon provides a platform for high performance complementary metal oxide semiconductor (CMOS) nanoelectronics that enables billions of transistors to be patterned on an area of 1 $cm^2$ using the most highly developed technologies in the world. However, the CMOS platform is reaching its scaling limits[1]. A viable continuation will require a paradigm shift in electronics, while preserving, as much as possible, current industrial fabrication methods[2]. The latter condition implies that this new platform should involve a single crystal substrate and a conventionally nanopatternable material. We show here that electronics based on the epigraphene edge state is perfectly suited: electronics grade single crystal SiC is already used in commercial electronics[3] and the edge state charge carrier is a new quasiparticle. In contrast, electronics approaches where chemically produced graphene ribbons are proposed to be interconnected with metal nanowires is based on conventional transistor concepts and still poses daunting technical challenges before they are technologically viable.[4, 5]

The high electronic mobility and long mean free paths in rolled up graphene sheets (i.e. carbon nanotubes) suggested a graphene nanoelectronics platform as a successor to the silicon nanoelectronics platform.[1, 6-11] The graphene platform promised to extend the Si miniaturization limit while at the same time introducing a new electronics paradigm by enabling quantum phase coherent devices[6, 12, 13] as first demonstrated in carbon nanotubes.[8, 9] The graphene platform requires graphene nanostructures to be lithographically patterned from two-dimensional graphene sheets to produce the seamlessly interconnected graphene device structures. This cannot be achieved with the alternative bottom-up approach where chemically produced graphene ribbons are interconnected with metal nano wires, which destroys quantum coherence and severely limits performance. While important for flexible electronics, it is not competitive with CMOS technology.[4, 11] Despite initial speculations that exfoliated graphene and deposited graphene could be used to realize this graphene nanoelectronics platform,[14] pervasive 10 nm scale edge disorder[15-18] dashed hopes of success[19] because an intact edge is required to produce a bandgap in armchair graphene ribbons and the edge state in chiral graphene. Moreover the edge state is highly desired in graphene interconnected electronics since it is the only transporting state in neutral graphene nanostructures.[20-23]

The high quality epigraphene layer that grows epitaxially on silicon carbide (SiC)[12, 13, 24-27] can be nanopatterned.[6, 13, 28-31] SiC is a mainstream electronic material[3, 32] that it is compatible with industrial nanoelectronics processing methods[32-34] that now can produce nanostructures as small as 5 nm[35]. This is relevant because a 5 nm wide graphene armchair ribbon has a predicted band gap greater than 0.2 eV.[36, 37] Moreover, SiC is compatible with THz electronics. Chiral graphene ribbons with intact edges are metallic due to a zero-energy edge state (zero-mode) in the gap.[21, 38, 39] With current technology, large scale graphene networks composed of seamlessly integrated metallic and semiconducting epigraphene nanoribbons can be produced.[6, 13] The absence of a bandgap in graphene is often cited as a showstopping problem,[19] however already in 2002

Wakabayashi demonstrated[22] that the gapless graphene edge state can also be gated in principle (see also Supplementary Information SI14).

The topologically protected edge state [20-23, 38-40] was predicted in 1996 and edge state transport was first observed[30, 41] in 2014 up to room temperature in self-assembled 40 nm wide graphene ribbons grown on the sidewalls[29, 31] of thermally annealed trenches etched in single crystal silicon carbide wafers[29-31, 41-43] (see also SI15). However, sidewall ribbons cannot be seamlessly integrated, and their topology severely limits conventional transport measurements, so that little was known experimentally prior to this work.

To verify the viability of the epigraphene zero-mode based platform we demonstrate seamlessly interconnected charge-neutral ballistic epigraphene ribbon networks using standard nanolithography methods [12, 13, 25, 27] on non-polar silicon carbide substrates. Exceptional edge state transport is observed with unprecedented mean free paths exceeding 50 microns even at room temperature. This is 5000 times greater than the mean free path of the bulk states on the same ribbon, which means that the epigraphene edge state survives the harsh nanolithographic processes and it is not affected by deposited amorphous dielectric coatings. Hence, the epigraphene zero mode platform[6, 12] is unique and it enables edge state nanoelectronics.

Theoretically, the graphene edge state,[20-23, 38-40, 44-46] which should not to be confused with quantum Hall states[14, 47-49] and other magnetically induced edge states,[50] is a zero-energy mode of graphene ribbons associated with the edge-localized flatband.[20, 21, 38-40, 46] Its basic electronic structure for zigzag ribbons,[20, 21] shown in Fig. 1a, generically applies to all chiral ribbons.[21, 38, 39] We find that the large density of states caused by the flatband (the 0-DoS peak)[21, 51] pins the Fermi level at energy $E=0$ at the edges (Fig. 1b).[51] Moreover a Schottky barrier[52] at the edge insulates the edge state from the bulk states resulting in an isolated 1D zero-energy ballistic edge state network with essentially elastic scattering at the ribbon-ribbon junctions. Consequently, heat is primarily generated at the metal-ribbon contacts and not at the ribbon-ribbon junctions[53]. We also find that the edge state does not generate a Hall voltage, which suggests that the edge state is an ambipolar single channel state that is half-electron and half-hole in contrast to theory that predicts that it is either an electron or a hole.

Since the edge state properties presented here defy current theory, they require rigorous experimental evidence. To that end, we first demonstrate that the epigraphene on the non-polar substrates used here is charge neutral graphene. We explain how the silicon carbide substrate stabilizes the edges. We next demonstrate that edge state transport involves a one-dimensional non-degenerate ballistic channel that scatters at graphene junctions. Then we demonstrate that the edge state is pinned at zero energy and that it does not generate a Hall voltage. We conclude with examples of device architectures.

## Results

### 2D nonpolar epigraphene and edge stabilization

The SiC wafers are cut and polished in-house from hexagonal SiC stock to expose nonpolar $(1\bar{1}0n)$, *n≈5*, surfaces. As shown in Fig. 2 the thermally grown epigraphene[25] has conventional neutral graphene properties (see also SI12). Scanning tunneling and electron microscopies confirm the graphene lattice structure and its epitaxial alignment with the SiC substrate (Fig. 2a).[12, 27, 54, 55] This is essential to accurately direct the lithography, however as discussed below, even an annealed edge is generally not straight (Fig. S20) so that the actual edges do not have a well-defined chirality. Raman spectroscopy shows a narrow 2D peak and a small D peak (Fig. 2b).[56] Angle resolved photoelectron spectroscopy (Fig. 2c) shows the characteristic Dirac cones with their apex at $E=0$, corresponding to neutral graphene with a Fermi velocity $v_F$=1.06x10$^6$ m/s.[57-59] Infrared magneto-spectroscopy (Fig. 2d) verifies the graphene electronic dispersion: $E(B) = v_F\sqrt{2|N_{LL}|e\hbar B}$, where $N_{LL}$ is the Landau level index. Scanning tunneling spectroscopy (STS) (Fig. 2e) verifies that the Fermi level $E_F$ is close to $E_F$=0 and STS measured at a series of locations across the edge of a graphene crystal shows the 0-DoS peak at the edge (Fig. 2f), similar to that observed in Ref [60].

We find that only thermally annealed epigraphene structures have an edge state (See SI16).[61] In Method 1 (see SI3), the epigraphene is patterned by reactive oxygen ion etching (RIE) followed by annealing in vacuum at 1200$^o$ C. Method 2 (SI3) involves inductively coupled plasma (ICP) etching,[62] that cuts through the graphene and 16 nm into the SiC substrate via an e-beam patterned resist mask. Like plasma cutting and welding, energetic ion collisions produce extremely high local temperatures ($T$>5000 K)[62, 63] that cause carbon and silicon to evaporate. During cooling, C-C and Si-C bands are formed,[63] fusing the graphene edges to the silicon carbide thereby chemically and mechanically stabilizing the edges.[30, 64-66] Cross-sectional electron microscopy indicates that thermally annealed sidewall ribbons are bonded to the substrate via acene (i.e. zigzag-like) edge atoms[67, 68] that have a single unoccupied state at $E=0$.[69] In graphene ribbons these edge atoms cause the 0-DoS peak that manifests as the $E=0$ flatband,[21, 39, 67, 68] which is essential for the graphene edge state.[38, 39] The fact that edge state is also seen in meandering ribbons (SI15)[30, 41] indicates that nominally zigzag (ZZ) and armchair (AC) edges are acene bonded in general. Here we use the ZZ and AC designation to distinguish these two directions and not to indicate the actual graphene edge morphology, which for annealed edges is determined by their stability that involves the bonding to the substrate, mostly likely favoring acene bonded edges (See SI17). The Si-C bond is slightly polar[63, 70] so that the edges will be very slightly n or p-doped depending on their bonding to the SiC substrate.

Three Hall bars labelled S1-S3 were prepared for this study of which S1 and S2 are discussed in detail. They have similar geometries (Fig. 3b, e and SI3) but with production and dimension variations. The long arm of Sample S1 prepared with ICP etching (Method 1, Fig. S4) is along the AC direction and Sample S2, prepared with Method 2 (Fig. S5), is along the ZZ direction.

Transport measurements were performed at temperatures from $T$=2 K to 300 K and magnetic fields $B$ up to 9 T. As shown in Figs. 3b, e and SI3, the Hall bar structures are composed of graphene ribbon segments, **A** through **J.** Adjacent segments are seamlessly connected by edgeless, square graphene junctions. The gate fully covers only segments **B** and **C**. While bulk epigraphene is intrinsically charge neutral (Fig. 2), processing charges the graphene, with a charge density $n_0$: $n_0 \approx 10^{12}$ cm$^{-2}$ in S1 and 2x10$^{11}$ cm$^{-2}$ in S2, so that the charge density of ungated segment sections is $n_0$. We normalize $V_G$ so that $V_G$=0 corresponds to the charge neutrality point (CNP), i.e. where the conductance has a minimum and the Hall voltage switches sign (see below). Resistances $R_{ij,kl}$ are determined with current $I_0$ flowing from contact *i* to *j*, and voltages measured between contacts *k* and *l*. Conductances $G_{ij,kl}$ are defined by $G_{ij,kl}=1/R_{ij,kl}$.

**Characterization of the single channel edge state**

The following systematic study (1) unambiguously identifies ballistic transport of the edge state from scaling; (2) reveals differences in AC and ZZ edges; (3) shows the insensitivity of the edge state to the gate voltage; (4) demonstrates isotropic scattering at the junctions; (5) shows the reduction of backscattering in a magnetic field. (6) Edgeless Corbino ring[71] measurements (SI2) confirm that the observed properties are due to the edge state.

The conductance of a diffusive graphene ribbon of length *L* and width *W* is $G=\sigma W/L$. The conductivity $\sigma=GL/W=ne\mu$ (*n* is the charge density and $\mu$ the mobility) is a material property of diffusive conductors that does not depend on size or shape and therefore imposes strict scaling rules. In contrast, in a single channel ballistic conductor[53] $G=G_0(1+L/\lambda)^{-1}$ so that the conductance is not inversely proportional to the length. Here, $G_0=e^2/h$, *e* is the electronic charge, *h* is Planck's constant, and $\lambda$ is the mean free path.

Figure 3a shows 4-point conductance measurements of ZZ ribbon segments **B** and **C** of S2: $G_{\boldsymbol{B}}(V_G)$ (blue) and $G_{\boldsymbol{C}}(V_G)$ (red). Scaling demands that $C_{\boldsymbol{CB}}=G_{\boldsymbol{C}}/G_{\boldsymbol{B}}$ is strictly constant ($C_{\boldsymbol{CB}} \approx L_B/L_C=2$), however $C_{\boldsymbol{CB}}(V_G)$ varies from ≈1 to 1.5. Similarly, for AC ribbon segments **B** and **C** of S1 (Fig. 3c), $C_{\boldsymbol{CB}}(V_G)$ varies from about 1.3 to 1.6. In both samples, the conductance reaches a minimum, but it does not vanish at the charge neutrality point (CNP). A disorder induced non-zero minimum conductivity $\sigma_{min}$ that obeys the scaling law is often seen in exfoliated graphene.[14, 72-74] In those cases, extrapolating the measured conductance away from the rounding at CNP to *n*=0 gives *G*=0 (SI1). But here (Fig. 3a, c) it does not, and a large residual conductance $G^{Res}$ remains (dashed lines): for S1, $G_{\boldsymbol{B}}^{Res}=0.62\ G_0$, and $G_{\boldsymbol{C}}^{Res}=0.77\ G_0$; for S2, $G_{\boldsymbol{B}}^{Res}=0.83\ G_0$, and $G_{\boldsymbol{C}}^{Res}=0.85\ G_0$.

The residual conductance is immediately explained in the Landauer formulism (see, for example Ref.[53]) where in general the conductance of a graphene ribbon with a ballistic edge state and a diffusive bulk (ignoring coherence effects), can be written as

$$G= G_{edge} + G_{bulk}= G_e(1+L/\lambda)^{-1} + \sigma W/L \qquad (1)$$

where $G_e$ is predicted to be 2 $G_0$ for a non-polarized graphene edge state and 1 $G_0$ for a polarized edge state[20, 21, 47, 48, 75-78]; the edge state mfp $\lambda$ is expected to depend on $E_F$.

Note that $GL/W$ (Eq. 1) depends on *L.* However, if we subtract $G_{\boldsymbol{B,C}}^{Res}$ we find a segment-independent conductivity $\sigma_{\boldsymbol{B,C}}=(G_{\boldsymbol{B,C}}-G_{\boldsymbol{B,C}}^{Res})L_{\boldsymbol{B,C}}/W$. The fact that $\sigma_{\boldsymbol{B}}$ and $\sigma_{\boldsymbol{C}}$ coincide (Fig. 3a and 3c insets) shows that scaling is restored by this subtraction, which implies from Eq. 1 that $G_{B,C}^{Res}$=$G_{edge}$ (see also SI4). Plugging the measured values in Eq. 1 gives $G_e$=$G_{ZZ}$=0.86 $G_0$ and $\lambda_{zz}$>50 µm for the ZZ direction (sample S2), and $G_e$=$G_{AC}$=0.95 $G_0$ and $\lambda_{AC}$=6 µm for the AC direction (sample S1). In contrast, the bulk mobility of S1, extracted from $\sigma$ is $\mu$=750 cm$^2$V$^{-1}$s$^{-1}$ which corresponds to a bulk mpf, $\lambda_{bulk}$=10 nm. This straightforward analysis of conventional 4-point measurements unambiguously shows that a 1 $G_0$ ballistic state exists at CNP.

The bulk conductivity $\sigma$ (Figs. 3a and 3c insets) is found by subtracting the residual conductance measured at $V_G$=0 (horizontal dashed lines), which implies that $\lambda$ in Eq. 1 does not depend on $V_G$ (if in fact it did, then $\sigma_B$ and $\sigma_C$ would not coincide). This independence contradicts the conventional band structure in Fig. 1a, where $V_G$ causes a rigid shift of the bands. For example, a $V_G$=1 V would shift $E_F$ by $\Delta$E≈110 meV so that $N_{0FI}$ would be $\Delta E$ below $E_F$ and would not participate in the transport which leaves only $N_{0Dis}$. However, if $E_F$ were 110 meV then $N_{0Dis}$ would mix with 4x$\Delta E$/d$E$=220 bulk subbands (d$E$≈2 meV is the subband spacing for this 740 nm wide ribbon, Fig. 1a). Therefore $\lambda$ should decrease rapidly with increasing $V_G$, so that for large $V_G$, $G_{edge}$ should essentially vanish. In that case we should have observed a pronounced ≈1 $G_0$ conductance peak at CNP, which is clearly not the case. Therefore Fig. 1a cannot represent the band structure for $V_G$≠0. This discrepancy is resolved below.

The measurements show that the edge state scatters at the junctions (see SI7).[79] If the scattering were isotropic then the transmission probability through the junction would be $T_J$=½ (see Refs.[30, 53, 80]) and a 4-point measurement would result in perfect 1 $G_0$ quantization. However at $B=0$, we find $T_J$≈0.45. The 10% discrepancy is discussed next. Figure 4a shows the conductance of Sample 2 ZZ Seg. **B** as a function of magnetic field $B$ for 4 values of $V_G$ at the vertical dashed lines in Fig. 3a. At CNP ($V_G$=0), the conductance increases to about 0.98 $G_0$ at $B$=3 T. Hence for $B$>3T, $T_J$≈½ (see SI11 for additional measurements). Furthermore, as shown in Fig. 4b, the zero-field dip is strongly reduced with increasing temperature tending towards exact 1 $G_0$ quantization at CNP (see also SI11).

The 3-point measurements of ZZ Segs. **E** and **H** of Sample 1 (Fig. 3d) are consistent with the 4-point measurements above, showing $G$=0.85±0.02 $G_0$ at CNP (taking into account the junction transmission $T_J$ and contact resistance, $R_c$≈1.5 kΩ). Figure 5a shows 2-point measurements of $G_{EH}$=$G_{26,26}$ as a function of $V_G$ (S1 along the ZZ direction) and Fig. 5b shows $G_{ABCD}$=$G_{15,15}$ (S1 along AC direction) for a range of magnetic fields. For $B=0$, the residual conductance $G^{Res}_{ABCD}$=0.15 $G_0$ ($R^{Res}_{ABCD}$=6.6 $R_0$). If $T_J$=½ then $R^{Res}_{ABCD}$=(4+$L_{ABCD}$/$\lambda_{AC}$)$R_0$,[30, 53] so that $\lambda_{AC}$=6.3 µm. For high magnetic fields the edge state conductances, $G_{EH}$ ($V_G$=0) and $G_{ABCD}$($V_G$=0), are very close to ½ $G_0$ and ¼ $G_0$ respectively, as expected for a 1 $G_0$ ballistic conductor with respectively 1 and 3 isotropic scattering centers[30, 53] (see also SI5). In addition, a magnetic field $B$≥2 T significantly increases $\lambda_{AC}$, i.e. it reduces backscattering at defects, reminiscently of the destruction of coherent backscattering by a magnetic field.[53] These observations confirm that the 1 $G_0$ ballistic edge state scatters at junctions, even in large magnetic fields, in contrast with quantum Hall edge states.

**1D edge state**

One might expect that edge state currents will follow the graphene periphery so that in general the conductance measured along the top edge of a segment should be unrelated to the bottom edge. However, the conductance measured along the top and bottom edges are identical, as shown for Seg. **B** of sample 2 (Fig. 6a), and Seg. **A** of sample 1 (Fig. 6b). Since only the edge state contributes to transport at CNP ($V_G$=0) this demonstrates that edge state transport involves both sides of the ribbons equally, consistent with its theoretically expected 1D nature. Hence, the edge state forms a network of 1D single channel ballistic conductors with nodes at the junctions (see SI7).

**Vanishing edge state Hall voltage**

The Hall resistance, $R_{Hall}=V_H/I_0$ (Fig. 7a) is determined from the Hall voltage $V_H=V_{26,15}$ and current $I_0$. Figure 7c plots $n^*=B/eR_{\text{Hall}}$ (see also SI6). For large $V_G$, $n^*$ is linear and independent of $B$, as expected in the diffusive limit where $n^*=n$. From this we derive the gate efficiency $C$=1.3x10[12] cm$^{-2}$V$^{-1}$ that agrees with the expected $C=\varepsilon_0\kappa/ed$=1.5x10[12] cm$^{-2}$V$^{-1}$, where $d=30$ nm is the aluminum oxide gate dielectric thickness and $\kappa$=9 is its dielectric constant. The thick black dashed line in Figs. 7a,b,c, corresponds to $(V_G - V_0^+) = n^*/C - \hbar v_F\sqrt{\pi n}/e$, where the second term represents the quantum capacitance correction.[81] The gate voltage $V_G=V_0^+$=0.196 V locates the conduction band edge (dashed blue line); it is significantly displaced from $V_G$=0. Likewise, the valence band edge is approximately located at $V_G=-V_0$ which indicates an energy gap.

The edge state is ballistic in the segments ($\lambda>>L$) and therefore it is effectively decoupled from the bulk. While edge state and the bulk thermalize at the contacts and back scattering occurs at the junctions, it is not clear whether inelastic edge state-to-bulk states scattering also occurs at the junctions. As shown below, Hall measurements indicate that this does not occur. This means that the edge state network is decoupled from the bulk network; the two only interact at the metal contacts. Consequently, an applied current $I_0$ is divided between the edge, $I_e$, and the bulk $I_b$:

$I_0=I_e+I_b$ ;
$I_e=I_0G_{edge}(G_{edge}+G_{bulk})^{-1}$. (2)

Since edge state is a quantized non-degenerate ballistic graphene state, conventional theory predicts that it should generate a Hall voltage $V_{He}=\pm I_eR_0\text{sgn}(B)$,[48, 82] where the sign depends on whether the edge state carrier is an electron or a hole. On the other hand, the diffusive bulk Hall voltage is $V_{Hb}=I_bB/ne$. As shown in SI8 the measured Hall voltage $R_{Hall}^m$ is easily calculated. Summarizing, $V_{He}$ and $V_{Hb}$ drive a circulating Hall current along the transverse arm, following the bulk, from the bottom to the top contact, and back, following the edge state. Consequently, for a cross with arms of equal length $L$ we expect that $R_{Hall}^m$ is given by Eq. 3a (see SI8 for details).

$$R_{Hall}^m = R_{Hb} + R_{He} = \frac{B}{ne}(R_b/R_e + 1)^{-2} + R_0(R_e/R_b + 1)^{-2} \quad \text{(a)}$$
$$R_{Hall}^m = \frac{B}{ne}(R_b/R_e + 1)^{-2} \quad (R_{He} = 0) \quad \text{(b)} \qquad (3)$$

where $R_e$=$R_0(1+L/\lambda)$ and $R_b$=$L/(\sigma W)$=$L/(ne\mu W)$ are the edge state and bulk resistances of the arms. Equation 3b corresponds to an edge state with no Hall effect ($R_{He}$=0).

Figure 7b plots $R_{Hall}/(R_0B)$, the Hall resistance in units of $R_0$, versus $V_G$, which should be independent of $B$ in the diffusive limit. Indeed, for $B$<0.5 T, we find that $R_{Hall}/(R_0B)$ is independent of $B$ and accurately follows the diffusive limit for $V_G$>0.4 V. However, it saturates at ≈0.2 $T^{-1}$ for $V_G$=0.18 V and reduces to 0 at CNP. Figure 7b inset shows that Eq. 3a does not agree with the data at all. On the other hand, Eq. 3b (Fig. 7b inset) reproduces all of the observed features. This shows that the edge state does not generate a Hall voltage (see also SI6). While a disordered electron-hole puddle state can produce a vanishing Hall voltage at CNP[47, 53] and occasionally a quasi-quantized conductivity,[14, 47] it does not produce a ballistic state and certainly not one with a 1 $G_0$ conductance. Hence, the edge state is not the conventional graphene subband that was predicted.[47, 48]

Summarizing, we have shown that the edge state is an unconventional one dimensional 1 $G_0$ ballistic state that is insensitive to the gate voltage and does not generate a Hall voltage at any magnetic field.

**The anomalous quantum Hall plateau**

Graphene ribbon theory[38] predicts that the $N$=0 subband (Fig. 1a) in general (not only for zigzag ribbons) is characterized by a flatband over 1/3 of the Brillouin zone and a linearly dispersing band, $N_{0Dis}$. The flatband produces a large density of states at $E$=0 at the ribbon edges, which has been experimentally observed in chemically produced ribbons,[60] in sidewall ribbons,[43] and again here in nonpolar epigraphene ribbons (Fig. 2f), however it is not seen in patterned exfoliated graphene, which correspondingly does not have an edge state (SI1). The graphene edge state is predicted to develop from the flatband ($N_{0Fl}$), while a 2 $G_0$ quantum Hall state[48, 82] corresponding to Landau level $LL_0$ is expected to evolve from $N_{0Dis}$.

Both the edge state and the bulk magnetoconductance are substantially insensitive to $B$ (note that $B\mu_{Bulk}$<1), therefore only $N_{0Dis}(V_G,B)$ can have a significant $B$ dependence. Since $G^{0Dis}$ is diffusive (see above), it is small at $B$=0 so that $G^{0Dis}(V_G,B)=G(V_G,B)-G(V_G,B=0)$. Figure 8a plots $G^{0Dis}(V_G,B)$ vs $V_G$ for various $B$ for Seg. **H**. Note that $G^{0Dis}{}_{H}(V_G,B)$ appears to saturate near 1.6 $G_0$ (solid black line). Likewise, Fig. 8b plots $2G^{0Dis}{}_{EH}(V_G,B)$ for Seg. (**H**+**E**) (the factor 2 helps to compare the conductance of the two segments in series, $G_{H+E}$, with that of the single segment, $G_H$, see also Fig. 5a). It saturates near 1.8 $G_0$. Both cases are consistent with the $2\,G_0$ conductance of the zeroth Landau level $LL_0$ in the quantum Hall regime, thereby demonstrating that $N_{0Dis}$ causes the conductance bumps (see also SI10) The associated anomalous quantum Hall plateau (SI6) is explained below.

If we assume that the edge state also shunts the quantum Hall voltage, then the calculation leading to Eq. 3b (SI8) also predicts that the measured Hall resistance will be:

$$R_{Hall}^{Pred} = R^{0Dis}\left(R^{0Dis}/R_0 + 1\right)^{-2} \qquad \text{Eq. 4}$$

Where $R^{0Dis}$=1/$G^{0Dis}$. In fact, $R_{Hall}^{Pred}$, calculated from Eq.4 using the experimental $G^{0Dis}{}_{\boldsymbol{H}}(V_G,B)$ and $G^{0Dis}{}_{\boldsymbol{EH}}(V_G,B)$ shown in Fig. 8c, 8d are remarkably similar and to the measured $R_{Hall}$ (Fig. 7a), which verifies Eq. 4. Hence we conclude that the quantum Hall-like plateau is the graphene $R_0$/2 quantum plateau that is shunted by the edge state (see also SI6).

**Pinning of the edge state at E=0**

In neutral graphene ribbons the 0-DoS peak is half filled with $n_0$≈5x10$^8$ empty states per cm$^2$ at $E=0$ at the graphene edge.[20] Conventional condensed matter principles[52] dictate that the 0-DoS peak will pin $E_F$ at $E_F=0$. As long as the 0-DoS peak is not saturated, it will deplete gate-induced charges near the edge as theoretically verified[51] in simulations of 5-10 nm wide ribbons with a 2 nm high-K dielectric top gate. to produce a Schottky barrier. The resulting electric fields bend the bands near the edge so that in the bulk of the ribbon the Dirac point is at $|E_D| = \hbar v_F\sqrt{\pi n}$ (Figs. 1b, d).[47, 82] The effective width of the depletion region is about half the dielectric thickness d=30 nm (Refs[51], Eq. 11). The 0-DoS peak saturates for $n_{max}$>10$^{13}$ cm$^{-2}$ (i.e. for $V_G>e/a_0\varepsilon_0\kappa$=8 V, with $a_0$=2.5 Å the graphene lattice constant) which is beyond our experimental range. Since we concluded that the annealed non-zigzag ribbons also terminate with acene atoms,[67] they should have a similar $n_{max}$.

This implies that the flatband pins the Fermi level that remains at E=0 along the entire graphene edge (excluding contacts), independent of $V_G$ for the gate voltages used here. These properties perfectly fit the characteristics of the edge state, which confirms that the edge state is pinned at $E=0$ (i.e. independent of $V_G$) and protected by a Schottky barrier.

An excited 1 $G_0$ edge state $\Delta E$≈10 meV above the ground state is also observed (SI10) which is probably related to the recently observed ≈0.1 eV gap in narrow sidewall ribbons[43] so that, as expected, this energy gap is tuned by the ribbon width.[30, 36, 37, 43, 45]

To qualitatively illustrate the band structure (Fig. 9), we use a staggered potential to produce a gap,[45] onsite potentials near the edge for the electrostatic edge potential, and couple the upper and lower $N=0$ states as for a ferromagnetic edge state.[83] For small $n$, the bulk Hall conductance (Fig.7c dashed line) vanishes at the conduction and valence band edges (Fig. 9 b, d) however, since it is shunted by the edge state, $R_{Hall}$ does not diverge. For large $n$ the bulk conductance dominates, and the Hall resistance converges to the diffusive limit (Fig. 7b dashed line). $N_{0Dis}$ becomes the $LL_0$ quantum Hall-like state in a magnetic field[82] (Fig. 7a, Fig. 8). Mixing with the bulk states (Fig. 9a, e) ultimately disrupts the quantum Hall-like state as $|V_G|$ increases.

**Discussion**

Several observed edge state properties seen here for the first time are consistent with theoretical predictions. However, the observation of a non-degenerate edge state that does not generate a Hall voltage is not predicted. Single-sided edge state transport, as recently proposed for sidewall ribbons[43, 84] is ruled out since both sides of the ribbons participate in the transport. Consequently, since the edge state comes from the flatband at *E=0* that is composed of electrons and holes, it is natural to consider a quasiparticle that is half electron and half hole, where both components follow the same path (even in a magnetic field) but in opposite directions. Interestingly, this quasiparticle is effectively charge neutral so that it does not generate a Hall voltage, but it will carry a current, half carried by the electron and half by the hole. If in addition this edge state is spin polarized as it is in sidewall ribbons, (see SI13), then the proposed quasiparticle will be a spin ½ fermion with no Hall voltage and with a conductance $G$=1 $G_0$, as observed. Note that the unusual quasiparticle described here is technically a spin polarized non-chiral fermion of which the Majorana fermion is an important example. In fact the edge state could be a Majorana fermion or a related new quasiparticle.

The excellent transport properties of the epigraphene edge state properties suggest a wide variety of device architectures, besides ultra-high frequency epigraphene field effect transistors.[85] For example, a graphene nanoribbon supplied with top and side gates is predicted to have impressive switching characteristics[22] (SI14). The spin polarized edge state (SI13)[86] provides an important step toward spintronics applications.[87, 88] Epigraphene ribbon constrictions have non-linear properties that may be modulated with a gate.[89, 90] The excited edge state observed here and in sidewall ribbons may be used in graphene ribbon tunneling transistors.[91] In fact the silicon carbide substrate itself, which is compatible with THz electronics and in which qubits may be realized,[92] may be incorporated in switching schemes.[93] Moreover, like carbon nanotubes[9] and graphene quantum Hall states,[94, 95] the zero-mode nanostructures will be phase coherent at low temperatures,[13] enabling graphene quantum interference devices[9, 95] so that ultimately quantum coherent structures may be realized that could be used for quantum computing.[94] Moreover, silicon-on-epigraphene methods[33] have already been demonstrated which allows integration with silicon electronics.

While more experimental and theoretical work needs to be done, the constellation of properties demonstrated here, on a nanopatternable substrate, shows that nonpolar epigraphene is an ideal platform for essentially dissipationless integrated zero-mode graphene nanoelectronics as originally envisionned[6, 12].

## Methods

*Sample fabrication*

Non-polar wafers were produced in-house from commercial bulk single crystal 4H-SiC rod, by cutting them along directions corresponding to the sidewall facets ($\bar{1}10n$), n=5. The wafers were then CMP polished. Graphene samples were prepared using the Confinement Controlled Sublimation method[25] in a graphite crucible provided with a 0.5 mm hole, under various growth conditions depending on the crucible condition, all producing monolayer epigraphene: Sample S1 (Figs. 3-8) in a 1atm Ar atmosphere at 1550°C for 30 min followed by 1650°C for 2 hours; S2 in 1atm Ar atmosphere at 1500°C for 30 min followed by 1600°C for 1 hour; the sample in Fig. 2b, c, e was grown in vacuum in a face-to-face configuration at 1550°C for 20 min, and the sample of Fig. 2d and SI2 in a 1 atm Ar atmosphere at 1550°C for 30 min followed by 1650°C for 15 min.

*Sample patterning*

Sample S1 (Figs 3-6) was patterned using conventional lithography methods. An alumina protecting layer was first deposited on graphene immediately after annealing in vacuum at 1000°C for 30 min, where Al was first evaporated (0.5 Å/sec) in a ≈5 $10^{-5}$ mb oxygen atmosphere. Hall bars were e-beam patterned with a bilayer MMA/PMMA resist and provided with an alumina coating. The graphene/alumina was dry-etched in $BCl_3$ plasma (ICP) and $O_2$ (RIE) through a PMMA/PPM e-beam patterned mask. Buffered HF was used to provide openings in the alumina for contacts. E-beam evaporated Pd/Au was used for contacts and the gate electrode (See Fig. S4). For sample S2 a Hall bar patterned $Al_2O_3$ coating serves as a mask in a RIE $O_2$ plasma etch followed by stripping $Al_2O_3$ in an Al etchant and a thermal anneal at 1200°C for 15 minutes. Following patterning provides Au/Pd side contacts and top gate (15nm of Al2O3 as dielectric), see Fig. S5.

*Experimental measurements*

Transport measurements were performed in a 1.6-420K cryocooler, provided with a 9 Tesla magnet. Voltages were sequentially measured by eight lock-ins (frequency <21Hz), with low current excitation (from 1 to 10nA). Cryogenic STM images were made in high resolution, AFM/STM[96] at the Néel institute, and at the TICNN using a RHK PanScan Freedom STM. Raman spectra were acquired with a high-resolution confocal Raman microscope system at an excitation wavelength of 532nm. Room temperature ARPES measurements were performed at the CASSIOPEE beam line of the Soleil synchrotron, equipped with a Scienta R4000 analyzer and a modified Peterson PGM monochromator with a resolution E/ΔE=70,000 at 100 eV and 25,000 for lower energies. The 6 axis cryogenic manipulator is motorized. The ARPES sample was prepared ex-situ and cleaned under ultra-high vacuum conditions by flash heating it at 700°C. The infrared (IR) magneto-spectroscopy measurements were carried out in reflection mode using a standard Fourier-transform IR spectroscopy technique (Bruker VERTEX 80v) at liquid helium temperature. The IR light from a Globar source was delivered to the non-polar epigraphene through an evacuated light pipe, and the reflected light was guided to a Si bolometer away from the magnetic field center. All measurements were performed in Faraday geometry with the field applied perpendicular to the graphene.

**Data availability:** Data are fully available upon request to the corresponding authors.

**References**

1. A. Chen, J. Hutchby, V. Zhirnov, G. Bourianoff, Emerging Nanoelectronic Devices, Wiley publisher2014.
2. F. Schwierz, Graphene transistors, Nat Nanotechnol, 5 (2010) 487-496.
3. C.R. Eddy, D.K. Gaskill, Silicon Carbide as a Platform for Power Electronics, Science, 324 (2009) 1398-1400.
4. O. Braun, J. Overbeck, M. El Abbassi, S. Käser, R. Furrer, A. Olziersky, A. Flasby, G. Borin Barin, Q. Sun, R. Darawish, K. Müllen, P. Ruffieux, R. Fasel, I. Shorubalko, M.L. Perrin, M. Calame, Optimized graphene electrodes for contacting graphene nanoribbons, Carbon, 184 (2021) 331-339.
5. H.S. Wang, L. Chen, K. Elibol, L. He, H. Wang, C. Chen, C. Jiang, C. Li, T. Wu, C.X. Cong, T.J. Pennycook, G. Argentero, D. Zhang, K. Watanabe, T. Taniguchi, W. Wei, Q. Yuan, J.C. Meyer, X. Xie, Towards chirality control of graphene nanoribbons embedded in hexagonal boron nitride, Nature Materials, 20 (2021) 202-207.
6. W.A. de Heer, C. Berger, P.N. First, Patterned thin films graphite devices and methods for making the same, US patent US7015142B2 (Provisional filed Jun. 12, 2003). This experimentally supported patent provides the foundation of graphene nanoelectronics in general and expliclty describes seamless zero-mode epigraphene electronics., DOI.
7. S. Frank, P. Poncharal, Z.L. Wang, W.A. de Heer, Carbon nanotube quantum resistors, Science, 280 (1998) 1744-1746.
8. H. Baughman Ray, A. Zakhidov Anvar, A. de Heer Walt, Carbon Nanotubes--the Route Toward Applications, Science, 297 (2002) 787-792.
9. W.J. Liang, M. Bockrath, D. Bozovic, J.H. Hafner, M. Tinkham, H. Park, Fabry-Perot interference in a nanotube electron waveguide, Nature, 411 (2001) 665-669.
10. S.J. Tans, A.R.M. Verschueren, C. Dekker, Room-temperature transistor based on a single carbon nanotube, Nature, 393 (1998) 49.
11. L.-M. Peng, Z. Zhang, C. Qiu, Carbon nanotube digital electronics, Nature Electronics, 2 (2019) 499-505.
12. C. Berger, Z.M. Song, T.B. Li, X.B. Li, A.Y. Ogbazghi, R. Feng, Z.T. Dai, A.N. Marchenkov, E.H. Conrad, P.N. First, W.A. De Heer, Ultrathin Epitaxial Graphite: 2D Electron Gas Properties And a Route Toward Graphene-Based Nanoelectronics, J Phys Chem B, 108 (2004) 19912-19916.
13. C. Berger, Z.M. Song, X.B. Li, X.S. Wu, N. Brown, C. Naud, D. Mayou, T.B. Li, J. Hass, A.N. Marchenkov, E.H. Conrad, P.N. First, W.A. de Heer, Electronic confinement and coherence in patterned epitaxial graphene, Science, 312 (2006) 1191-1196.
14. A.K. Geim, K.S. Novoselov, The rise of graphene, Nature Materials, 6 (2007) 183.
15. M.Y. Han, B. O¨ zyilmaz, Y. Zhang, P. Kim, Energy Band-Gap Engineering of Graphene Nanoribbons, Physical Review Letters, 98 (2007) 206805
16. C. Stampfer, J. Gutttinger, S. Hellmueller, F. Molitor, K. Ensslin, T. Ihn, Energy Gaps in Etched Graphene Nanoribbons, Physical Review Letters, 102 (2009) 056403.
17. P. Gallagher, K. Todd, D. Goldhaber-Gordon, Disorder-induced gap behavior in graphene nanoribbons, Physical Review B, 81 (2010) 115409.
18. A. Epping, C. Volk, F. Buckstegge, K. Watanabe, T. Taniguchi, C. Stampfer, Insulating State in Low-Disorder Graphene Nanoribbons, physica status solidi (b), 256 (2019) 1900269.

19. K.S. Novoselov, V.I. Fal'ko, L. Colombo, P.R. Gellert, M.G. Schwab, K. Kim, A roadmap for graphene, Nature, 490 (2012) 192-200.
20. M. Fujita, K. Wakabayashi, K. Nakada, K. Kusakabe, Peculiar localized state at zigzag graphite edge, J Phys Soc Jpn, 65 (1996) 1920-1923.
21. K. Nakada, M. Fujita, G. Dresselhaus, M.S. Dresselhaus, Edge state in graphene ribbons: Nanometer size effect and edge shape dependence, Physical Review B, 54 (1996) 17954-17961.
22. K. Wakabayashi, T. Aoki, Electrical conductance of zigzag nanographite ribbons with locally applied gate voltage, Int J Modern Physics B, 16 (2002) 4897-4909.
23. D. Gunlycke, H.M. Lawler, C.T. White, Room-temperature ballistic transport in narrow graphene strips, Physical Review B, 75 (2007) 085418.
24. K.V. Emtsev, A. Bostwick, K. Horn, J. Jobst, G.L. Kellogg, L. Ley, J.L. McChesney, T. Ohta, S.A. Reshanov, J. Rohrl, E. Rotenberg, A.K. Schmid, D. Waldmann, H.B. Weber, T. Seyller, Towards wafer-size graphene layers by atmospheric pressure graphitization of silicon carbide, Nature Materials, 8 (2009) 203-207.
25. W.A. de Heer, C. Berger, M. Ruan, M. Sprinkle, X. Li, Y. Hu, B. Zhang, J. Hankinson, E.H. Conrad, Large Area and Structured Epitaxial Graphene Produced by Confinement Controlled Sublimation of Silicon Carbide, Proc Nat Acad Sci, 108 (2011) 16900-16905.
26. S. Forti, U. Starke, Epitaxial graphene on SiC: from carrier density engineering to quasi-free standing graphene by atomic intercalation, Journal of Physics D-Applied Physics, 47 (2014) 094013.
27. C. Berger, E. Conrad, W.A. de Heer, Epigraphene, in: P.C. G. Chiarotti (Ed.) Physics of Solid Surfaces, Landolt Börstein encyclopedia Springer-Verlag, Germany, 2018, pp. 727-807. ArXiv:1704.00374.
28. W.A. de Heer, C. Berger, X. Wu, P.N. First, E.H. Conrad, X. Li, T. Li, M. Sprinkle, J. Hass, M. L. Sadowski, M. Potemski, G. Martinez, Epitaxial graphene, Solid State Communications, 143 (2007) 92-100.
29. M. Sprinkle, M. Ruan, Y. Hu, J. Hankinson, M. Rubio-Roy, B. Zhang, X. Wu, C. Berger, W.A. de Heer, Scalable templated growth of graphene nanoribbons on SiC, Nat Nanotechnol, 5 (2010) 727-731.
30. J. Baringhaus, M. Ruan, F. Edler, A. Tejeda, M. Sicot, A. Taleb-Ibrahimi, A.P. Li, Z.G. Jiang, E.H. Conrad, C. Berger, C. Tegenkamp, W.A. de Heer, Exceptional ballistic transport in epitaxial graphene nanoribbons, Nature, 506 (2014) 349-354.
31. M. Ruan, Y. Hu, Z. Guo, R. Dong, J. Palmer, J. Hankinson, C. Berger, W.A. de Heer, Epitaxial graphene on silicon carbide: Introduction to structured graphene, MRS Bulletin, 37 (2012) 1138-1147.
32. J.C. Zolper, M. Skowronski, Advances in Silicon Carbide Electronics, MRS Bulletin, 30 (2011) 273-278.
33. R. Dong, Z.L. Guo, J. Palmer, Y.K. Hu, M. Ruan, J. Hankinson, J. Kunc, S.K. Bhattacharya, C. Berger, W.A. de Heer, Wafer Bonding Solution to Epitaxial Graphene-Silicon Integration, Journal of Physics D-Applied Physics, 47 (2014) 094001.
34. C.-M. Zetterling, Integrated circuits in silicon carbide for high-temperature applications, MRS Bulletin, 40 (2015) 431-438.
35. IBM news, https://www.ibm.com/blogs/think/2017/06/5-nanometer-transistors/, DOI (2017).

36. Y.W. Son, M.L. Cohen, S.G. Louie, Energy gaps in graphene nanoribbons, Physical Review Letters, 97 (2006) 216803.
37. L. Yang, C.H. Park, Y.W. Son, M.L. Cohen, S.G. Louie, Quasiparticle energies and band gaps in graphene nanoribbons, Physical Review Letters, 99 (2007) 186801.
38. A.R. Akhmerov, C.W.J. Beenakker, Boundary conditions for Dirac fermions on a terminated honeycomb lattice, Physical Review B, 77 (2008) 085423.
39. D.A. Areshkin, D. Gunlycke, C.T. White, Ballistic transport in graphene nanostrips in the presence of disorder: Importance of edge effects, Nano Lett, 7 (2007) 204-210.
40. K. Wakabayashi, Y. Takane, M. Sigrist, Perfectly conducting channel and universality crossover in disordered graphene nanoribbons, Physical Review Letters, 99 (2007) 036601.
41. M. Ruan, Structured epitaxial graphene for electronics, School of Physics, PhD - Georgia Institute of Technology, Atlanta, 2012.
42. A. De Cecco, V.S. Prudkovskiy, D. Wander, R. Ganguly, C. Berger, W.A. de Heer, H. Courtois, C.B. Winkelmann, Non-Invasive Nanoscale Potentiometry and Ballistic Transport in Epigraphene Nanoribbons, Nano Lett, 20 (2020) 3786-3790.
43. T.T.N. Nguyen, N. de Vries, H. Karakachian, M. Gruschwitz, J. Aprojanz, A.A. Zakharov, C. Polley, T. Balasubramanian, U. Starke, C.F.J. Flipse, C. Tegenkamp, Topological Surface State in Epitaxial Zigzag Graphene Nanoribbons, Nano Lett, 21 (2021) 2876-2882.
44. K. Wakabayashi, Electronic transport properties of nanographite ribbon junctions, Physical Review B, 64 (2001) 125428.
45. J.L. Lado, N. Garcia-Martinez, J. Fernandez-Rossier, Edge states in graphene-like systems, Synthetic Met, 210 (2015) 56-67.
46. L. Brey, H.A. Fertig, Electronic states of graphene nanoribbons studied with the Dirac equation, Physical Review B, 73 (2006) 235411.
47. S. Das Sarma, S. Adam, E.H. Hwang, E. Rossi, Electronic Transport in Two-Dimensional Graphene, Review of Modern Physics, 83 (2011) 407-466.
48. A.H. Castro Neto, F. Guinea, N.M.R. Peres, K.S. Novoselov, A.K. Geim, The electronic properties of graphene, Rev Mod Phys, 81 (2009) 109-162.
49. Y.B. Zhang, Y.W. Tan, H.L. Stormer, P. Kim, Experimental observation of the quantum Hall effect and Berry's phase in graphene, Nature, 438 (2005) 201.
50. A. Aharon-Steinberg, A. Marguerite, D.J. Perello, K. Bagani, T. Holder, Y. Myasoedov, L.S. Levitov, A.K. Geim, E. Zeldov, Long-range nontopological edge currents in charge-neutral graphene, Nature, 593 (2021) 528-534.
51. J. Guo, Y. Yoon, Y. Ouyang, Gate Electrostatics and Quantum Capacitance of Graphene Nanoribbons, Nano Lett, 7 (2007) 1935-1940.
52. W. Harrison, Electronic Structure and the properties of solids, Dover1980.
53. S. Datta, Electronic transport in mesoscopic systems, Cambridge University Press, Cambridge, 1995.
54. A.J. Van Bommel, J.E. Crobeen, A. Van Tooren, LEED and Auger electron observations of the SiC(0001) surface, Surface Science, 48 (1975) 463-472.
55. A. Charrier, A. Coati, T. Argunova, F. Thibaudau, Y. Garreau, R. Pinchaux, I. Forbeaux, J.M. Debever, M. Sauvage-Simkin, J.M. Themlin, Solid-state decomposition of silicon carbide for growing ultra-thin heteroepitaxial graphite films, Journal of Applied Physics, 92 (2002) 2479-2484.

56. A.C. Ferrari, J.C. Meyer, V. Scardaci, C. Casiraghi, M. Lazzeri, F. Mauri, S. Piscanec, D. Jiang, K.S. Novoselov, S. Roth, A.K. Geim, Raman spectrum of graphene and graphene layers, Physical Review Letters, 97 (2006) 187401
57. E. Rollings, G.H. Gweon, S.Y. Zhou, B.S. Mun, J.L. McChesney, B.S. Hussain, A. Fedorov, P.N. First, W.A. de Heer, A. Lanzara, Synthesis and characterization of atomically thin graphite films on a silicon carbide substrate, J Phys Chem Sol, 67 (2006) 2172-2177.
58. A. Bostwick, T. Ohta, T. Seyller, K. Horn, E. Rotenberg, Quasiparticle dynamics in graphene, Nature Physics, 3 (2007) 36-40.
59. M. Sprinkle, D. Siegel, Y. Hu, J. Hicks, A. Tejeda, A. Taleb-Ibrahimi, P. Le Fevre, F. Bertran, S. Vizzini, H. Enriquez, S. Chiang, P. Soukiassian, C. Berger, W.A. de Heer, A. Lanzara, E.H. Conrad, First Direct Observation of a Nearly Ideal Graphene Band Structure, Physical Review Letters, 103 (2009) 226803.
60. C.G. Tao, L.Y. Jiao, O.V. Yazyev, Y.C. Chen, J.J. Feng, X.W. Zhang, R.B. Capaz, J.M. Tour, A. Zettl, S.G. Louie, H.J. Dai, M.F. Crommie, Spatially resolving edge states of chiral graphene nanoribbons, Nature Physics, 7 (2011) 616-620.
61. Y. Hu, The edge states of epitaxial graphene on SiC, PhD dissertation School of Physics, Georgia Tech, 2021.
62. C. Cardinaud, M.-C. Peignon, P.-Y. Tessier, Plasma etching: principles, mechanisms, application to micro- and nano-technologies, Applied Surface Science, 164 (2000) 72-83.
63. H. Ito, T. Kuwahara, K. Kawaguchi, Y. Higuchi, N. Ozawa, M. Kubo, Tight-binding quantum chemical molecular dynamics simulations for the elucidation of chemical reaction dynamics in SiC etching with SF6/O2 plasma, Phys Chem Chem Phys, 18 (2016) 7808-7819.
64. I. Palacio, A. Celis, M.N. Nair, A. Gloter, A. Zobelli, M. Sicot, D. Malterre, M.S. Nevius, W.A. de Heer, C. Berger, E.H. Conrad, A. Taleb-Ibrahimi, A. Tejeda, Atomic Structure of Epitaxial Graphene Sidewall Nanoribbons: Flat Graphene, Miniribbons, and the Confinement Gap, Nano Lett, 15 (2015) 182−189.
65. W. Norimatsu, M. Kusunoki, Growth of graphene from SiC{0001} surfaces and its mechanisms, Semicond Sci Tech, 29 (2014) 064009.
66. G. Nicotra, Q.M. Ramasse, I. Deretzis, A. La Magna, C. Spinella, F. Giannazzo, Delaminated Graphene at Silicon Carbide Facets: Atomic Scale Imaging and Spectroscopy, Acs Nano, 7 (2013) 3045-3052.
67. Z. Sun, J.S. Wu, Open-shell polycyclic aromatic hydrocarbons, J Mater Chem, 22 (2012) 4151-4160.
68. F. Plasser, H. Pasalic, M.H. Gerzabek, F. Libisch, R. Reiter, J. Burgdorfer, T. Muller, R. Shepard, H. Lischka, The Multiradical Character of One- and Two-Dimensional Graphene Nanoribbons, Angew Chem Int Edit, 52 (2013) 2581-2584.
69. S.E. Stein, R.L. Brown, Pi-Electron Properties of Large Condensed Polyaromatic Hydrocarbons, J Am Chem Soc, 109 (1987) 3721-3729.
70. J. Ristein, S. Mammadov, T. Seyller, Origin of Doping in Quasi-Free-Standing Graphene on Silicon Carbide, Physical Review Letters, 108 (2012) 246104.
71. Y. Zhao, P. Cadden-Zimansky, F. Ghahari, P. Kim, Magnetoresistance Measurements of Graphene at the Charge Neutrality Point, Physical Review Letters, 108 (2012) 106804.
72. S. Cho, M.S. Fuhrer, Charge transport and inhomogeneity near the minimum conductivity point in graphene, Physical Review B, 77 (2008) 081402 R

73. Y.W. Tan, Y. Zhang, K. Bolotin, Y. Zhao, S. Adam, E.H. Hwang, S. Das Sarma, H.L. Stormer, P. Kim, Measurement of scattering rate and minimum conductivity in graphene, Physical Review Letters, 99 (2007) 246803.
74. J.H. Chen, C. Jang, S. Adam, M.S. Fuhrer, E.D. Williams, M. Ishigami, Charged-impurity scattering in graphene, Nature Physics, 4 (2008) 377-381.
75. Y.W. Son, M.L. Cohen, S.G. Louie, Half-metallic graphene nanoribbons, Nature, 444 (2006) 347-349.
76. O.V. Yazyev, Emergence of magnetism in graphene materials and nanostructures, Reports on Progress in Physics, 73 (2010) 056501.
77. J. Li, Y.-M. Niquet, C. Delerue, Magnetic-Phase Dependence of the Spin Carrier Mean Free Path in Graphene Nanoribbons, Physical Review Letters, 116 (2016) 236602.
78. L. Yang, M.L. Cohen, S.G. Louie, Magnetic Edge-State Excitons in Zigzag Graphene Nanoribbons, Physical Review Letters, 101 (2008) 186401.
79. J. Wurm, M. Wimmer, İ. Adagideli, K. Richter, H.U. Baranger, Interfaces within graphene nanoribbons, New J Phys, 11 (2009) 095022.
80. R. de Picciotto, H.L. Stormer, L.N. Pfeiffer, K.W. Baldwin, K.W. West, Four-terminal resistance of a ballistic quantum wire, Nature, 411 (2001) 51-54.
81. T. Fang, A. Konar, H.L. Xing, D. Jena, Carrier statistics and quantum capacitance of graphene sheets and ribbons, Appl Phys Lett, 91 (2007) 092109.
82. L. Brey, H.A. Fertig, Edge states and the quantized Hall effect in graphene, Physical Review B, 73 (2006) 195408.
83. J. Jung, A.H. MacDonald, Carrier density and magnetism in graphene zigzag nanoribbons, Physical Review B, 79 (2009) 235433.
84. F.D.M. Haldane, Model for a Quantum Hall Effect without Landau Levels: Condensed-Matter Realization of the "Parity Anomaly", Physical Review Letters, 61 (1988) 2015-2018.
85. Z.L. Guo, R. Dong, P.S. Chakraborty, N. Lourenco, J. Palmer, Y.K. Hu, M. Ruan, J. Hankinson, J. Kunc, J.D. Cressler, C. Berger, W.A. de Heer, Record Maximum Oscillation Frequency in C-Face Epitaxial Graphene Transistors, Nano Lett, 13 (2013) 942-947.
86. J. Hankinson, Spin dependent current injection into epitaxial graphene nanoribbons, School of Physics, PhD - Georgia Institute of Technology, Atlanta, 2015.
87. W. Han, R.K. Kawakami, M. Gmitra, J. Fabian, Graphene spintronics, Nat Nano, 9 (2014) 794-807.
88. B. Dlubak, M.B. Martin, C. Deranlot, B. Servet, S. Xavier, R. Mattana, M. Sprinkle, C. Berger, W.A. De Heer, F. Petroff, A. Anane, P. Seneor, A. Fert, Highly efficient spin transport in epitaxial graphene on SiC, Nature Physics, 8 (2012) 557-561.
89. J. Baringhaus, M. Settnes, J. Aprojanz, S.R. Power, A.P. Jauho, C. Tegenkamp, Electron Interference in Ballistic Graphene Nanoconstrictions, Physical Review Letters, 116 (2016) 186602.
90. J. Palmer, Pre-growth structures for nanoelectronics of EG on SiC, School of Physics - Georgia Inst. Techn, 2014
91. W.S. Hwang, P. Zhao, S.G. Kim, R. Yan, G. Klimeck, A. Seabaugh, S.K. Fullerton-Shirey, H.G. Xing, D. Jena, Room-Temperature Graphene-Nanoribbon Tunneling Field-Effect Transistors, npj 2D Materials and Applications, 3 (2019) 43.

92. E.M.Y. Lee, A. Yu, J.J. de Pablo, G. Galli, Stability and molecular pathways to the formation of spin defects in silicon carbide, Nat Commun, 12 (2021) 6325.
93. M.T. Schlecht, S. Preu, S. Malzer, H.B. Weber, An efficient Terahertz rectifier on the graphene/SiC materials platform, Sci Rep-Uk, 9 (2019) 11205.
94. Y. Ronen, T. Werkmeister, D. Haie Najafabadi, A.T. Pierce, L.E. Anderson, Y.J. Shin, S.Y. Lee, Y.H. Lee, B. Johnson, K. Watanabe, T. Taniguchi, A. Yacoby, P. Kim, Aharonov–Bohm effect in graphene-based Fabry–Pérot quantum Hall interferometers, Nat Nanotechnol, 16 (2021) 563-569.
95. M. Jo, P. Brasseur, A. Assouline, G. Fleury, H.S. Sim, K. Watanabe, T. Taniguchi, W. Dumnernpanich, P. Roche, D.C. Glattli, N. Kumada, F.D. Parmentier, P. Roulleau, Quantum Hall Valley Splitters and a Tunable Mach-Zehnder Interferometer in Graphene, Physical Review Letters, 126 (2021) 146803.
96. T. Quaglio, F. Dahlem, S. Martin, A. Gérardin, C.B. Winkelmann, H. Courtois, A subKelvin scanning probe microscope for the electronic spectroscopy of an individual nano-device, Review of Scientific Instruments, 83 (2012) 123702.

**Acknowledgements**
Financial support was provided by the U. S. National Science Foundation-Division of Electrical, Communications and Cyber Systems (No 1506006) and NSF-Division of Material Research No 1308835. C.B, and V.P acknowledge funding from the European Union grant agreements No. 696656 and No 785219. This work was also made possible by the French American Cultural Exchange council through a Partner University Fund project and a Thomas Jefferson grant and by the Agence Nationale de la Recherche (No ANR-19-CE24-0025). Financial support is acknowledged from the National Natural Science Foundation of China (No 11774255), the Key Project of Natural Science Foundation of Tianjin City (No 17JCZDJC30100), and the Double First-Class Initiative of Tianjin University from the Department of Education in China. WdH and LM thank Prof. Jiajun Li for his unwavering support in creating the TICNN center. The magneto-infrared spectroscopy measurement was supported by the U.S. Department of Energy (grant No. DE-FG02-07ER46451) and performed at the National High Magnetic Field Laboratory, which is supported by NSF Cooperative Agreement No. DMR-1644779 and the State of Florida.
We thank Evangelos Papalazarou as well as François Bertrand and Patrick Lefebvre for their help with the ARPES measurement at the synchrotron Soleil-Cassiopée beam line. Nikolay Cherkashin is thanked for polishing preliminary nonpolar SiC chips, and John Hankinson and Owen Vail for spin transport measurements, Noel Dudeck for Fig S21c, b. WdH thanks M. Crommie, C. Delerue, P.N First, L. Levitov, G. Mele, M. Pustilnik, E. Rossi and O. Yazyev for helpful discussions.

**Authors' contributions**
VP, CB, Yiran H, Yue H, GN, performed transport measurements and device patterning (Figs. 3-8, S2-14, S20), graphene growth (Figs. 2-8, S2-S20) and sample characterization (Figs. 2a, 2b, S15c). LM, KZ, PJ, JZ and CS, fabricated and characterized nonpolar face SiC wafers, performed graphene growth (Figs. 2a-inset, 2f) and high resolution STM and STS measurements (Fig. 2a-inset, 2f); ZJ, YJ and TZ performed the IR spectroscopy (Fig. 2d). CB, VP, and AT performed the ARPES experiments (Figs. 2c, S15a-b); DW, AdC and CW performed STM and STS experiments (Fig. 2e and Fig.SI15e-f). KW provided theoretical support and Figs. 1b, S17 and WdH Fig.9. CB and WdH directed the Atlanta based experiments. LM and WdH directed the TICNN efforts. WdH is primarily responsible for the analysis and interpretation.

**Competing interests**: The authors declare no competing interests.

**Supplementary Information** is available for this paper.

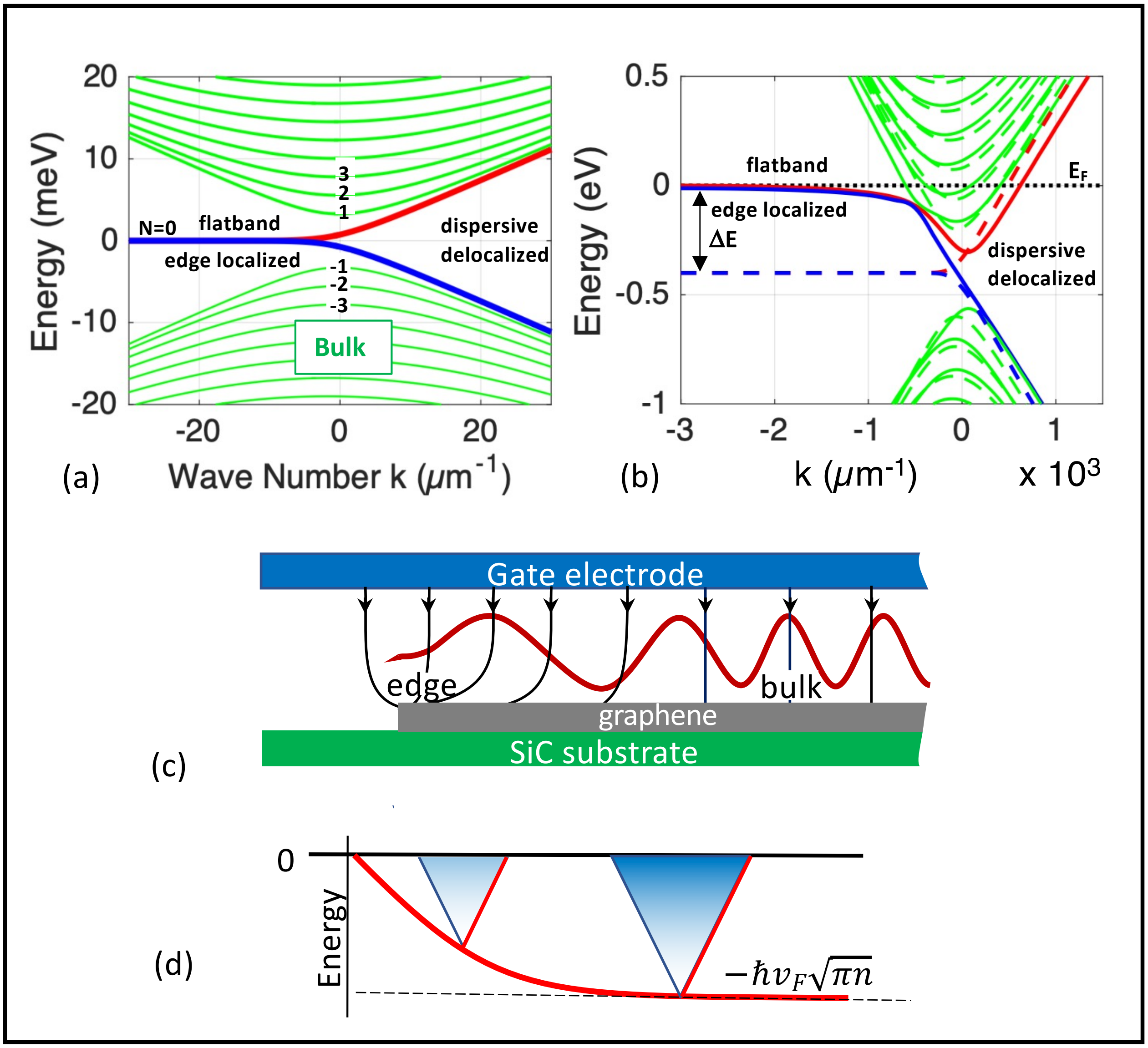

**Fig.1** *The epigraphene edge state*. **(a)** Tight binding band structure of one valley of a 740 nm wide zigzag graphene ribbon that consists of the hyperbolic bulk 1D subbands and the edge state. The edge state is composed of a flat band at E=0 that is localized at the ribbon edge that merges into two delocalized linear dispersing bands. Original theory predicted that the delocalized branches of the *N*=0 subband is a protected edge state for energies between the *N*=±1 bulk subbands.[40] **(b)** Tight binding band structure where the flat band at energy *E*=0 pins the Fermi level $E_F$ at *E*=0 at the edge. Charges induced near the edge will be depleted by the flatband to produce a Schottky barrier between the edge and the bulk. The resulting electric fields cause band bending so that the Dirac point will be at $\hbar v_F\sqrt{\pi n}$ below the Fermi level, as schematically depicted in **(d)**. **(c)** Schematically shows the divergence of the Fermi wavelength at the edge.

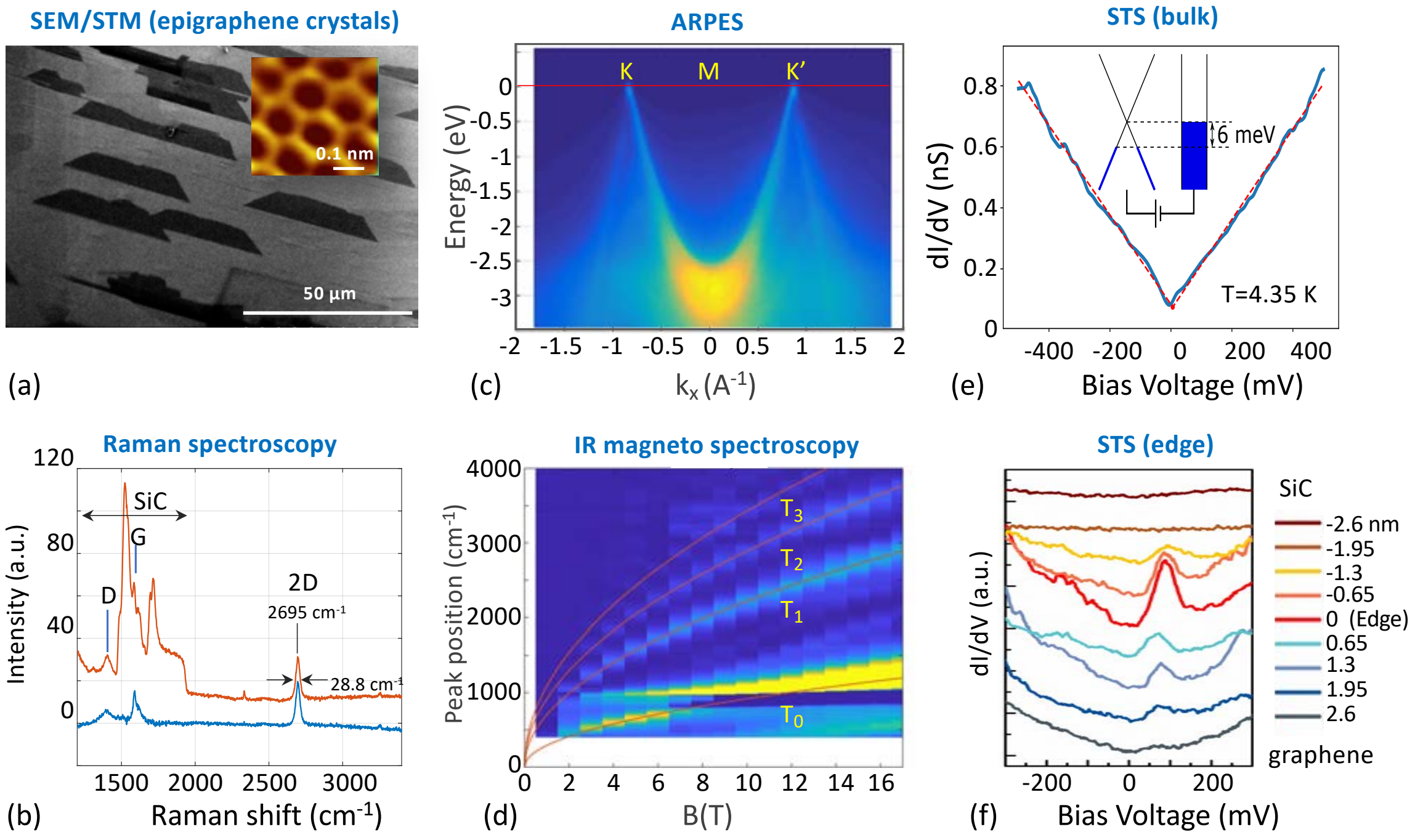

**Fig.2** *Neutral epigraphene characterization*. **(a)** SEM micrograph of trapezoidal graphene islands that form early in the growth and ultimately coalesce to produce a uniform graphene layer. (Inset) STM image of the epigraphene showing the characteristic hexagonal lattice of graphene at $T$=12.5 K. **(b)** Raman spectroscopy. Measured spectrum (red) and SiC subtracted spectrum (blue). The 2D peak is typical of a graphene monolayer. **(c)** ARPES (beam energy=200 eV, $E_F$=197.4 eV) taken at room temperature along K-M-K' showing characteristic graphene Dirac cones with $v_F$=1.06x10$^6$ m/s, with an apex at $E$=0 confirming charge neutrality and no detectable anisotropy. **(d)** Infrared magneto-spectroscopy. The transitions follow the expected characteristic graphene $\sqrt{B}$ dispersion (indicated by the red lines) confirming its monolayer character. **(e)** Typical scanning tunneling spectrum (T=4.4 K, $I_{set}$=400 pA at $V_{bias}$=500 mV), showing the characteristic graphene density of states. A linear fit (dashed lines) indicates a doping level $|E_F$-$E_D|$<6 meV, showing that the graphene is charge neutral. **(f)** STS image at a graphene island edge (T=12.5 K, $I_{set}$=250 pA at $V_{bias}$=2V) taken at various distances from the edge from SiC to inside the ribbon with a lateral resolution of about 2 nm (traces are displaced vertically for clarity). Note the 0-DoS peak at the edge, similar to that observed in sidewall ribbons.[43.]

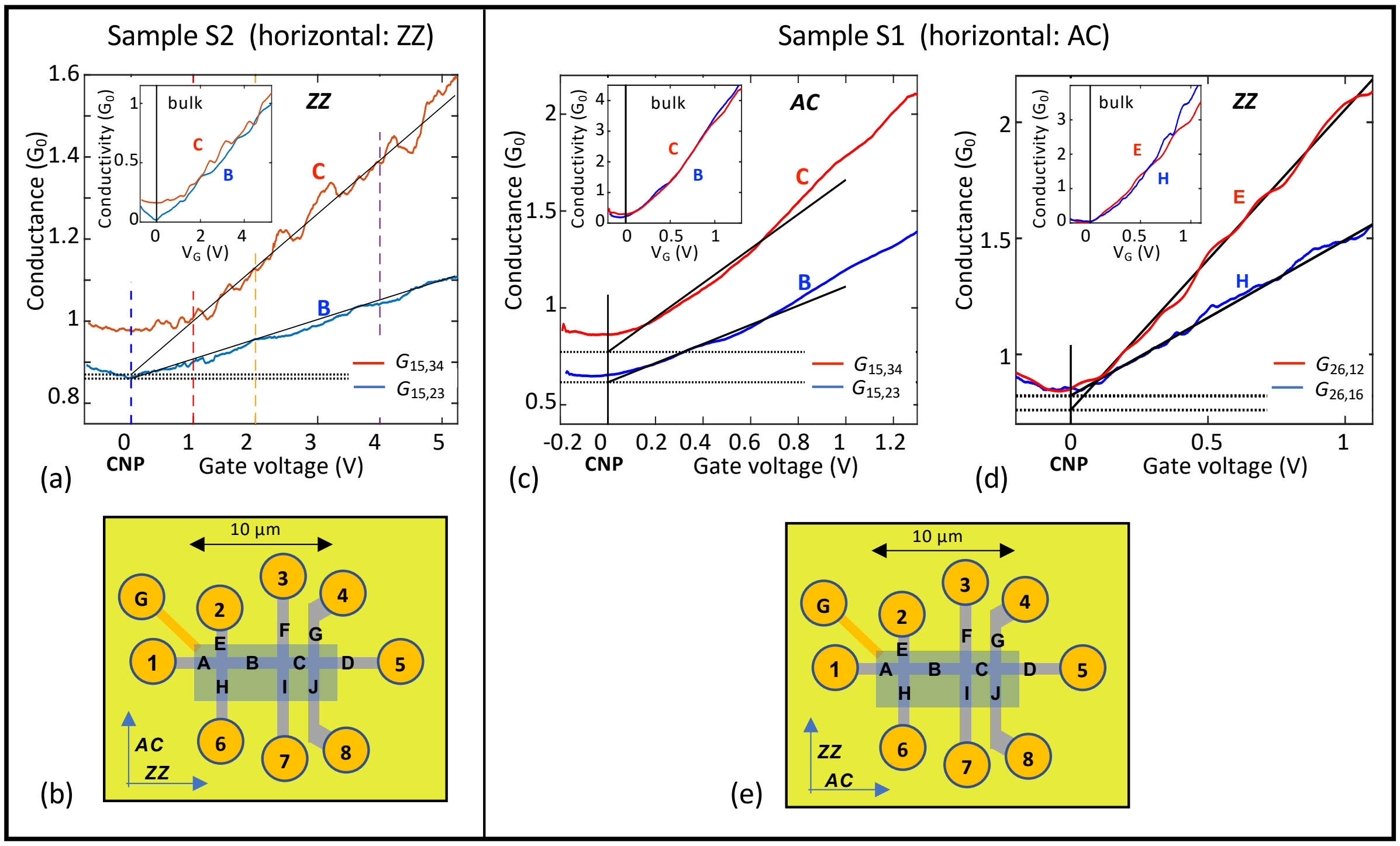

**Fig.3** *Demonstration of the 1 $G_0$ edge state.*

**(a-c)** Sample S1, width=740 nm, axis along AC direction. **(a)** 4-point conductances $G_B$, $G_C$ of segments **B** and **C** (AC segments, $L_B$=3.7 µm, $L_C$=1.7 µm ). The residual conductance $G^{res}$ (i.e. the edge state, dashed lines) is determined from extrapolation of $G(V_G)$ to $V_G$=0 (black lines) from which we find $\lambda_{AC}$=6 µm, $R_J$ =0.08 $R_0$ (see text). Inset shows that $\sigma_{AC}$ for segments **B** and **C** are essentially identical as expected. **(b)** 2-point conductances $G_E$ and $G_H$ (vertical ZZ segments, $L_E$=0.77 µm, $L_H$=1.9µm) $G_E$ and $G_H$ converge at CNP which indicates that $\lambda_{ZZ}$>>segment lengths $L_{E,H}$. Inset indicates that there is no significant anisotropy in $\sigma_{ZZ}$ and $\sigma_{AC}$. See text for discussion of contact resistances and junction resistances **(c)** Schematic diagram of S1; contacts (gold), segments (bold letters), and gate (blue rectangle). Ungated graphene ribbon sections have a charge density $n_0$≈-$10^{12}$ cm$^{-2}$ ($\sigma_0$≈1.5 $G_0$). **(d)-(f)** Sample S2, axis along ZZ orientation, width=1 µm. **(d)** 4-point conductances $G_B$, $G_C$ of segments **B** and **C** ($L_B$=4.5 µm, $L_C$=2.5 µm). **(e)** Schematic diagram of S2. Ungated graphene ribbon sections have a charge density and $n_0$≈-2x$10^{11}$ cm$^{-2}$ ($\sigma_0$≈1 $G_0$).

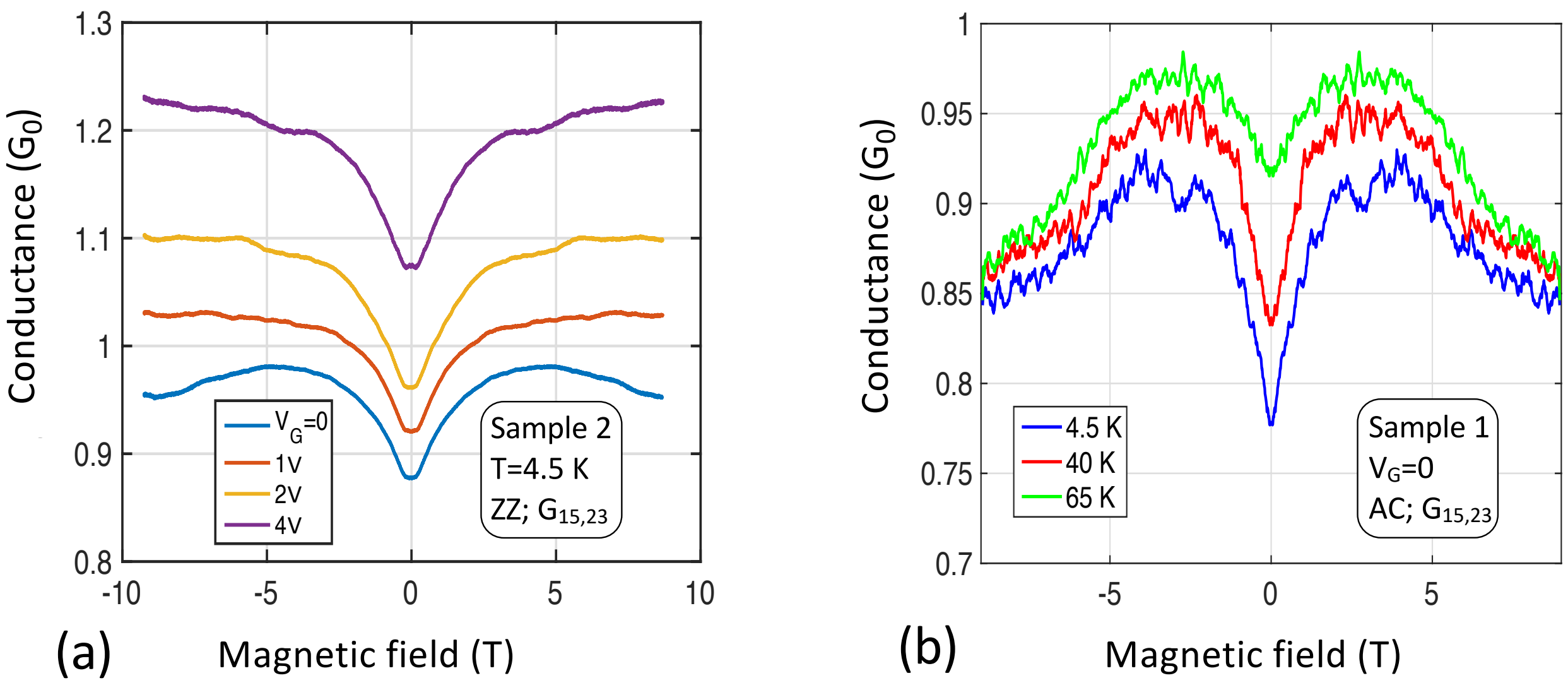

**Fig.4** *Magnetic field effect on the edge state.*

**(a)** Conductance of Sample 2, Seg. **B** versus *B* for 4 values of $V_G$ (Fig. 3d vertical dashed lines ) showing a ≈0.1 $G_0$ dip at *B*=0 that is independent of $V_G$ (similar to Seg, **C** see SI5), indicating that the dip is related to edge state scattering at the junction (see text). **(b)** Conductance of Sample 1, Seg. **B** at CNP for 3 temperatures, which shows that the conductance dip vanishes with increasing temperature. These properties suggest coherent scattering of the edge state at the junctions as explained in the text.

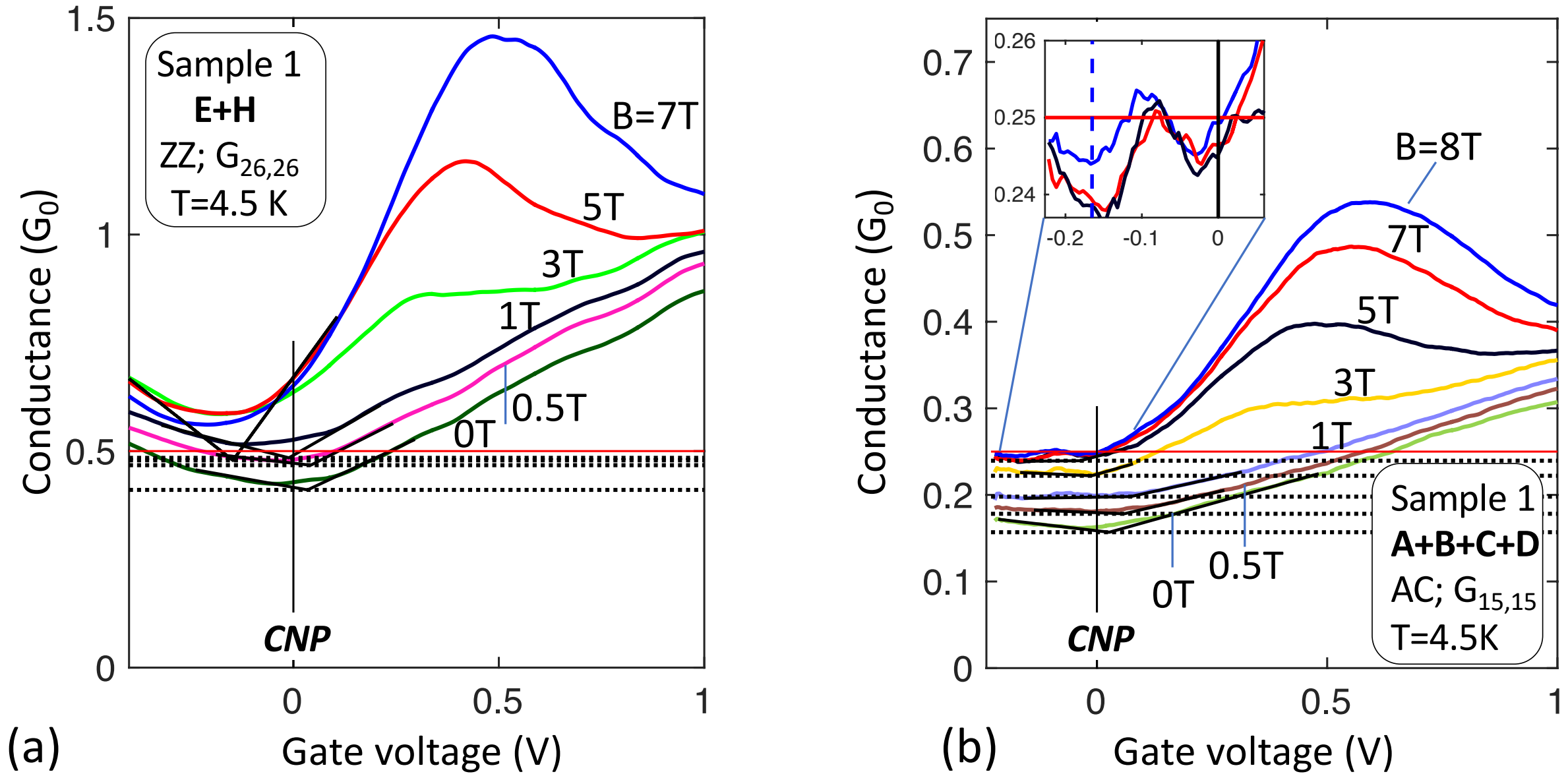

**Fig.5** *Segmentation of the edge state.*

The edge state conductance of a sequence of *N* segments in series approaches $G_0/N$. **(a)** 2-point conductance $G_{EH}(V_G)$, ZZ, for various magnetic fields B. At CNP ($V_G$=0), for large B, $G_{EH}$≈½ $G_0$, is consistent with a ballistic conductor with 1 isotropic scattering center. **(b)** 2-point conductance $G_{ABCD}(V_G)$, AC, for various *B*. At CNP ($V_G$=0) the conductance at large *B* is consistent with 4 conductors (each of conductance ≈1$G_0$) in series, as expected for a 1 $G_0$ ballistic conductor with 3 isotropic scattering centers. In a magnetic field, the conductance increases to ¼ $G_0$ (see inset) indicating essentially perfect quantization, which implies that $\lambda_{AC}$ diverges and $R_J$ vanishes.

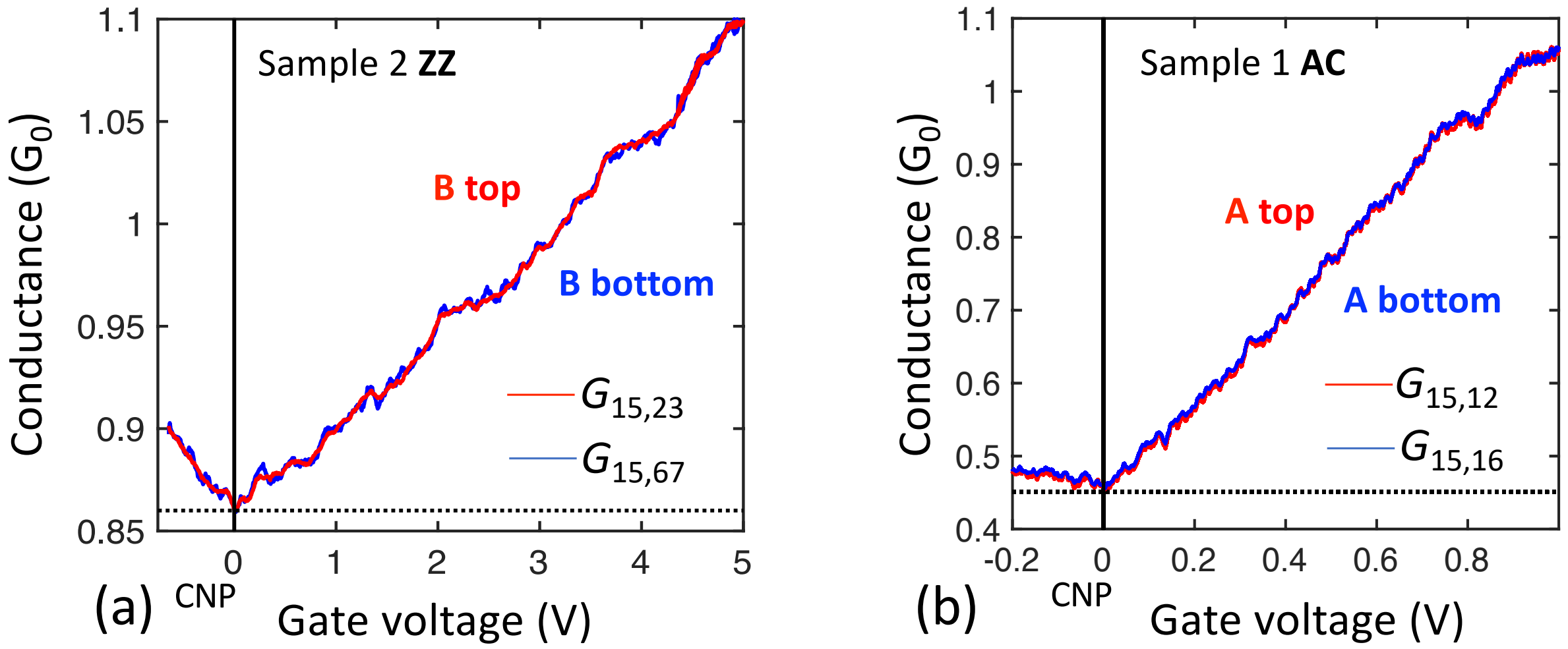

**Fig. 6** *Demonstration of 1D edge state transport.*

**(a)** 4-point conductances of sample 2, Seg. **B** measured from the top ($G_{15,23}$; red) and the bottom ($G_{15,67}$; blue); **(b)** 3-point conductances of sample 1, Seg. **B** measured from the top ($G_{15,12}$; red) and the bottom ($G_{15,16}$; blue). The perfect overlap shows in the two cases CNP shows that is consistent with a 1D edge state that involves both edges coherently and not with a top edge that is independent of the bottom edge which would cause the conductances, especially at CNP to be very different (see text).

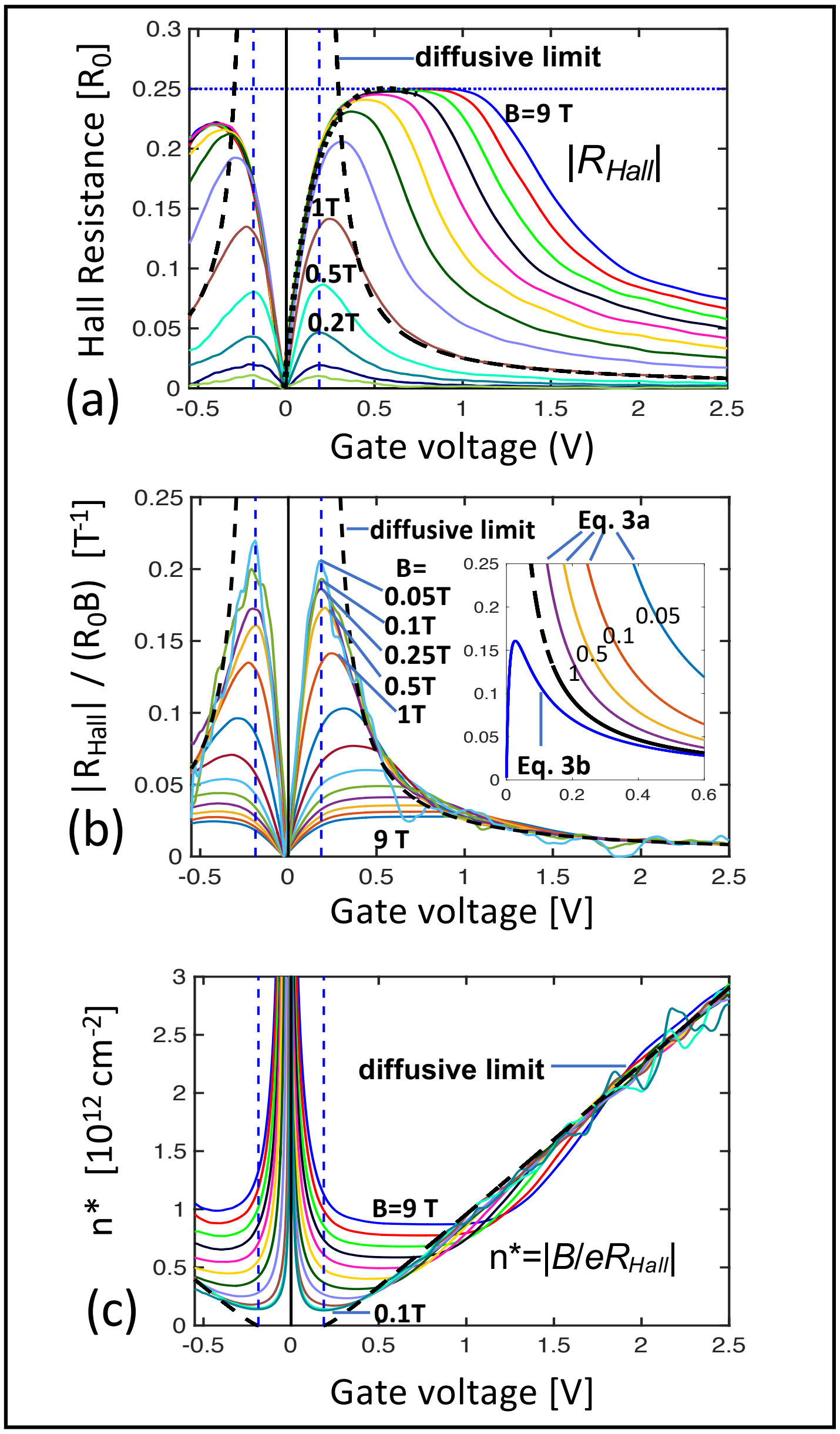

**Fig.7** *Absence of an edge state Hall voltage.*

**(a)** Hall resistance $R_{Hall}$=|$R_{26,15}$| (in units of $R_0$) versus $V_G$ for 0.02 T≤$B$≤9 T shows an anomalous quantum Hall plateau at ≈0.25 $R_0$ for $B$>3 T, consistent with a quantum Hall plateau from the dispersing branch $N$=0 subband that is shorted by the edge state (see Eqs. 3b, 4). The black dashed lines is the diffusive bulk limit for $B$=1 T. **(b)** Normalized Hall resistance |$R_{Hall}/R_0B$| converges to the diffusive bulk limit for $B$<0.5 T and $V_G$>0.4 V. For $V_G$ <0.4 V and $B$<0.1 T it becomes independent of $B$ and saturates at ≈0.2 $T^{-1}$ and then decreases to 0 independent of $B$ as $V_G$ decreases to 0. This behavior is predicted in Eq. 3b (inset); Eq. 3a (inset) does not correspond at all, showing that the edge state does not generate a Hall voltage. **(c)** Effective charge density $n^*$=|$B/eR_{Hall}$|. For small $B$ and/or large $V_G$, $n^*$ converges to the diffusive limit (thick black dashed line) which is corrected for the quantum capacitance. A significant gap $\Delta V_G$=0.17 V is observed which indicates a band gap which however cannot be quantified by this measurement.

**s**

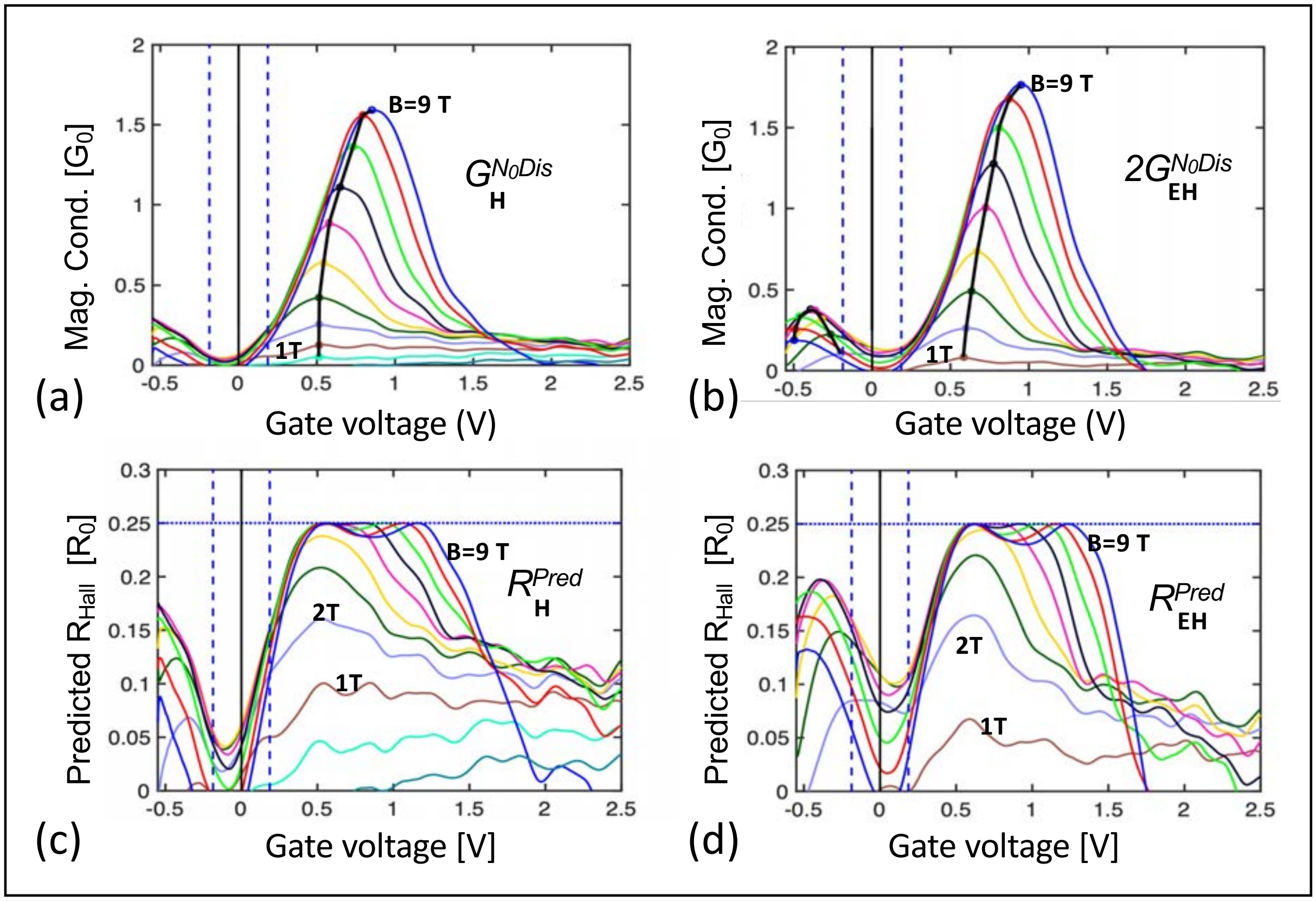

**Fig. 8** *Anomalous quantum Hall plateau*

**(a)** Magnetoconductance of $N_{0Dis}$, in units of $G_0$ for Seg. **H**. **(b)** Twice the magnetoconductance of $N_{0Dis}$ for Seg. (**E+H**), see text. Note the saturation near $2G_0$ as expected for $N_0$ in the quantum Hall regime. **(c, d)** Calculated quantum Hall resistance of Seg **H** and Seg. (**E+H**) from Eq. 4, that predicts the anomalous 0.25 $R_0$ plateau from the magnetoconductance, see text. Note the remarkable resemblance between **(a)** and **(c)**, and **(b)** and **(d)**, which further strongly supports Eq. 4 and sence the vanishing of the edge state Hall voltage.

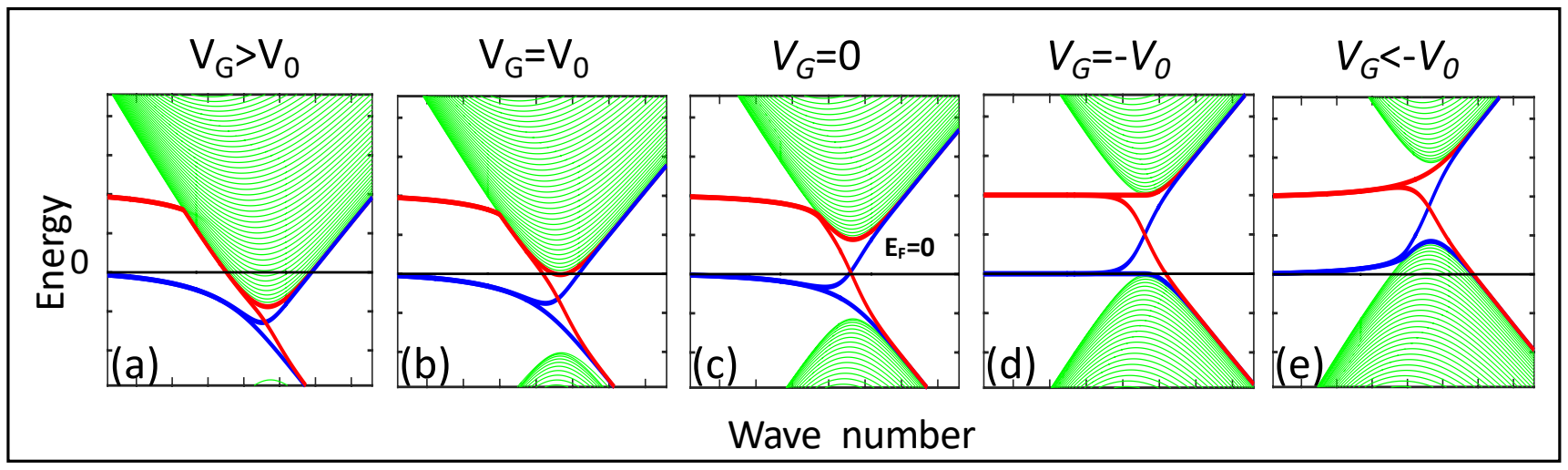

**Fig. 9** *Tight-binding model for gapped graphene with an edge state pinned at E=0 for various gate voltages, $V_G$.*

**(a)-(e)** Conduction and valence band (green) and split edge state (blue and red) for increasing $V_G$**.** For $V_G=V_0$ in **(b),** the Fermi level grazes the bottom of the conduction band, which is where the Hall voltage of the bulk conduction band vanishes. For $V_G=-V_0$ in **(d)** it grazes the top of the conduction band where the bulk valence band Hall voltage vanishes. The Hall voltage of the N=0 subband crosses 0 midway between these two in **(c)**.

# Supplementary Information

## SI1. *Charge puddles and residual conductivity in exfoliated graphene flakes*

**A non-zero conductivity at CNP is observed in many graphene samples which normally manifests as a rounding at CNP: a linear fit beyond the rounding extrapolates to the baseline ($\sigma$=0) as in Fig. S1a. This rounding has been explained in terms of various forms of disorder (eg. charge puddles[1,2]). However, occasionally, a residual conductivity $\sigma_{res}$ is also observed, where the extrapolation is significantly above the baseline (Fig. S1c, e, g). Since a residual conductivity is observed in exfoliated virgin graphene flakes (which have intact edges) but not in lithographically patterned exfoliated flakes the residual conductance is apparently caused by the edges, as it is in epigraphene ribbons, as explicitly demonstrated in the main text.**

In their highly cited paper,[3] Chen, Fuhrer et al. experimentally demonstrated that charge impurity scattering causes the minimum conductivity and provided an empirical equation $\sigma(n)=Ce|n/n_{imp}|+\sigma_{res}$, where n is the charge density, $n_{imp}$ is the impurity concentration and $C=5\text{x}10^{15}\ V^{-1}s^{-1}$. They cite Trushin et al.[4] who predict that charge impurity scattering will give a conductivity $\sigma(n)\approx\sigma_{res}(1+2\pi R^2 n)$, where R is the impurity screening radius and $\sigma_{res}$ is proportional $1/(n_{imp}\ R^2)$. This explains $\sigma_{res}$ but not the rounding.

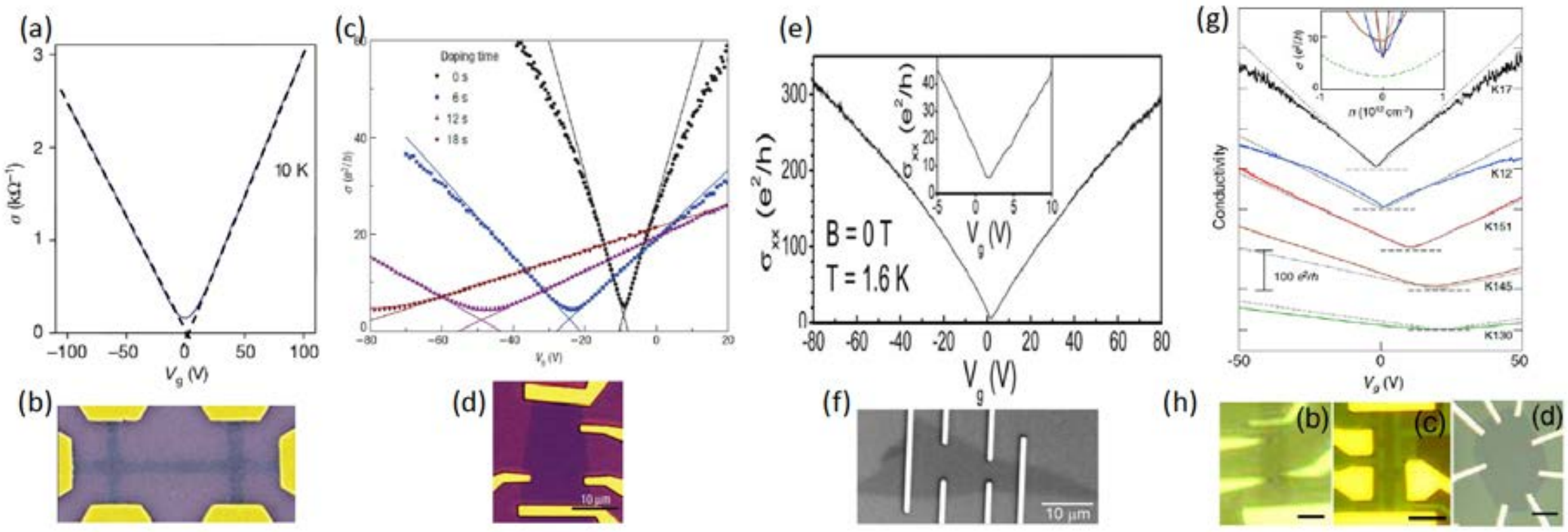

**Figure S1.** Conductivity as a function of gate voltage and corresponding optical images of the measured devices adapted from **(a)(b)** Ref. [5]; **(c)(d)** Ref. [3]: potassium doping of graphene for a pristine sample and three different doping concentrations taken at 20 K in UHV; **(e)(f)** Ref. [6]; **(g)(h)** Ref. [7]: the scale bars in the optical images in (h) represent 1 µm and 10 µm for (b) and (c),(d), respectively.

Cho and Fuhrer[6] concurrently provided compelling evidence that the rounding is due to relatively large P and N charge puddles and stated that $\sigma_{res}$ is not understood (micron scale charge puddles are commonly observed in exfoliated graphene[2] but not in epitaxial graphene, see Fig.S2). They give several possible explanations for $\sigma_{res}$ including experimental artifacts due to the non-ideal sample structures. In fact, in both papers[3,6] and for some of Kim et al's samples[7], the measured devices are unpatterned exfoliated graphene flakes that are supplied with contacts (see Fig. S1d, f, h). In contrast the patterned Hall bar[5] from Novoselov et al. (Fig. S1a-b) does not have a residual conductivity.

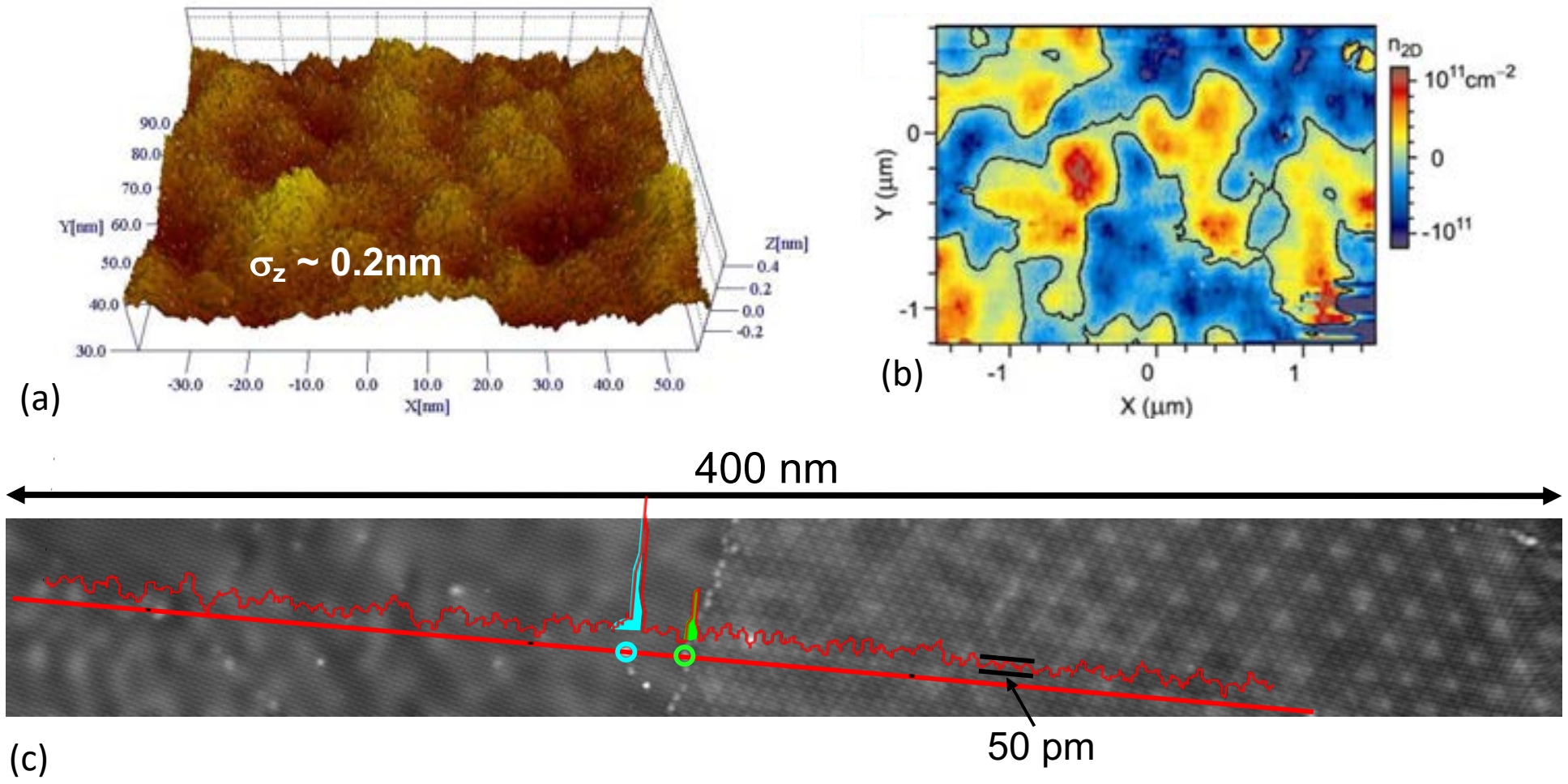

**Figure S2.** Surface characterization. **(a)-(b)** exfoliated graphene flake on $SiO_2$ substrate. **(a)** STM topographic image[8] **(b)** Color map of the spatial density variations in the graphene flake extracted from surface potential measurements at high density and when the average carrier density is zero. Blue regions correspond to holes and red regions to electrons.[2] **(c)** STM image of epigraphene (multilayer graphene on C-face), across a boundary between two different Moiré regions.[9] The layer is atomically flat with a roughness less than 50pm (red line - courtesy Joseph Stroscio).

The conductivity $\sigma$ of a diffusive conductor is derived from the measured conductance G: if the conductance of a ribbon of length L and width W is G, then its conductivity is $\sigma$=GL/W, which is independent of L and W. It has been shown that inhomogeneous graphene (i.e. with charge puddles) and with charged impurities has a non-zero conductivity at the charge neutrality point: $\sigma$(CNP)≠0. Nevertheless the conductance scales with L: $\sigma$(CNP)=G(CNP)L/W for measurements at several lengths L on the same sample. Therefore, ballistic transport at CNP cannot be concluded simply from a non-zero conductance at CNP (even if it appears to be quantized). An unambiguous way to demonstrate a ballistic edge state is to measure conductance G(CNP) at several lengths L to show that $G(CNP)=G_e(1+L/\lambda)^{-1}$. In this way, we have definitively demonstrated that in our samples the residual quantized conductance $G_e$ is due to the edge state and that it cannot be incorporated in the conductivity of the sample because of length scaling.

It may well be that the edge state in fact did contribute to the conductance in many of the early, unpatterned graphene flakes but this possibility was not investigated and a residual *conductivity* was assumed, which had different values on samples produced from the same unpatterned flake but with different geometries[7] which is in fact a sign of a ballistic edge state. On the other hand, lithographically patterned exfoliated graphene flakes like Fig. S1a have disordered edges and therefore have no edge state and no residual conductance so that this problem went away when patterned samples were used.

## SI2. *Corbino ring measurements*

**A Corbino ring is an edgeless device specifically designed to eliminate properties caused by sample edges. Therefore differences in properties observed in a Corbino ring and in a Hall bar can be attributed to the edge state. These measurements show that the residual conductance, the anomalous magnetic properties, and the anomalous temperature dependence observed in the epigraphene ribbons are all due to the edge state, consistent with the conclusions in the main text.**

The Corbino ring was produced using the process below (Fig. S4). It consists of a conducting ring with electrodes contacting its inner and outer rims (Fig S3a-d). Comparing the transport in the Corbino ring and that of the Hall bar, we can further separate the transport properties related to the edge from those of the bulk.

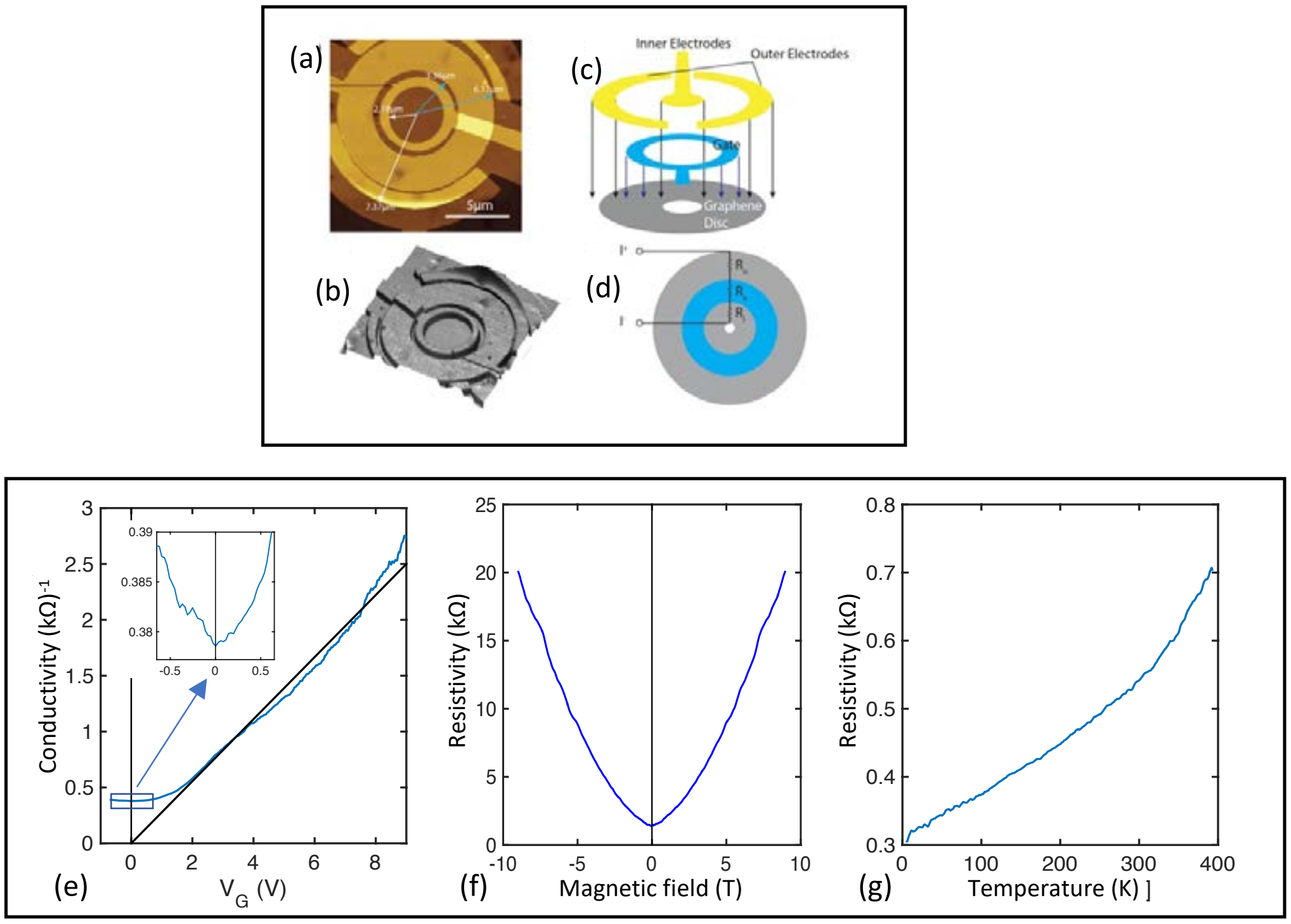

**Figure S3.** Corbino ring. **(a)** AFM image of the Corbino ring. **(b)** 3d rendering of the device. **(c)** Construction diagram showing electrodes (yellow), $Al_2O_3$ dielectric (blue), graphene (gray). **(d)** The Corbino ring is modeled as three resistors in series. **(e)** Conductivity at 4.5K as a function of gate voltage showing a significant non-zero minimum conductivity at CNP, however there is no significant residual conductivity, in contrast to that observed in the S1 and S2 Hall bars where it is due to the edge state. **(f)** A large positive magnetoresistance is observed at 4.5K. In contrast, the S1 and S2 Hall bar resistance decreases with increasing magnetic field. **(g)** The bulk resistivity increases with increasing temperature. In contrast the S1 and S2 Hall bar resistance decreases with increasing temperature (see SI9).

Figure S3e shows the conductivity $\sigma$ versus gate voltage. A minimum conductivity, which properly extrapolates to $\sigma$=0 at CNP, is observed. In contrast to the Hall bar, the Corbino ring does not have a significant residual conductance.

Figure S3f shows the large *positive* magnetoconductance, which is consistent with $\rho=(1+(\mu B)^2)/ne\mu$ with $\mu$=4300 $cm^2V^{-1}s^{-1}$ and n=1.2x$10^{12}$ $cm^{-2}$ at $V_G$=2.5 V. In contrast, the Hall bars S1 and S2 have a large negative magnetoconductance that saturates at about B=3T, which therefore can be attributed to the edge state.

The graphene to Pd/Au contact resistance is 890 $\Omega$, similar to the S1 and S2 Hall bar contact resistance. Similarly to the S1 and S2 Hall bars, charging due to the $Al_2O_3$ dielectric is observed: $n_0 \approx 3 \times 10^{12} cm^{-2}$.

Figure S3g shows the resistivity of the Corbino ring at n=4x$10^{12}$ $cm^{-2}$ as function of temperature, showing an approximately linear increase for T<200 K, followed by a non-linear increase at higher temperatures, consistent with scattering from acoustic phonons.[10] In contrast the resistance of the Hall bars *decreases* with increasing temperature so that this property also can be attributed to the edge state (see SI9).

Hence, these measurements show that the edge state causes the residual conductance, the anomalous magnetic properties, and the anomalous temperature dependence.

**SI3.** ***Device production***

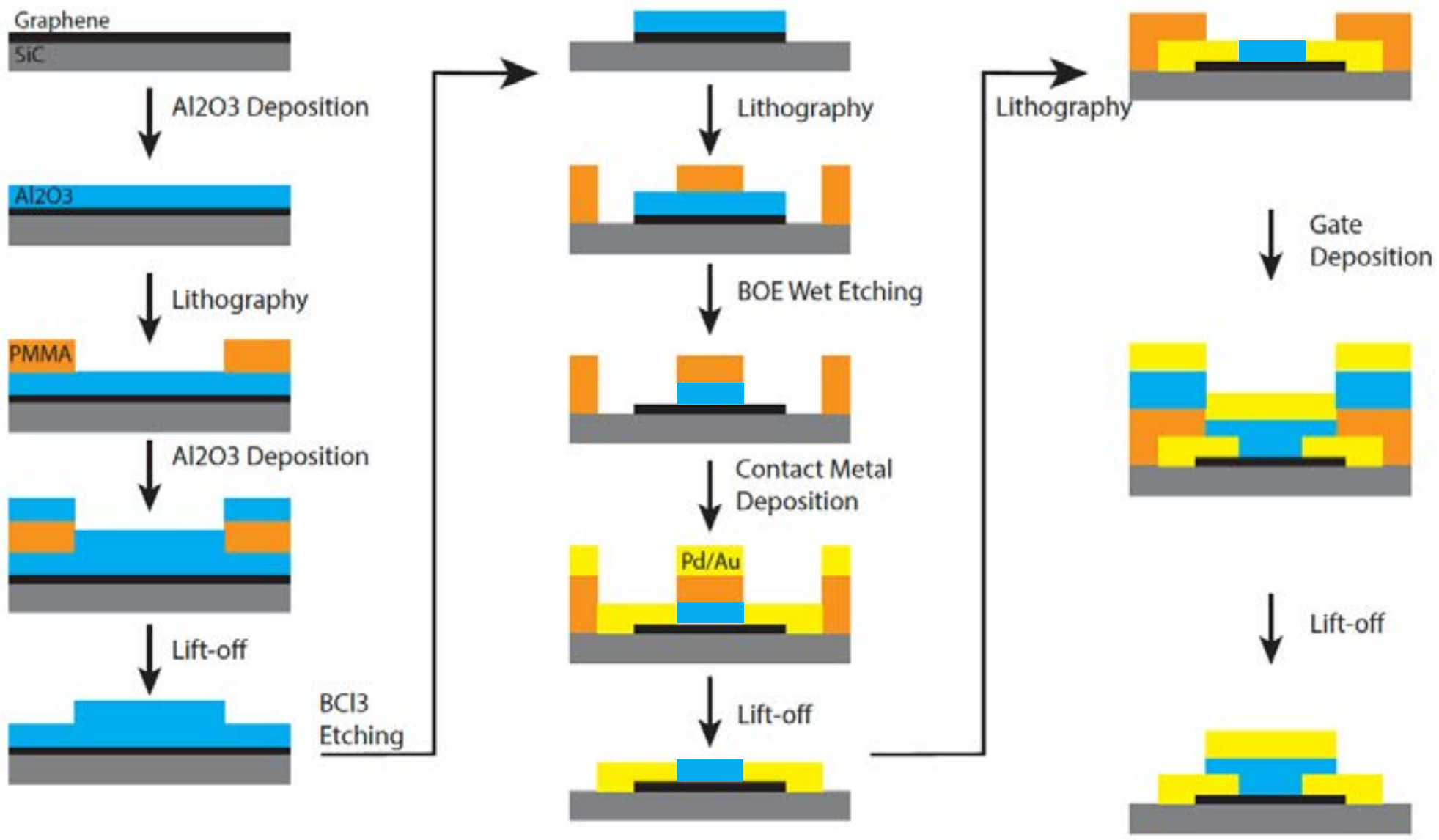

**Figure S4.** Device production process flow Sample S1 (Method 1).

Hall bar S1 was produced using the process flow described in Fig. S4, however the dielectric patterning failed in the first attempt and was removed using hydrogen fluoride (HF), after which it was reapplied. In this process the graphene was subjected to hydrogen fluoride and this Buffered Oxide Etch wet etching significantly reduced the graphene mobility. This turned out to be fortuitous because it greatly enhanced the contrast between the edge state transport and the bulk transport. Moreover, it further demonstrates that the edge state is not affected by disorder that greatly reduces the mobility of the bulk. This immunity of the edge state to processing damage is important for nanoelectronics fabrication.

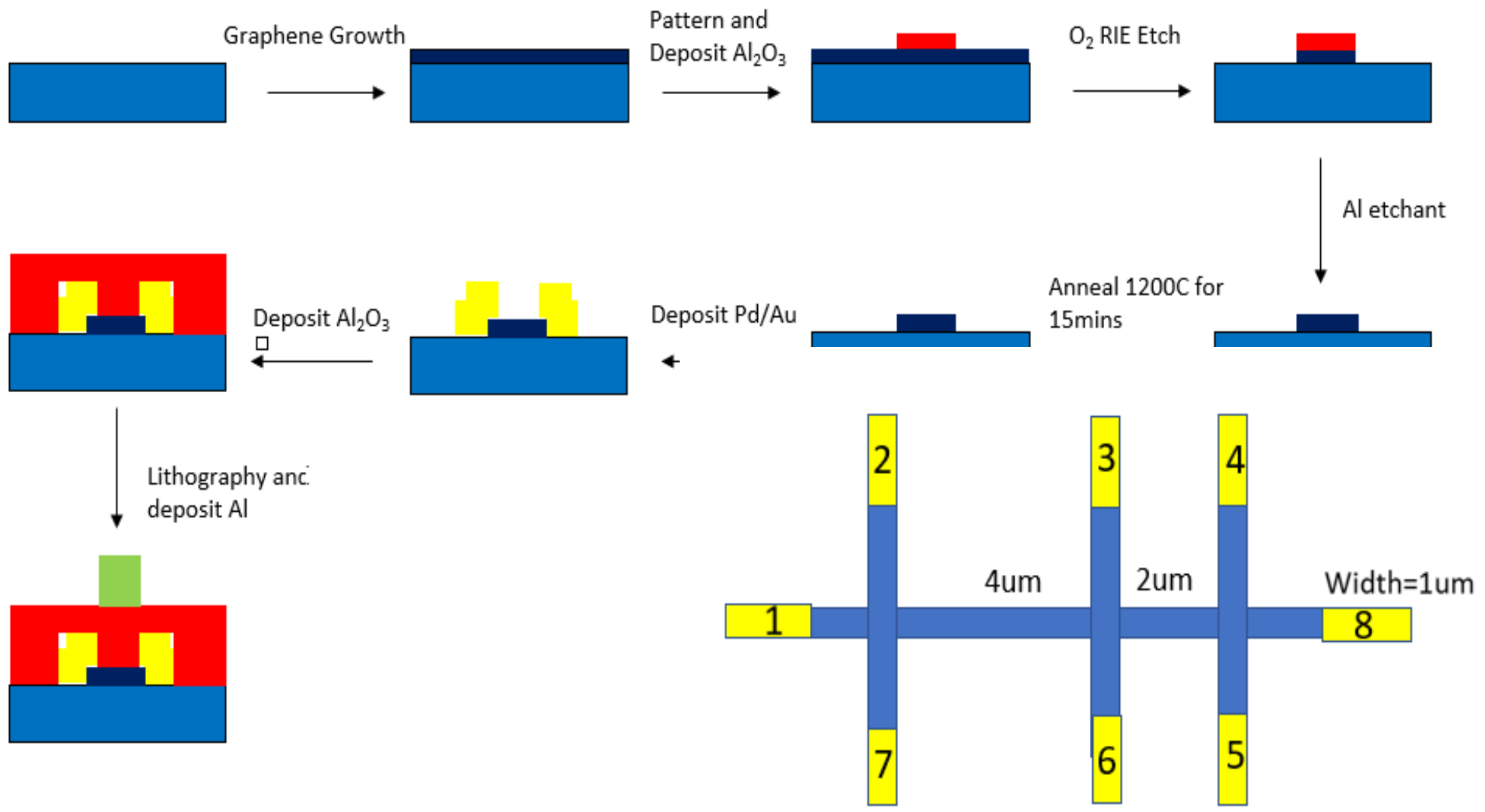

**Figure S5.** Device production process flow for Sample S2 (Method2)

Hall bar S2 was produced using the process flow described in Fig. S5. The process is simplified compared with Sample S1 (Fig. S4). A patterned $Al_2O_3$ layer was used as a mask for the oxygen plasma etch. The mask was removed with Al etchant and the patterned graphene was annealed at 1200°C for 15 min. Pd/Au contacts were then deposited and the $Al_2O_3$ dielectric was deposited followed by the deposition of the Al gate electrode.

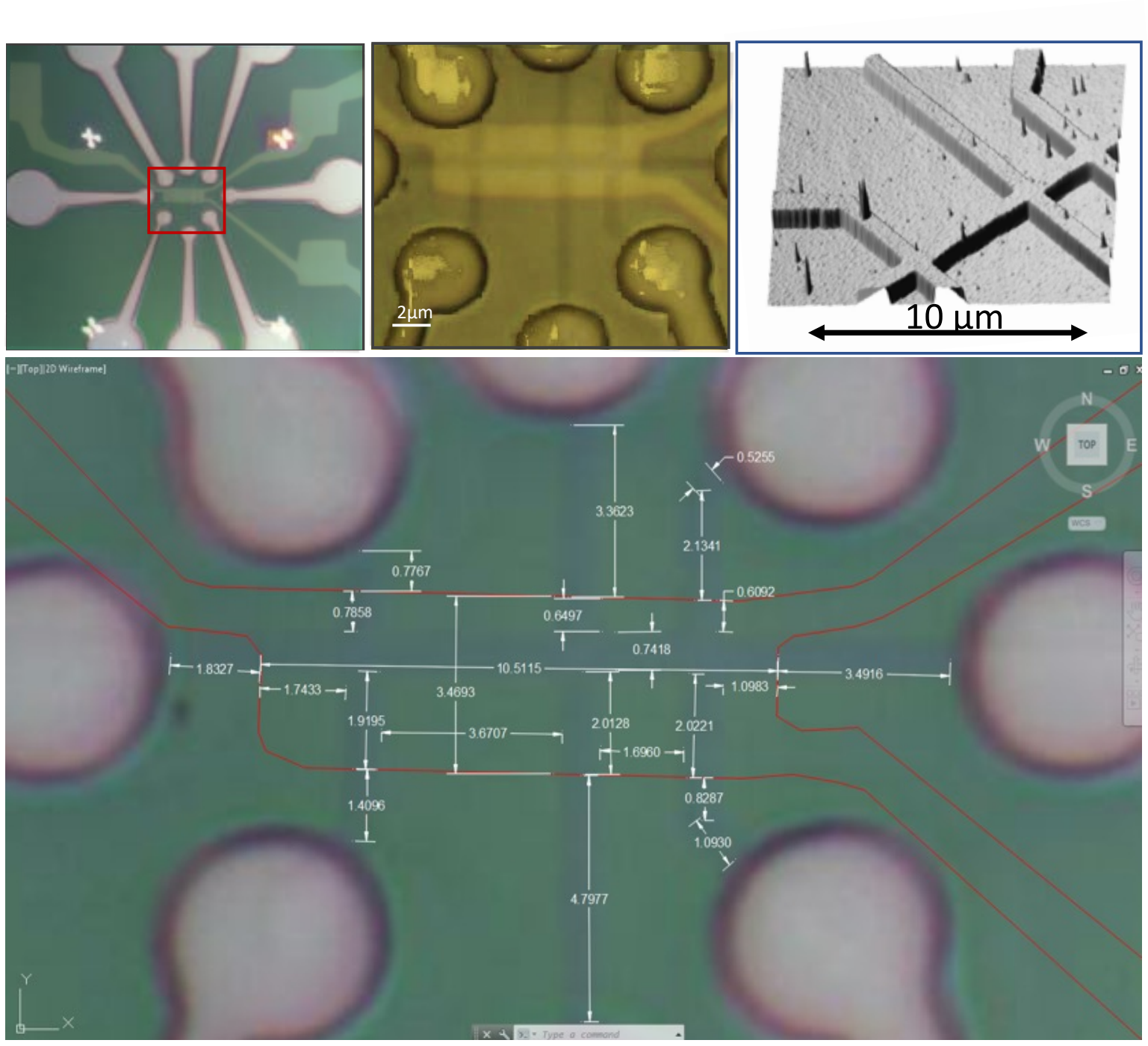

**Figure S6.** *Optical images of the Hall bar S1.* (Top left) Superposition of an image taken before gate deposition, with the resist in place, to outline the geometry of the Hall bar, and an image of the completed device. (Top middle) zoom in of the red square in the left image. (Bottom) Dimension measurements. Red line indicates the perimeter of the gate. (Top right) 3D rendition of the AFM image of sample S1.

## SI4. *Partially gated graphene ribbons*

**Partially gated graphene ribbons show that the edge state conductance is not affected by the bulk charge density step at the interface between gated and ungated portions of the ribbon segment. They also show that the bulk conductivity along the ZZ and AC directions are identical, however that the edge state mfp along the ZZ direction is much larger than along the AC direction.**

Graphene ribbons that are partially gated have been measured using 2-point and 3-point measurements (4-point measurements are for fully gated segments). For these, the conductance of the bulk component consists of two conductors in series: $G_{gated}^{bulk}$ and $G_{ungated}^{bulk}$ corresponding to the measured gated and ungated segment lengths $L_{gated}$ and $L_{ungated}$. Hence the net conductance of a single segment after subtraction of the residual conduction (i.e. the edge state) is

$$G_x^{bulk} = \left(\frac{L_{gated}}{W}\frac{1}{\sigma(n)} + \frac{L_{ungated}}{W}\frac{1}{\sigma(n_0)} + R_{contact}^{bulk}\right)^{-1} \qquad \text{Eq.S1}$$

Where σ(n) is determined from the 4-point measurements and $\sigma(n_0)$ is the ungated conductivity. The edge state conductance is $G_x^{edge}$, see Eq. 2 in main text:

$$G_x^{edge} = 1/\left(R_0(1 + L/\lambda) + R_{junc} + R_c\right) \qquad \text{Eq.S2}$$

So that the total conductance is (Eq. 1)

$$G = G_x^{edge} + G_x^{bulk} \qquad \text{Eq.S3}$$

As shown in Fig. S7, Eq. S1 applied to 2, 3, and 4 point measurements gives excellent agreement for $\sigma(V_G)$ over the entire $V_G$ range, using $\sigma(n_0)$=2.7 $G_0$ ($n_0$=1.2x10$^{12}$cm$^{-2}$) and $R_{contact}^{bulk}$=0.015 $R_0$ (≈400 Ω). However if the residual conductance is not subtracted there is no agreement at all. This again clearly demonstrates that the residual conductance cannot be lumped together with the minimal conductivity, since this combination will not property scale with length. It also shows that the residual conductance does not depend on $V_G$.

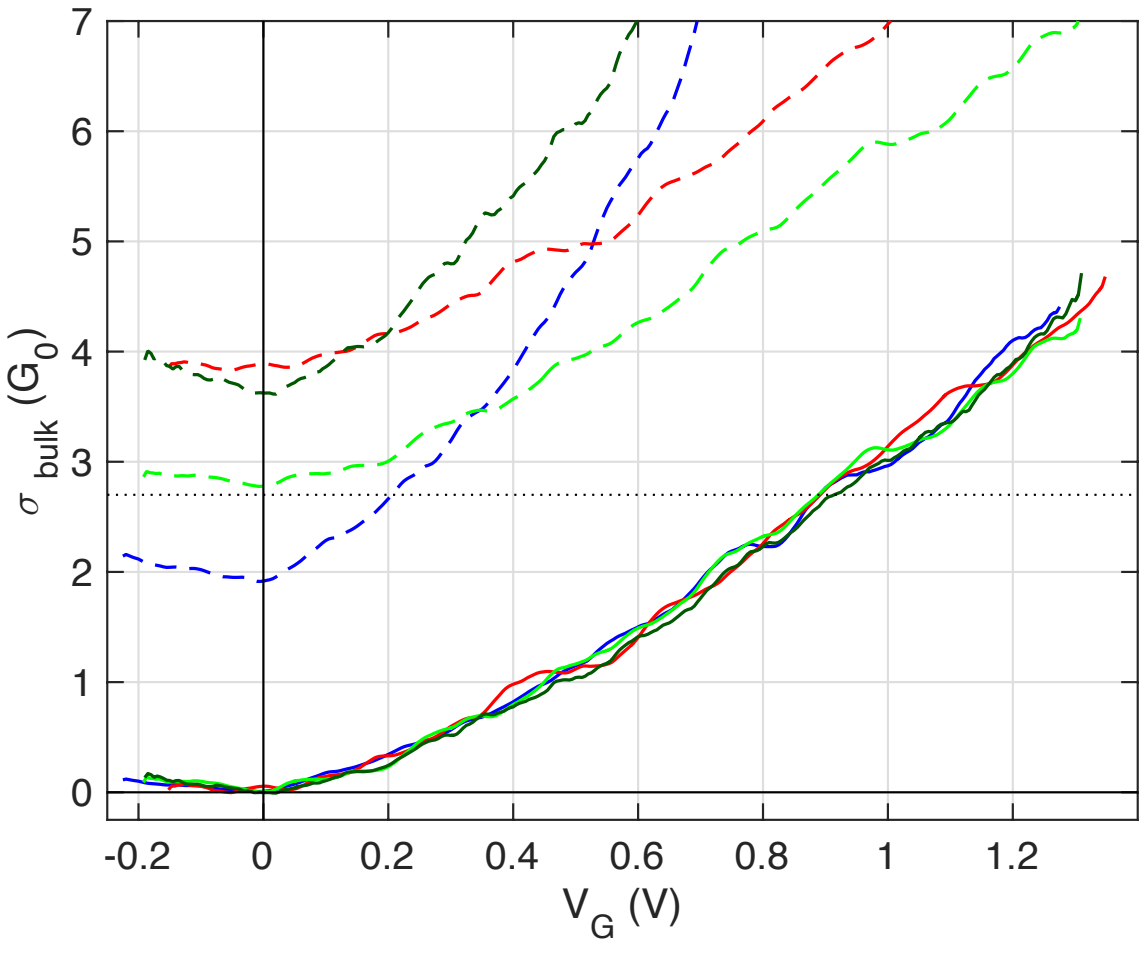

**Figure S7**. Solid lines show bulk conductivity (Eq. S1) after subtraction of the residual conductance, taking into account the ungated lead resistance and contact resistances (4.5K measurements). The conductivities superimpose showing excellent agreement for 2, 3 and 4 point measurements. Seg. **A**: $G_{15,12}$, 3-point (blue); Seg. **B**: $G_{15,23}$, 4-point (red); Seg. **C**: $G_{15,34}$, 4-point (green); Seg. **A**+**B**+**C**+**D**: $G_{15,15}$, 2-point (black). If the residual conductance is not subtracted (dashed lines) then no scaling factor can superimpose them which unambiguously demonstrates that the residual conductance does not scale like a conductivity and that it does not depend on the gate voltage.

These measurements confirm that the bulk conductivity is the same in different ribbon regions, as expected for a homogeneous material; it is explicitly identical along the ZZ and the AC directions: $\sigma_{AC}=\sigma_{ZZ}$. It also shows that the edge state conductance is not affected by the bulk charge density step at the interface between gated and ungated portions of the ribbon segment.

## SI5. *Magnetoconductance dip*

**A broad dip (width≈3 T) in the magnetoconductance at B=0 is observed in epigraphene ribbons and in sidewall ribbons, however not in Corbino ring measurements. The dip vanishes at increased temperatures. The dip is associated with edge state scattering at junctions, contacts and edge defects.**

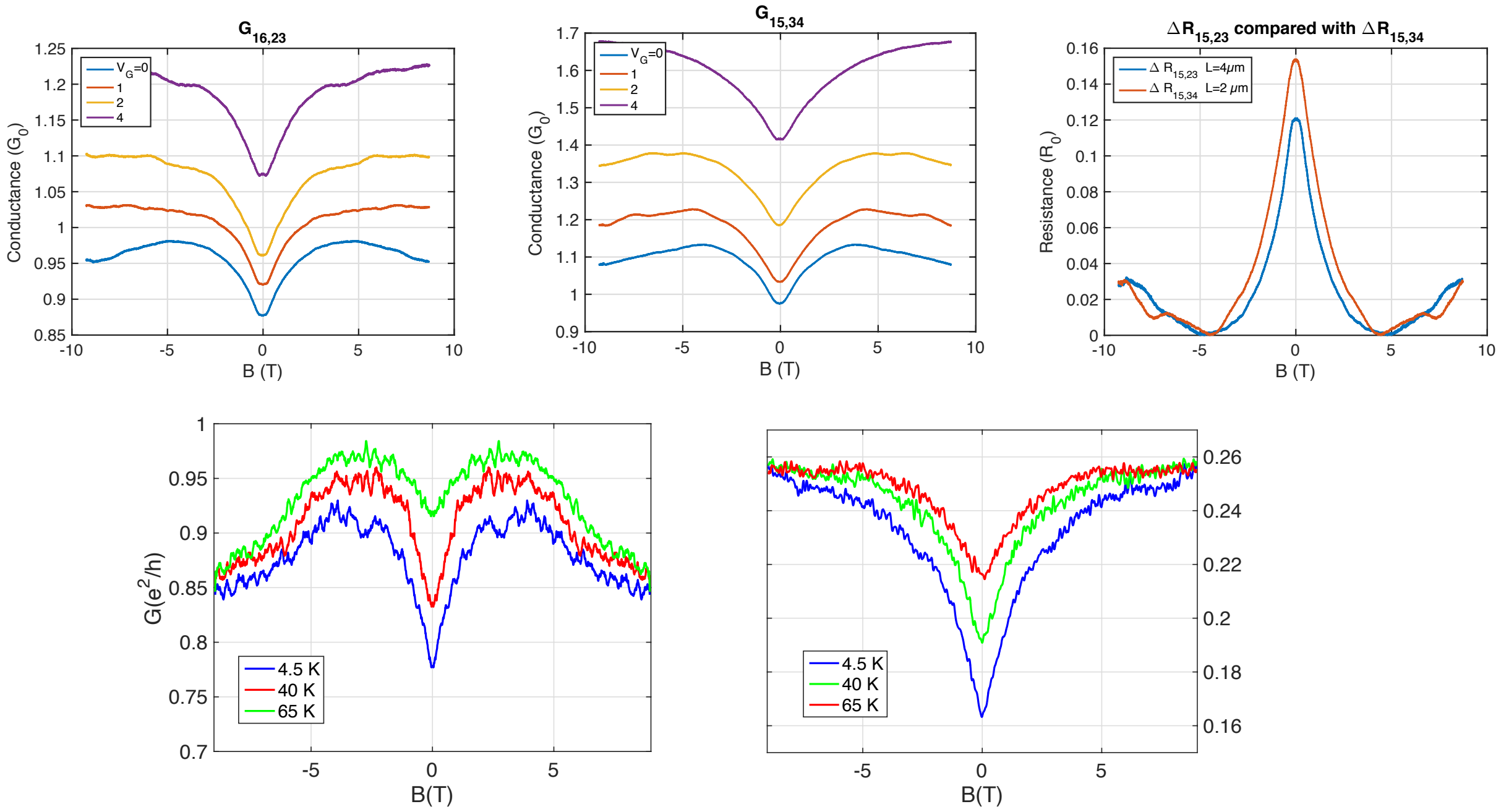

**Figure S8**. The conductance of graphene ribbons with an edge state shows a dip in the magneto-conductance at B=0 and 4.5K (top panels and for various temperature (bottom panels). **Top left panel** is $G_{15,23}$ (Sample S2-Seg. **B**, 4µm, ZZ segment, 4-point), **top middle panel** is $G_{15,34}$ (Sample S2-Seg. **C,** 2 µm long ZZ segment, 4-point) measured for 3 gate voltages. **Top right panel** compares the resistance of the dip $\Delta R=R(B)-R_{min}$ at $V_G=0$ of $R_{15,23}$ and $R_{15,34}$; $\Delta R_{15,23}$ and $\Delta R_{15,34}$ are almost identical, indicating that the dip is primarily caused by scattering of the edge state (as it is observed at CNP) at the junction and not due to scattering along the ribbon because the resistance difference ΔR would scale linearly with the length difference ΔL=2 µm. If the small residual ΔR=0.033 $R_0$ is attributed to ΔL, then ΔR corresponds to a mean free path $\lambda_{ZZ}=R_0\Delta L/\Delta R$=60 µm, which is consistent with other measurements (see main text). **Bottom left panel** is $G_{15,45}$ (Sample-S1, Seg **D**, 4.5 µm, AC direction, symmetrized, 3-point) at CNP; it shows the decrease of the dip with increasing temperature and the saturation near $G=G_0$. All individual segments show similar behavior. **Bottom right panel**, same measurement but for $G_{15,15}$ (Sample S1-Seg. **A+B+C+D**, 15 µm, AC direction, 2-point). The conductance saturates at $G_0/4$, consistent with four 1 $G_0$ ribbon segments in series. Note that the dip amplitude is considerably larger along the AC direction than the ZZ direction, indicating that scattering at AC edge defects and at the junctions is reduced both in a magnetic field and at increased temperatures.

## SI6. *Pseudo-plateaus*

**Non-quantized quantum Hall resistance plateaus are observed close to CNP that saturate at about R≈0.25 $R_0$ for $|V_G|$>0.3 V. The anomalous ≈0.25 $R_0$ quantum Hall plateau and the pseudo plateaus are caused by the shunting effect of the edge state.**

Near CNP, the Hall resistance of sample S1 measured for various gate voltages $V_G$, has non-quantized plateaus that converge to ≈0.25 $R_0$ for $V_G$=0.3 V as shown here for $G_{15,26}$, Sample S1. The pseudo-plateaus converge to $R_H$=0 at CNP, rather than diverging which is caused by the shunting effect of the edge state. The value of the plateau is similarly explained by the shunting effect.

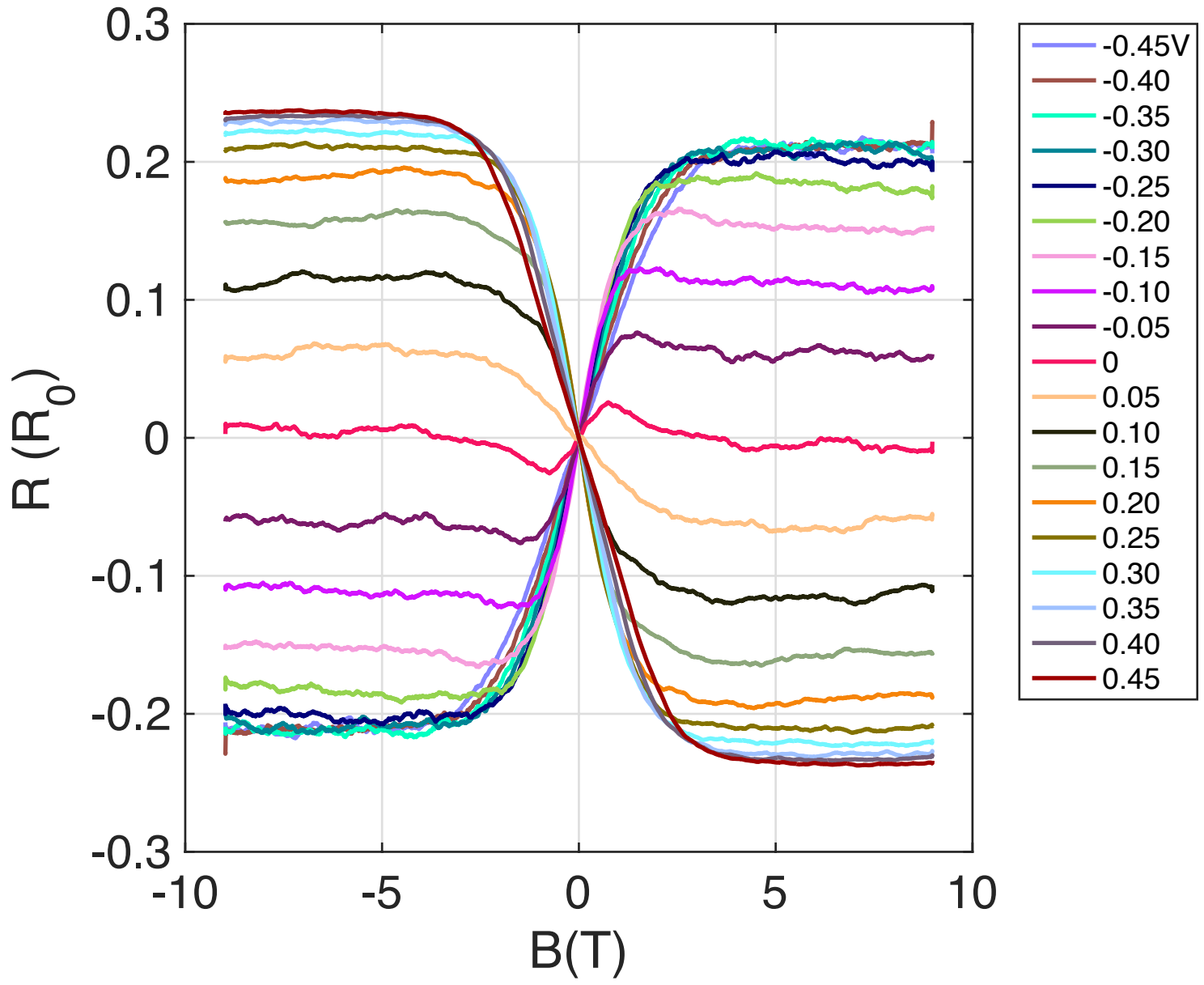

**Figure S9**. Resistance $R_{15,26}$ (Hall measurement configuration at 4.5 K) of sample S1 for various gate voltages, showing non quantized plateaus for B>3T.

**SI7.** ***Schematic diagram of the 1D edge state network***

**The edge state forms a 1D network that can be schematically represented by ballistic segments with a resistance of 1 $R_0$ between contacts and junctions.**

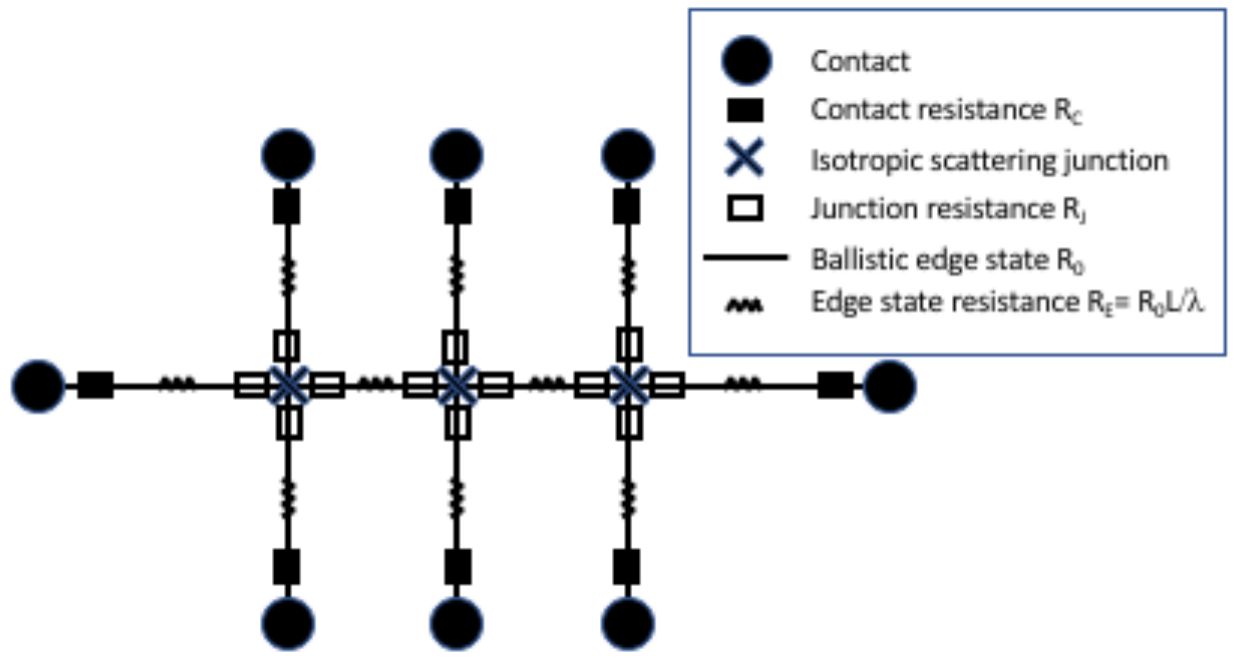

**Figure S10.** Schematic diagram of the 1D edge state network of S1 and S2.

The resistance is increased by the junction resistance $R_J$, the contact resistance $R_C$, and scattering along the edge state, $R_E=R_0L/\lambda$ so that each segment has an effective resistance $R_{Seg}=R_0+R_E+R_J+R_C$. Note that $R_E$ is larger for the AC direction compared with the ZZ direction, whereas defects at segment corners probably cause $R_J$. Both $R_E$ and $R_J$ reduce in a magnetic field and at increased temperatures (see above). The effects are reminiscent of weak localization. The large width of the magnetoconductance dip may be due to the 1d nature of the edge state that cannot enclose a magnetic flux.

**SI8.** ***Two component Hall effect***

**The edge state network and the bulk network are independent and only interact at the contacts. Therefore the measured Hall voltage is a combination of the edge state Hall voltage and the bulk Hall voltage that involve currents in the voltage arms of the Hall bar.**

We have shown that the edge state network and the bulk network are independent and only interact at the contacts. This also applies to the Hall voltage, as shown in Fig. S11, where the bulk network is labeled 1 and the edge state network is labeled 2. A current $I_0$ is applied between contact A and B and is divided between the two networks: $I_{1,2}=I_0G_{1,2}/(G_1+G_2)$ where $G_{1,2}$ are the conductances. These currents generate Hall voltages $V_{1,2}=R_{H1,2}I_{1,2}$ that cause a current i to circulate along the vertical arms of the Hall bar: $i=(V_1-V_2)/(R_1+R_2)$, so that the resulting measured Hall voltage $V_{CD}=V_1-iR_1=I_0R_{H1}(R_1/R_2+1)^{-1}(G_2/G_1+1)^{-1} + I_0R_{H2}(R_2/R_1+1)^{-1}(G_1/G_2+1)^{-1}$. For a diffusive conductor $R_{H1}=B/en$. For a chiral 1 $G_0$ edge state $R_{H2}=R_0$. For the edge state we find that $R_{H2}=0$. For a cross of equal arm lengths $G_1=1/R_1$ and $G_2=1/R_2$, which results in Eq. 3a in the main text; setting $R_{He}=0$ gives Eq. 3b, main text.

$$R^m_{Hall} = R_{Hb} + R_{He} = \frac{B}{ne}(R_b/R_e + 1)^{-2} + R_0(R_e/R_b + 1)^{-2} \quad \text{(Eq. 3a, main text)}$$
$$R^m_{Hall} = \frac{B}{ne}(R_b/R_e + 1)^{-2} \qquad (R_{He} = 0) \quad \text{(Eq.3b, main text)}$$

On the other hand, for a cross with equal arms in the quantum Hall limit of the $LL_0$ state, we expect for the edge state, $R_1=R_e=R_0$; $G_1=G_e=G_0$ and $R_2=1/G_2=R^{0Dis}$ where $G_2$ is the measured 2 point conductance (note that in the quantum Hall regime the Hall resistance equals the two point resistance). With these substitutions and assuming that the Hall voltage of the flatband vanishes, we find Eq. 4 of the main text:

$$R^{Pred}_{Hall} = R^{0Dis}\left(R^{0Dis}/R_0 + 1\right)^{-2} \quad \text{(Eq.4, main text)}$$

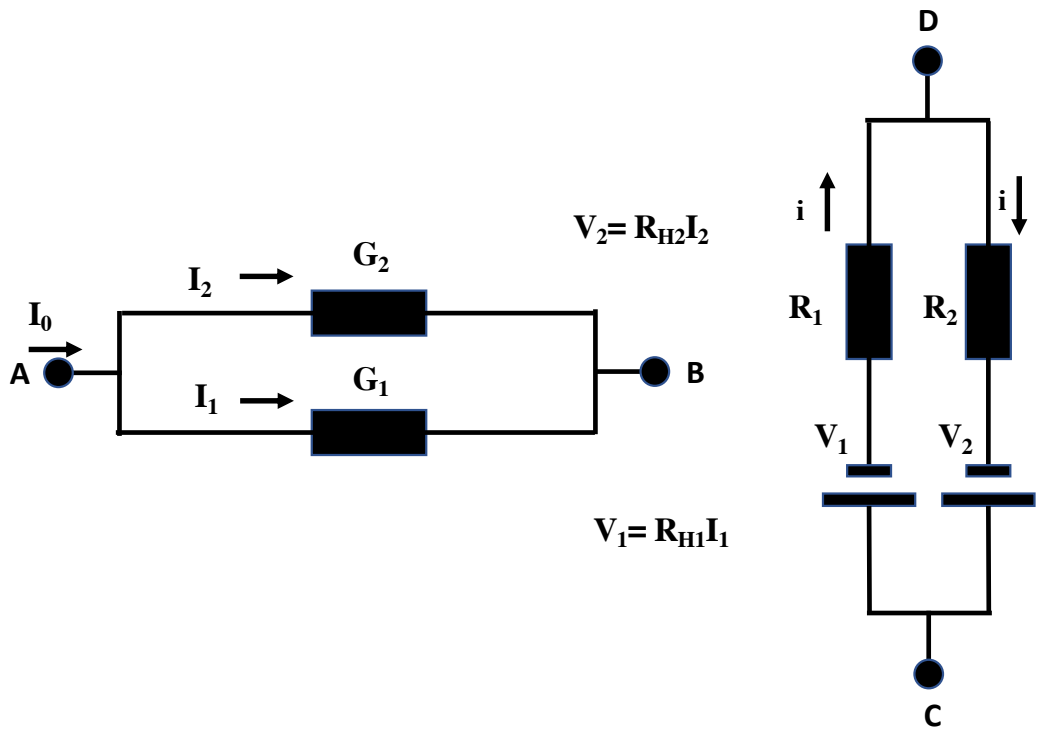

**Figure S11.** Schematic diagram voltages and currents in the interacting edge state and bulk networks for a 4-contact cross A-B-C-D where a current $I_0$ is applied horizontally along AB, generating Hall voltages $V_{1,2}$ in the vertical arms (CD).

## SI9. *Temperature dependence*

**The conductance of epigraphene ribbons *increases* with increasing temperature, rather than *decreasing* as in Corbino ring measurements and for 2D graphene in general, which shows that the conductance increase is an edge state property. In a magnetic field this increase is suppressed (see below).**

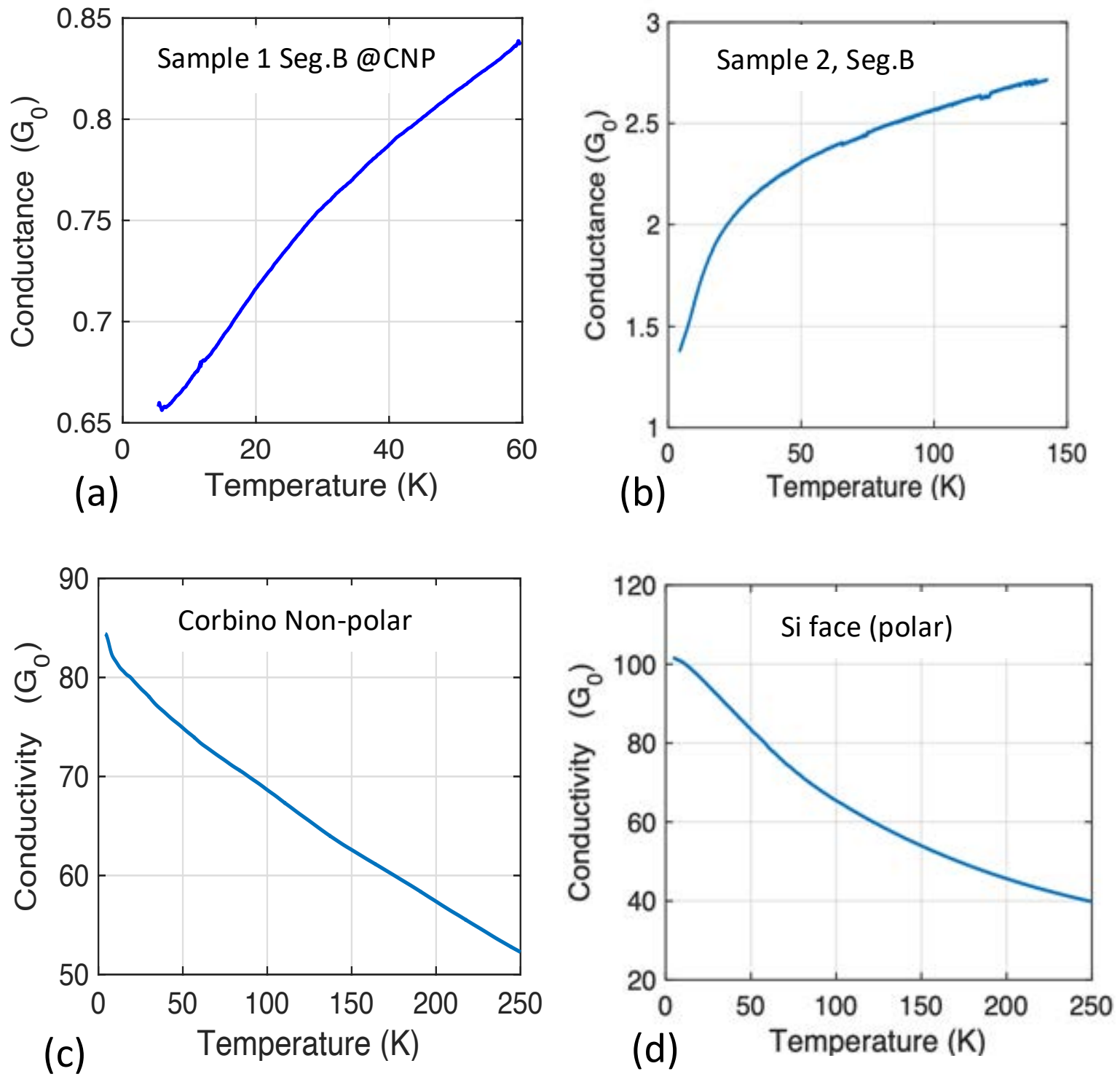

**Figure S12.** Temperature dependences. **(a)** Conductance of Sample 1-Seg. B (L=4 µm) measured at $V_G$=0; **(b)** Conductance of Sample 2-Seg. B (L=4 µm), measured at $V_G$=4V; **(c)** Corbino ring conductivity of non-polar epigraphene; **(d)** conductivity of Si face, polar 2D epigraphene.

## SI10. *High temperatures and the excited edge state*

**High temperature measurements show that the anomalous quantum Hall plateau persists and is consistent with the edge state shunted $R_0/2$ quantum Hall plateau that is derived from the high mobility $N_{0Dis}$ subband. The anomalous temperature and magnetic field dependence is consistent with a 1 $G_0$ excited edge state on the order of 10 meV above the ground state, with ballistic properties that resemble the ground state. This state can only be accessed thermally and not with a gate voltage, which implies that it is located at the edge that is pinned at E=0, making it insensitive to $V_G$. The excited state is suppressed in a magnetic field.**

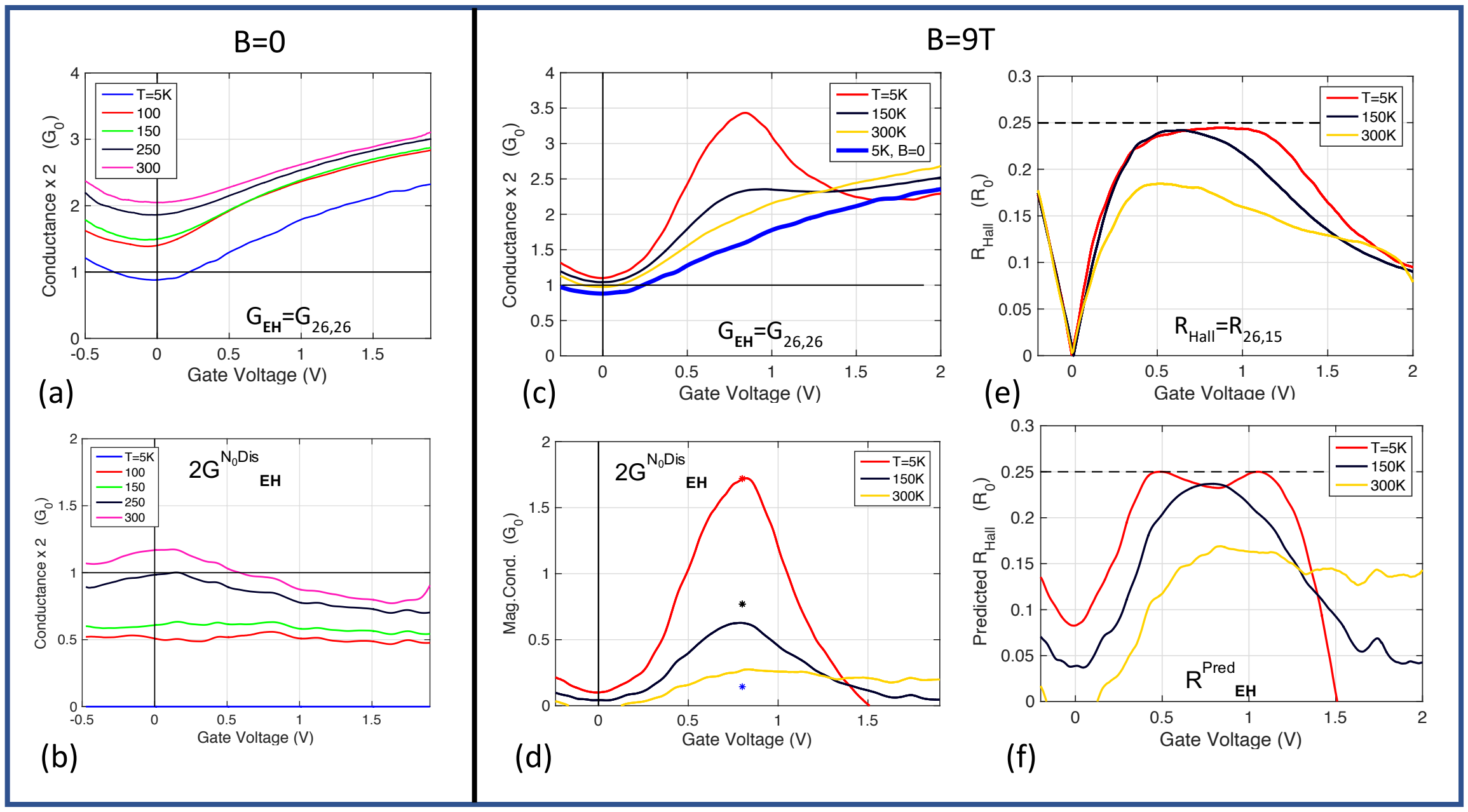

**Figure S13.** *Excited edge state and $N_{0Dis}$ for the 2 segments* ***E+H*** *inseries, Sample 1, 2points* **(a)** $2G_{EH}$ at B=0 T for T=5 K, 50 K and 300 K (the factor 2 accounts for 2 segments in series like in Fig. 4e, main text). **(b)** $2G^{N0Dis}(T,V_G)$ at B=0 T for T=5 K, 50 K and 300 K, where $G^{N0Dis}(T,V_G)=G_{EH}(T,V_G)-G_{EH}(T=5K,V_G)$ Note that $G^{N0Dis}$ increases essentially uniformly with increasing temperature suggesting an excited edge state of the edge state with a conductance of essentially 1 $G_0$. **(c)** $2G_{EH}$ at B=9 T for T=5 K, 50 K and 300 K. **(d)** $2G^{N0Dis}(T,V_G)$ at B=9 T for T=5 K, 150 K and 300 K. The edge state conductance reverts to ≈1 $G_0$ indicating that the excited state is suppressed in a magnetic field. The Shubnikov -de Haas oscillation has a maximum conductance of about 2 $G_0$, i.e. consistent with the conductance of $LL_0$. The reduction with increasing T is consistent with the Shubnikov-de-Haas peak of the $LL_0$ state of graphene ($E_{LL0}/k_B$ =1250 K at B=9 T) as indicated by the asterisks. **(e)** Hall resistance $R_{Hall}$ =$R_{26,15}$ at T=5 K, 150 K, 300 K showing the anomalous plateau persists at high temperatures, consistent with a high mobility N=0 dispersive graphene subband. **(f)** Hall resistance as predicted in Eq.4 of the main text, using measured $G^{N0Dis}$ plotted in (d) as in Fig.4g of the main text.

Figure S13a shows $2G_{EH}$ =$2G_{26,26}$ for several temperatures from 5 K to 300 K at B=0 T. Figure S13b shows $2G_{EH}(V_G,T)$-$2G_{EH}(V_G,T=5K)$ which corresponds to the conductance increase with increasing temperature. Note that $G_{EH}=(G_E^{-1}+G_H^{-1})^{-1} \approx G_{Seg}/2$ i.e. half of a single segment conductance which is the reason for the factor of 2 in the y axis.

Since the bulk conductance only weakly decreases with increasing temperature, this increase is due to the edge state (see Corbino data, SI2). Note that the increase is close to 1 $G_0$ and relatively independent of $V_G$, which is similar to the low temperature edge state behavior. Since this behavior is not seen in sidewall ribbons, we can reasonably assume that an excited state is involved with an energy gap on the order of 10 meV. This gap value is consistent with the observed 0.1 eV gap observed in 40 nm wide sidewall ribbons[11] assuming a 1/W width dependence, which is often observed in calculated energy gaps in edge states. Moreover, the fact that the edge state conductance at high temperatures is 2 $G_0$ is consistent with a 2-fold degenerate edge state as predicted for the edge state in general.

Figure S13c shows $2G_{EH}(V_G,T, B=9T)$ Note that now the conductance at CNP only slightly increases with increasing temperature which is explained in the main text as due to reduced scattering at the junction. If we subtract the conductance at T=5K, B=0: $G_{EH}(V_G,T,B=9T)-G_{EH}(V_G,T=5K,B=0T)$, then we recover the temperature and field dependence of the $N_{0Dis}$ (as explained in the main text) which shows a clear Shubnikov de Haas oscillation that corresponds to $LL_0$ (see main text).

The amplitude $A_{SdH}(B,T)$ of the G(B,T) oscillations (Shubnikov-de Haas oscillations) are given by the Lifshitz-Kosevich equation: $A_{SdH}(B,T)$ is proportional to u/sinh(u), where $u=2\pi^2 k_B T/E_{LL1}(B)$ and $E_{LL1}(B)$ is the energy of the $LL_1$: $LL1(B) = v_F c^* \sqrt{2e\hbar B}$. Hence, normalizing to $G_{EH}$(T=5K, B=9T ) gives $A_{SdH}$(T=150 K, B=9 T)=0.75 $G_0$ and $A_{SdH}$(T=300 K, B=9 T)=0.15 $G_0$ which corresponds well with the observed amplitudes (Fig. S13d), leaving little doubt that this is the correct interpretation.

Figure S13e shows the measured Hall resistance showing the anomalous Hall plateau associated with $LL_0$ (see main text) for B=9T at various temperatures. Note that the Hall plateau is still visible at T=150 K, and is decreased at T=300 K. This trend is expected for $LL_0$ however the anomalous value (i.e. 0.25 $R_0$ rather than 0.5 $R_0$) implies that the shunting due to the edge state persists up to high temperatures. This interpretation is confirmed in Fig. S13f which shows that the Hall resistance continues to be well predicted from the measured $G^{N0Dis}$ (Fig. S13d) using Eq.4 (main text) up to T=300 K.

**SI11.** ***Resistance segmentation and quantization***

**Resistance (in units of $R_0=h/e^2$) measured at CNP closely equals the number of segments, so that the resistance per segment is very close to 1 $R_0$ (at most 12% above $R_0$ at B=0) for all the segments measured. For B>0, the resistance decreases, resulting in essentially exact 1 $R_0$ quantization at high field, reminiscent of decreased coherent backscattering in magnetic field.[12] Similarly, at CNP and B=9 T, $R_E \approx 1\ R_0$, $R_{E+H} \approx 2\ R_0$ for all temperatures studied. Hence, at CNP the resistance per segment is approximately 1 $R_0$ at all temperatures both for nearly zigzag and armchair segments.**

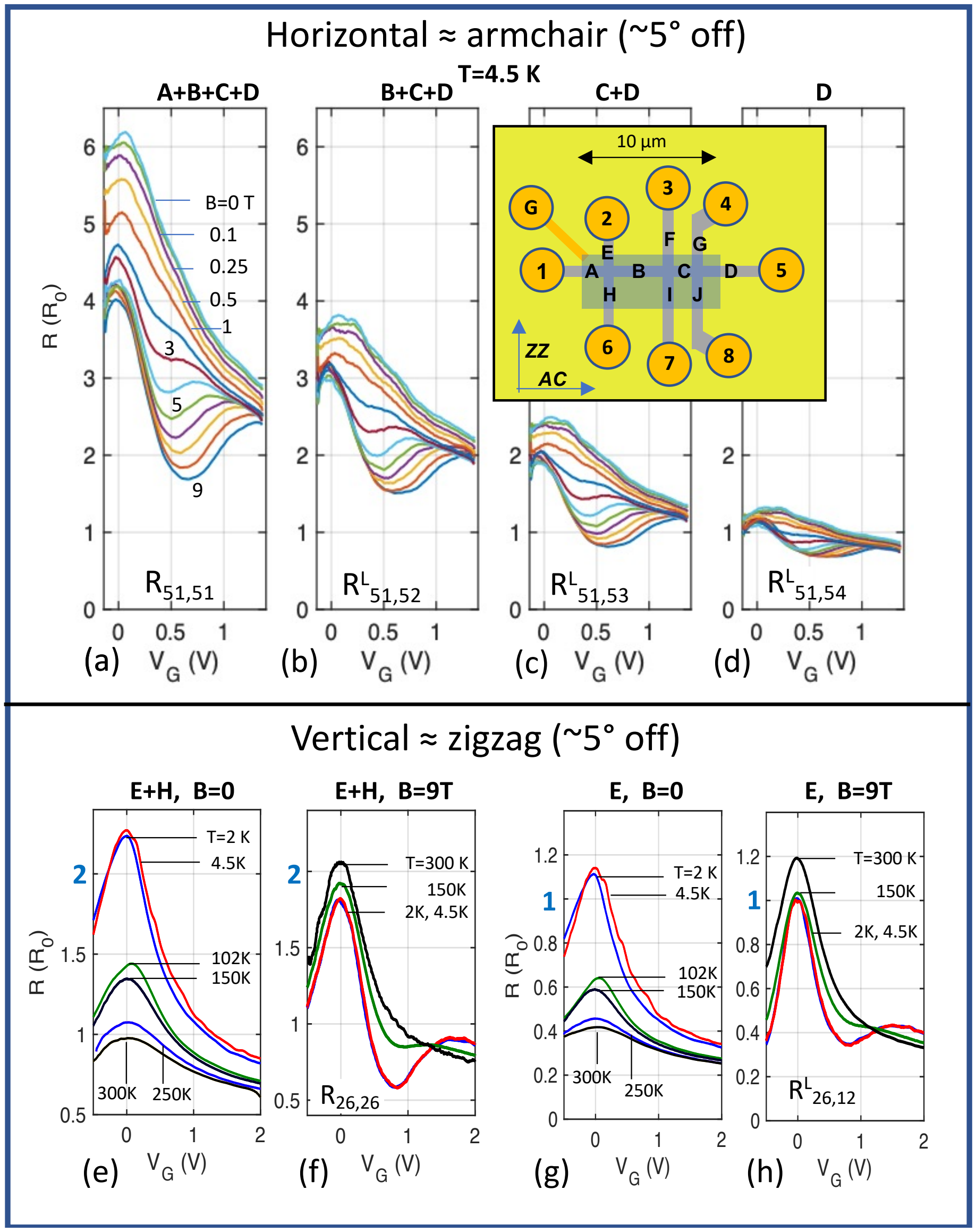

**Figure S14.** *Quantization and segmentation of the resistance.*

Three-point and two-point resistance (in units of $R_0$=h/e$^2$) for sample S1, width=740 nm, armchair orientation. $R^L_{ij,kl}$ denotes the longitudinal resistance $R^L$=(R(B)+R(-B))/2 with current flowing from contact i to j, and voltages measured between contacts k and l (**a**)-(**d**) *Horizontal segmentation*. $R^L_{51,5X}$ (X=1, 2, 3, 4) of the horizontal segments **A**, **B**, **C**, **D** (T=4.5 K, magnetic field |B| from 0 to 9 T and $V_G$ from -0.1 to 1.3 V). At CNP and |B|≥2 T, $R^L_{51,5X}$ = 4, 3, 2, 1 $R_0$ for X=1, 2, 3, 4 demonstrating 1 $G_0$ conductance quantization of the conductance in the segments and scattering at the junctions. At CNP and |B|≤2 The resistance increase is caused by reduced backscattering, reminiscently of coherent backscattering[12] **(e)-(h)** *Vertical segmentation*. **(e) (f)** segments **E+H** ($R_{11',11'}$); **(g) (h)** segment **E** ($R^L_{11',01}$). For **(e) (g)** B=0 T: resistance reduction at CNP with increasing temperature is due to thermal population of the bulk subbands. **(f) (h)** B=9 T: Edge state resistance quantization at CNP is observed up to T=300 K. Inset shows the device composed of a 15 µm long 740 nm wide horizontal ribbon crossed with 3 vertical ribbons giving 10 segments (white letters), 3 junctions, 8 low resistance Ohmic Pd-Au contacts. Measured segment lengths, in µm: $L_A$=3.6; $L_B$=3.3; $L_C$=1.7; $L_D$=4.5; $L_E$=1.6; $L_F$=4.0; $L_G$=3.3; $L_H$=3.6; $L_I$=6.8; $L_J$=3.8; (see Fig. S4). Vertical is 5° from the zigzag edge direction. The nominal gate efficiency (shaded rectangle) is $dn_c/dV_G$=-0.9x10$^{12}$ V$^{-1}$cm$^{-2}$. Gate voltages $V_G$ are reported with respect to CNP.

**SI12.** ***Supplemental structural and spectroscopic characterization of monolayer non-polar epigraphene***

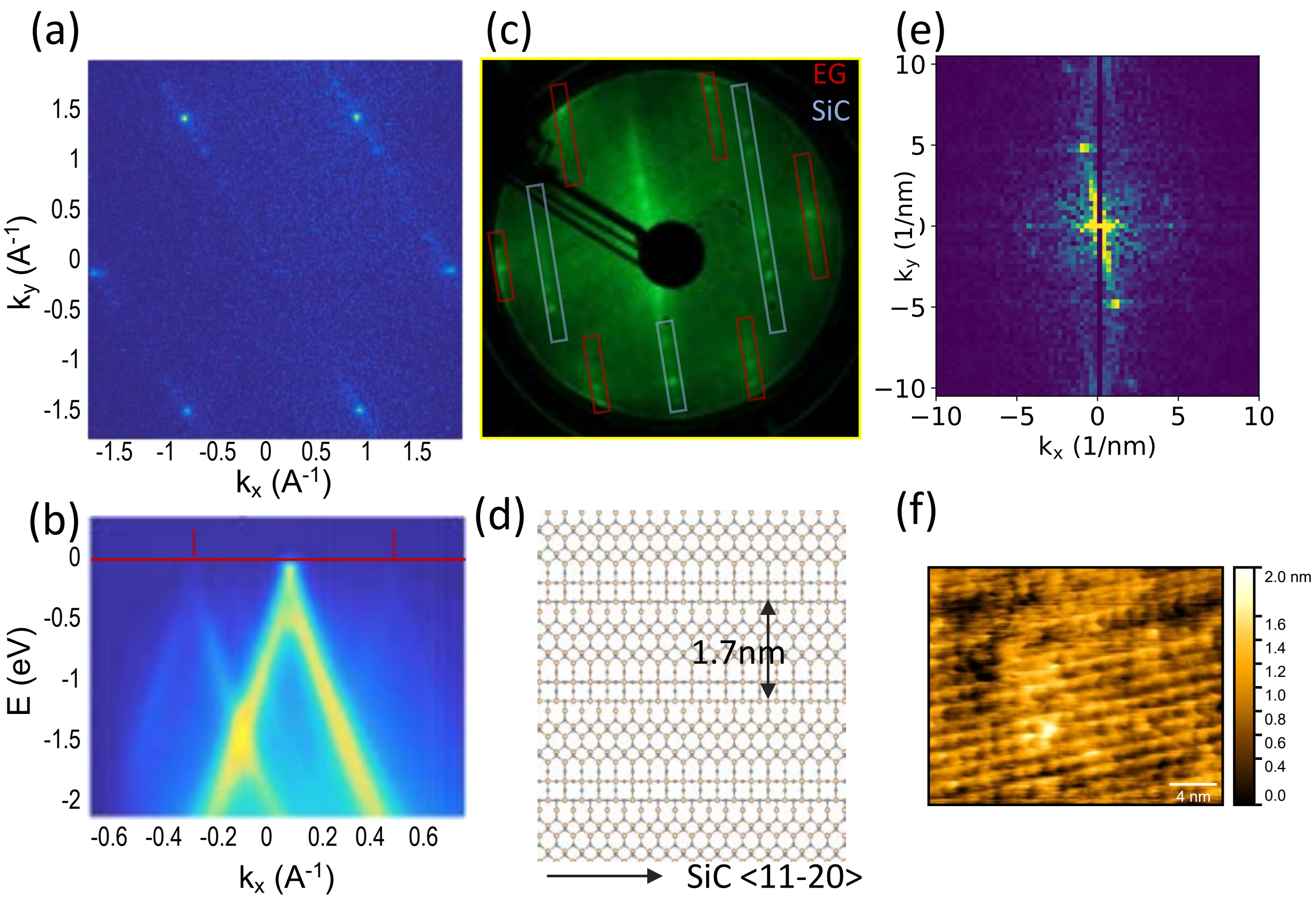

**Figure S15.** Sample S4 (non-patterned 2d epigraphene grown on non-polar facet 4HSiC$(1\bar{1}0n)$, n≈5) presents characteristics of monolayer graphene, with a Raman 2D peak at 2758 $cm^{-1}$ ($\lambda_{laser}$=532 nm) that may indicate interaction with the substrate. **(a)** Fermi surface measured by ARPES (Cassiopée beam line at the Soleil synchrotron) at room temperature at E=0 eV (Beam energy 200 eV, $E_F$=197.4 eV), showing the expected hexagonal Brillouin zone for graphene. No distortion of the Brillouin zone is observed. The Dirac point is at $E_F$; replicas are observed in one direction only, with periodicity 0.4±0.01 $Å^{-1}$, consistent with LEED. Note that the $k_x$ axis is oriented 30° from the <11-20> SiC direction shown in (d). **(b)** Energy vs $k_x$ map at $k_y$=1.624$Å^{-1}$ from ARPES measurements (beam energy=3 eV, $E_F$=32.1 eV); the sample was rotated so that the $k_x$ axis is oriented along the replica dots, i.e. perpendicularly to the <11-20>SiC direction shown in (d). The plot shows the linear graphene dispersing band and the Dirac point at $E_F$=0 (red line) and two (faint) replica band on each side (red arrows). **(c)** LEED pattern (E=73 eV) showing the graphene (outlined by red rectangles) and SiC (outlined by light blue rectangles) diffraction spots. Rectangles are oriented perpendicularly to the <11-20> SiC direction shown in (d). Replica spots aligned with the graphene indicate uniaxial modulation of the graphene by the substrate. The spot separation $\Delta$k=0.39±0.02 $Å^{-1}$ agrees with ARPES. **(d)** Example of the SiC (1-105) facet (top view), that presents a periodicity of 1.7 nm perpendicular to the <1-105> direction. **(e)** Fast Fourier transform of the STM image in (f) showing two main peaks corresponding to a modulation of periodicity 1.3±0.26 nm. **(f)** STM image (bias voltage =90 mV, I =900 pA). The parallel set of lines is along the <1-10n> direction, as expected from (d).

**SI13.** ***Spin polarization in sidewall nanoribbons***

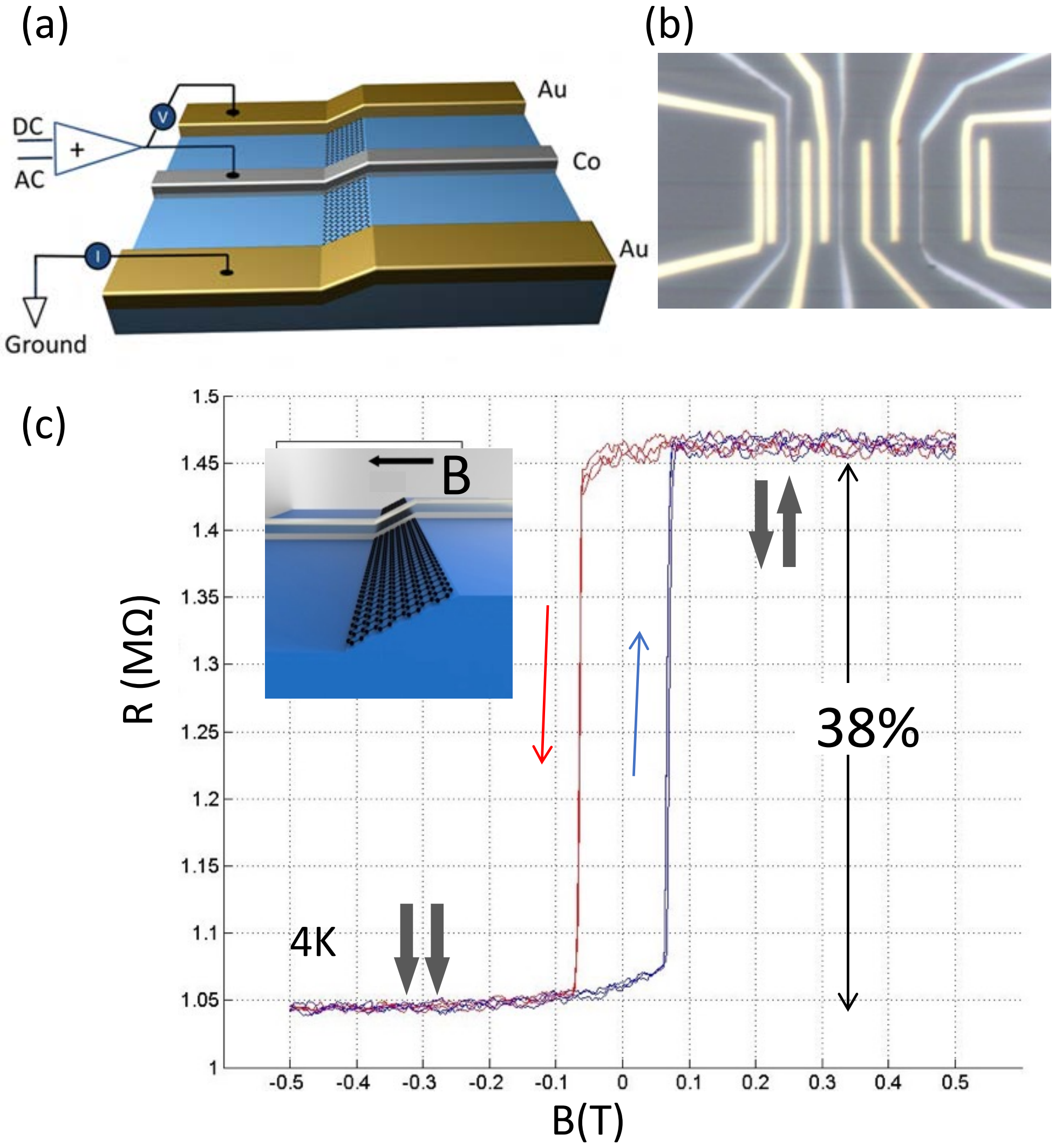

**Figure S16.** *Epigraphene sidewall nanoribbon spin polarized transport* (from Refs [13,14]). **(a)** Schematics of a three terminal non-local device, including a slopping epigraphene sidewall nanoribbon connected with a single spin polarizing tunnel contact (Cobalt on Alumina) and two Pd/Au contacts. An AC+DC voltage is applied between the tunnel contact and a Pd/Au contact, where current is measured, while the non-local voltage is measured on the opposing Pd/Au contact. **(b)** Optical image of a multi-terminal device (grey: Co/$Al_2O_3$, gold: Pd/Au). **(c)** Reproducible non-local tunnel resistance (2mV AC and 8mV DC applied) showing switching between a high and a low resistance value as the cobalt magnetization is reversed by aligning with the applied magnetic field. The hysteresis is expected for a Cobalt magnetic polarizer. The magnetic field is oriented parallel to the basal (0001) plane, as indicated in the diagram (inset). Note that a single resistance jump is observed, indicating magnetization of the ribbon itself.

## SI14. *Prediction of electronic conductance in graphene ribbons with side-gates*

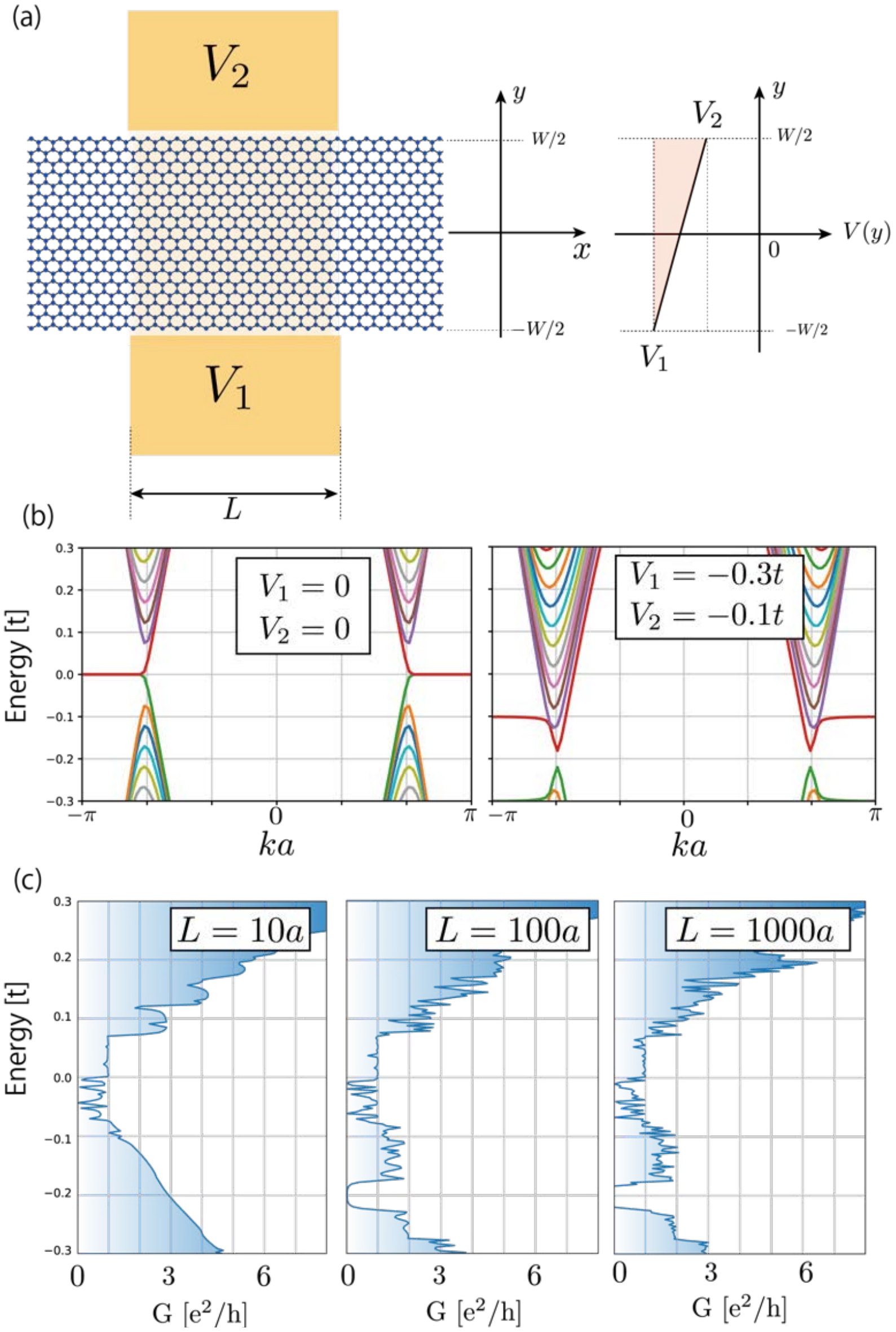

**Figure S17.** *Theoretical calculations for electronic conductance of graphene ribbons with side-gates*, on the basis of nearest-neighbor tight-binding model, showing that the quantum interference based top-gating effect of the edge state, found in Ref. [15] also applies to side gates. **(a)** Schematic setup of graphene ribbons with side gate. The transverse electric field is locally applied by using side gates, where lower and upper gates have bias voltages of $V_1$ and $V_2$, respectively. In this model, the linear slope potential with $V(y) = \left[\frac{(V_2-V_1)}{W}y + \frac{V_1+V_2}{2}\right]t$ is included. Here $t = 2.7$ eV is the transfer integral between nearest-neighbor carbon atoms of tight-binding model, $W$ is the ribbon width and $L$ is the length of side gates. **(b)** Energy band structure of ribbon with *W*=53*a*, where $a = 0.246$ nm is the lattice constant of graphene. (left) Energy band structure in absence of side gate voltage, i.e. $V_1 = V_2 = 0$. Flat bands appear owing to the edge localized states at zero energy. (right) Energy band structure in presence of side gate voltage, i.e. $V_1 = -0.3t$, $V_2 = -0.1t$. Under side gate bias, the energy bands lift downwards and opening small gap at Dirac cones. **(c)** Landauer conductance through graphene ribbons with side gate bias for several different length of side gates, i.e. (left) *L*=10*a*, (middle) 100*a* and (right) 1000*a*. The parameters for side gate bias are $V_1 = -0.3t$, $V_2 = -0.1t$. Ribbon width *W*=53*a*.

**SI15.** ***First observation of the edge state in transport***

Graphene research was originally motivated by the imminent need for a successor for silicon for nanoelectronics.[16,17] Earlier work on carbon nanotubes demonstrated spectacular ballistic transport properties,[18] and theory predicted that the edge state in graphene nanoribbons could have similar ballistic properties.[19-21] In fact, in graphene nanostructures the 1d edge state dominates the electronic properties and the 2d bulk does not play an essential role.

Epigraphene electronics research, pioneered at the Georgia Institute of Technology,[22-24] was motivated by these considerations and has developed independent of exfoliated graphene research. In fact, it predates exfoliated graphene by several years[22,25]. From the outset, and even now, only epigraphene (graphene grown on electronics grade silicon carbide crystals) provides the alignment and stability that is ideal for nanoelectronics fabrication.[26] In contrast lithographically patterned exfoliated graphene invariably has highly disordered graphene edges[27-29] which effectively makes it unsuited for nanoelectronics.

Since 2008 epigraphene electronics research has focused on sidewall ribbons that were grown by the thermal sublimation method on the natural and lithographically produced steps on the (0001) face of 4H-SiC and methods were developed to produce large arrays of epigraphene transistors.[30-32]

The 1 $G_0$ edge state was first observed in natural steps on 4H SiC in a natural sidewall that connected two lithographically graphene contacts (Fig. S15a, b).[31,32] Room temperature 1 $G_0$ conductance was confirmed in experiments of sidewall graphene ribbons where the conductance was measured as a function of length using an AFM probe (Fig. S15c-e).[31,32] Large arrays of graphene sidewall ribbons were produced[30,33] and the 1 $G_0$ edge state conductance was definitively confirmed using 4-probe methods in UHV (Fig. S16).[34]

Unexpectedly, 1 $G_0$ ballistic conductance is also observed meandering natural ribbons[31,32,34] (Fig. S15a) which implies that the general chiral ribbons also support the edge state. This led to the conclusion that the high temperature annealing prefers acene edges thereby significantly enhancing their applications potential.

The sidewall graphene was also found to be charge neutral in contrast to the highly charged graphene ($n \approx 10^{13}$ $cm^{-2}$) on the polar faces.[31,32,34] Since such large charge densities are challenging for graphene nanoelectronics in general, we initiated research on non-polar face graphene where we could use standard nanolithography methods on 2d graphene.

However, the required SiC wafers with surfaces corresponding to the sloping sidewall facets are not commercially available, so that we initiated a SiC production facility at the Tianjin International Center for Nanoparticles and Nanosystems.

Epigraphene grown on non-polar surfaces is indeed epitaxial and charge neutral as shown here (see also Fig. S14). Moreover, surprisingly, we also found that the graphene ribbons again had an edge state with a 1 $G_0$ conductance even without applying a high temperature annealing step. It was clear that the plasma etching itself not only cut the graphene but also annealed the edges so that the annealing step is not required. Moreover, the 2d epigraphene itself has a low mobility which is of no consequence for the 1d edge state. These spectacular results clearly show a route to edge state 1d nanoelectronics. Moreover, as we show here, the edge state is pinned at E=0 due to the 0-DoS at E=0. This adds another unexpected aspect to epigraphene electronics and furthermore sheds light on the edge state itself which may turn out to be a Majorana fermion.

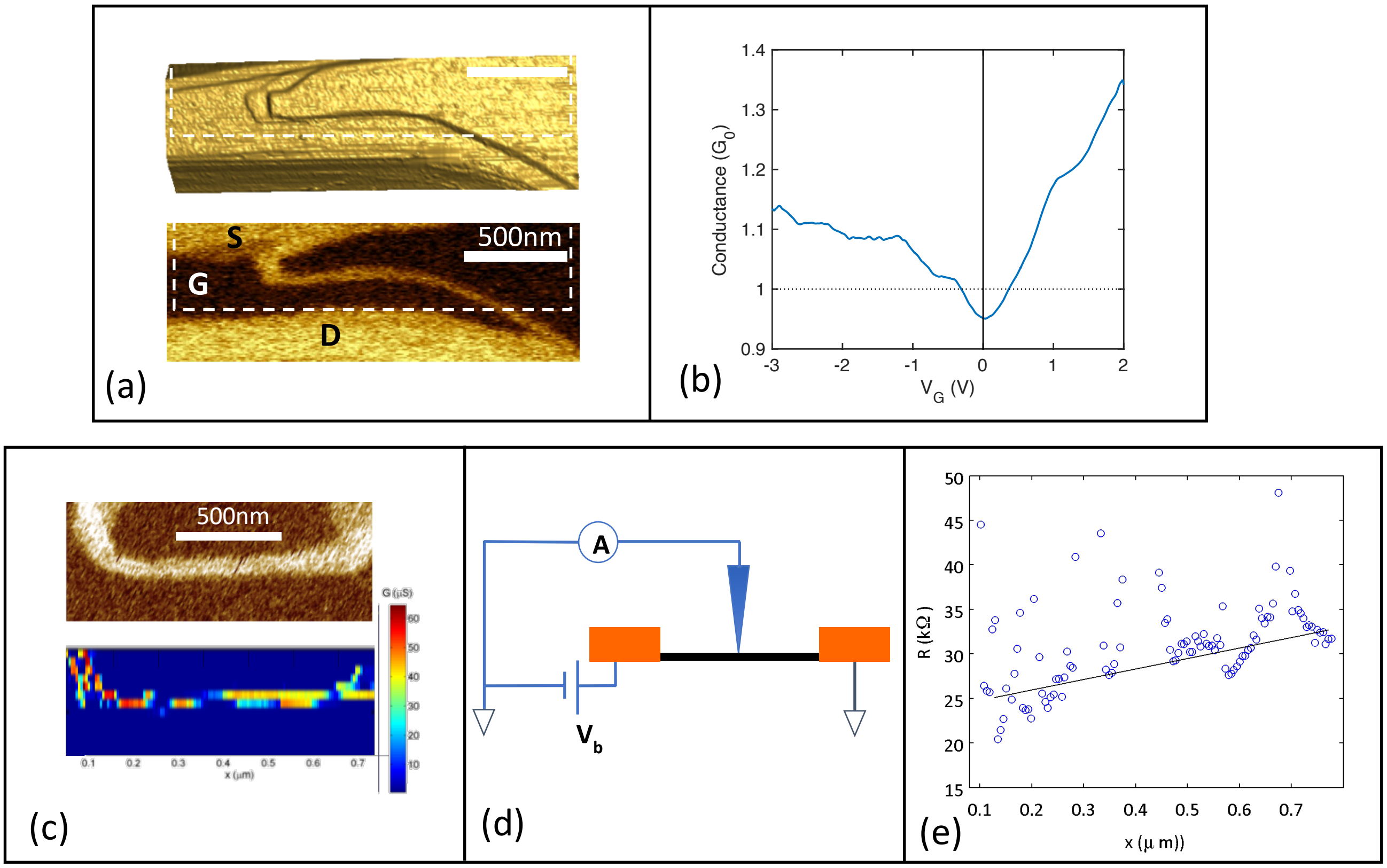

**Figure S18.** *First observations of the edge state*[31,32] **(a)** Top: AFM image of meandering natural step on the (0001) face of 4H-SiC. Bottom: EFM image of the same step; bright regions correspond to graphene, large bright areas are prepatterned graphene source (S) and drain (D) pads, narrow bright line is sidewall graphene decorating the step edge; white dashed line indicates outline of the gate (G). **(b)** Conductance measurements of device in (a) at T=4.5 K, showing the 1 $G_0$ edge state conductance **(c)** Top: EFM of natural graphene sidewall ribbon between two prepatterned graphene pads. Bottom: Local conductance trace using a conducting AFM tip as in (d). **(d)** Schematic diagram of the electrical measurement method of the device using a conducting AFM tip (blue arrow). Orange rectangles represent the graphene pads, black line represents the sidewall ribbon. **(e)** Resistance measurements, x is the distance of the tip to the left pad. The resistance is <2 kΩ when the tip is placed on the left pad. The black line is a linear fit to the data.

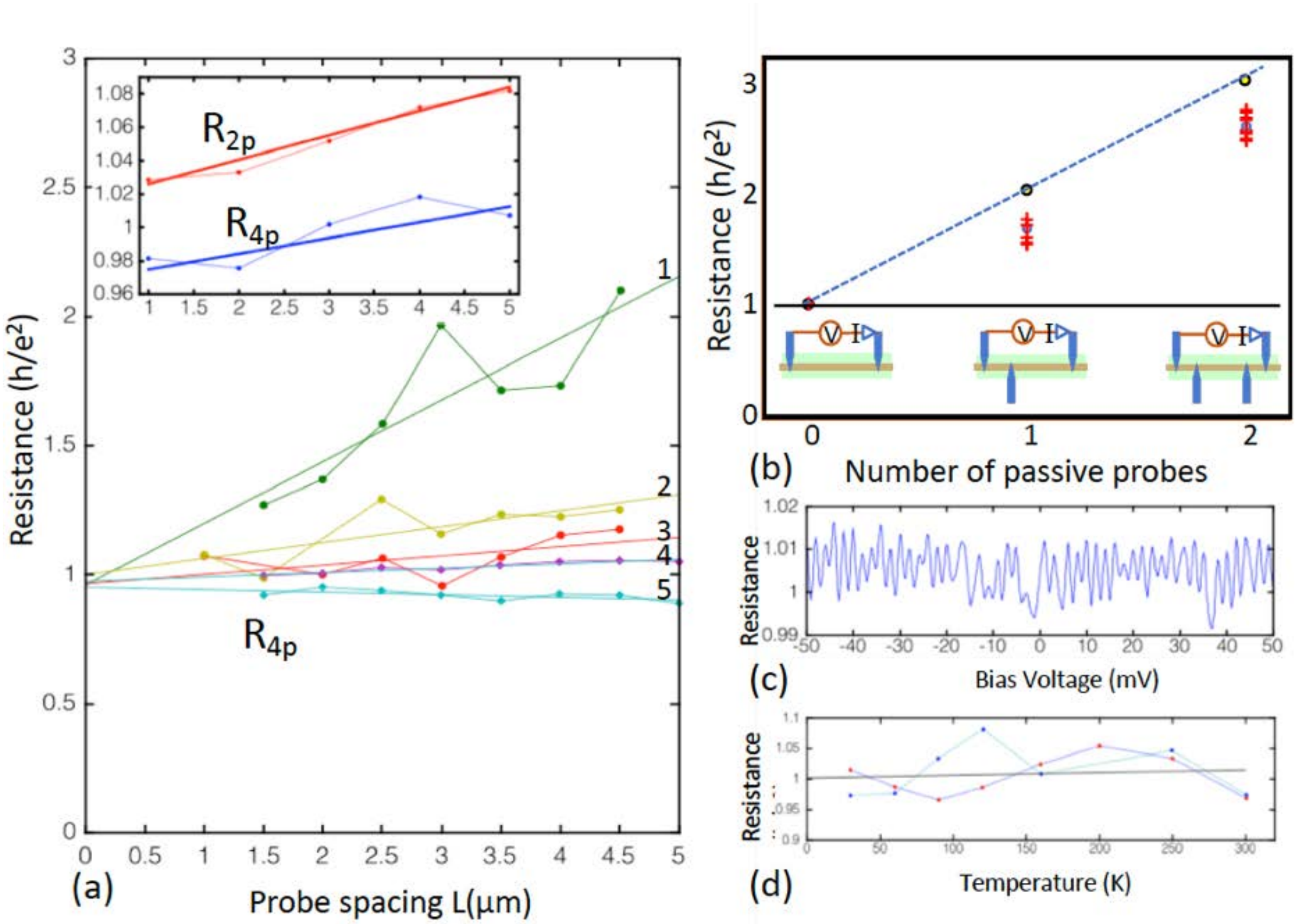

**Figure S19.** *Multi probe in-situ transport measurements of several 40 nm wide graphene sidewall ribbons* (adapted from Ref.[34]). **(a)** Resistances as a function of voltage probe spacing L. Linear fits extrapolate to 1 $R_0$ within a few percent at L=0. Slopes from 1 to 5 correspond to mean free paths $\lambda$=4.2, 28, 16, 58, >70 μm, respectively. (Inset) Sidewall ribbon with $\lambda$=106 μm; two-point measurement (red) and 4-point measurement (blue) differ by (only) 4%, indicating a probe contact resistance ≈500 Ω. **(b)** Segmentation of a sidewall ribbon caused by scattering at non-current carrying passive probes placed on the ribbon. A single non-current carrying passive probe, approximately doubles the 2-point resistance of sidewall ribbon. Two passive probes approximately triple the resistance consistent with ballistic transport theory.[12] **(c)** Resistance as a function of bias voltage $V_b$ showing essentially no effect for -50 mV≤$V_b$≤50 mV **(d)** Resistance as a function of temperature for two 5 μm long ribbons showing no significant temperature dependence.

**SI16.** ***Effect of annealing on epigraphene nanoribbons***

**Following Eq. 1 in the main text,** $G(L)=1/R(L)= G_{edge} + G_{bulk}= G_e(1+L/\lambda_{edge})^{-1} + \sigma W/L$. **For a diffusive ribbon with a ballistic edge state, the resistance as a function of length of a nanoribbon provided with increasingly spaced probes will intercept the origin in absence of an edge state and will intercept at R=1 $R_0$ for the edge state with $\lambda_{edge}$>>L. The significant slope change after high temperature annealing indicates a decrease in the resistivity of the bulk. The concomitant increase in the L=0 intercept resistance indicates an increase in the edge state mean free path.**

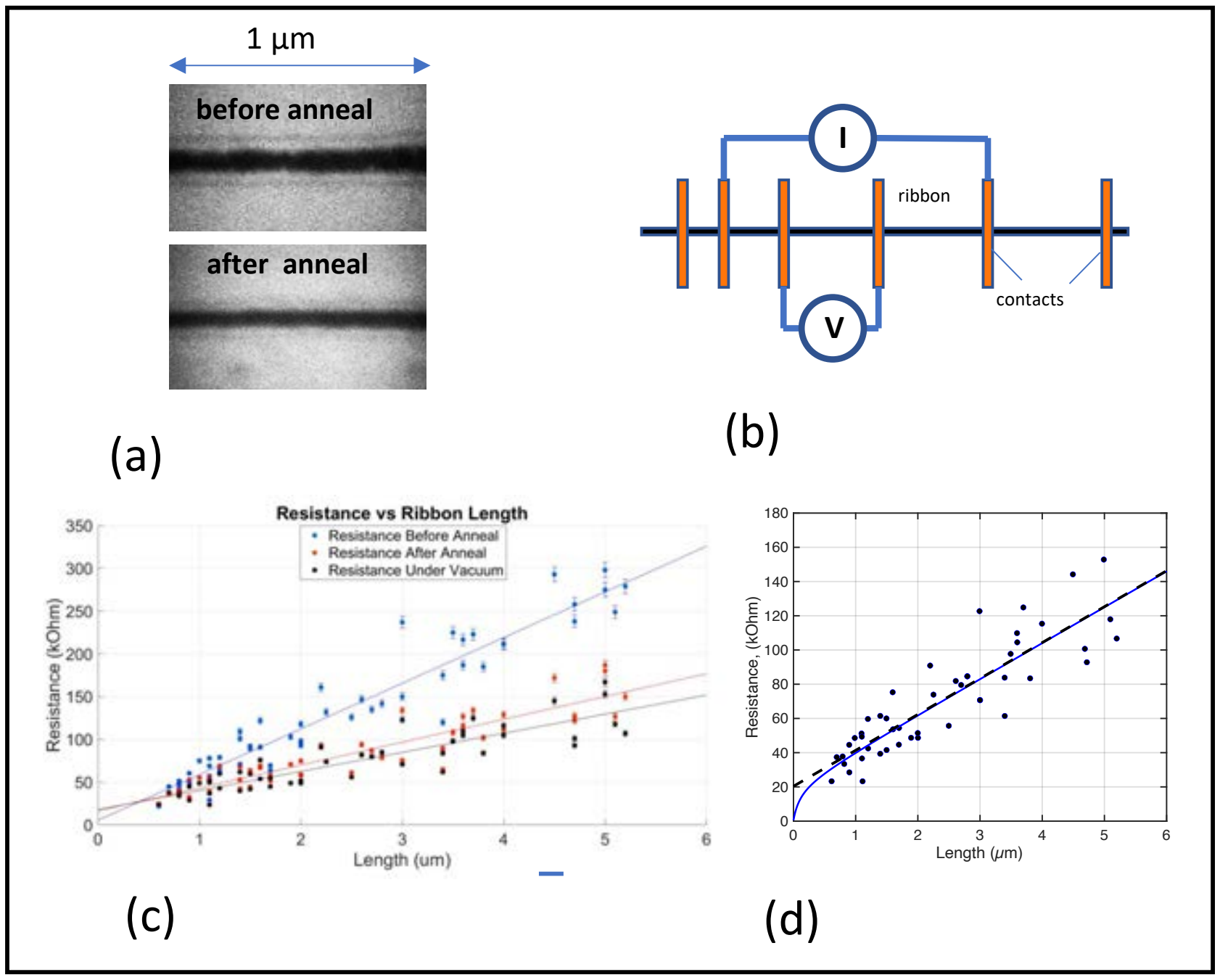

**Figure S20.** *Room temperature 4-point probe station resistance measurements R(L) of 90 ± 10 nm wide oxygen plasma etched Si-face graphene nanoribbons (on the Si-face) before (charge density ≈5 x $10^{12}$ $cm^{-2}$) and after vacuum annealing* at 1200°C for 15 minutes[35] **(a)** SEM image of graphene a ribbon before annealing (top) and after annealing (below), showing that annealing significantly smooths the edges and also slightly narrows the ribbon. **(b)** Schematic diagram of the device showing a nanoribbon (black) that is crossed with contact strips (orange) spaced at various intervals L, on which probes are placed for 4-point measurement. Voltage measurements are performed on single segments (uninterrupted by probes) **(c)** 4-point resistances of 50 sections at various lengths made of the unannealed device $R_1$, (blue); after annealing and measured in air, $R_2$ (orange); after annealing and measured in vacuum, $R_3$ (black); The linear fits are (R in units of $h/e^2$ and L in µm): $R_1(L) = 53.38(\pm3.21) \times L + 5.51(\pm4.23)$; $R_2(L) = 26.7(\pm2.84) \times L + 16.76(\pm2.39)$; $R_3(L) = 22.31(\pm2.29) \times L + 18.01(\pm2.56)$. More quantitatively, **(d)** shows the same annealed ribbon data as the black squares in **(b)**. The fit given by the blue line (Eq.1) is consistent with $G_{edge}$ =1 $G_0$, $\lambda_{edge}$=1.1 µm, and $\lambda_{bulk}$=4 nm. The dashed line is a linear fit to the data, similarly to (a), showing that Eq.1 is dominated by the edge contribution at large length.

The data[35] in Fig. S20 clearly demonstrate the efficacy of annealing and are consistent with a 1 $G_0$ edge state in these narrower conventionally patterned ribbons. The relatively large spread in the data is due to charge inhomogeneities and variations in the ribbon widths. These measurements were performed on significantly charged Si-face substrates rather than non-polar substrates at low charge densities, which contributes to relatively small $\lambda_{edge}$. Further studies on charge neutralized Si face substrates are ongoing.

### SI17. *Possible low energy edge structures*

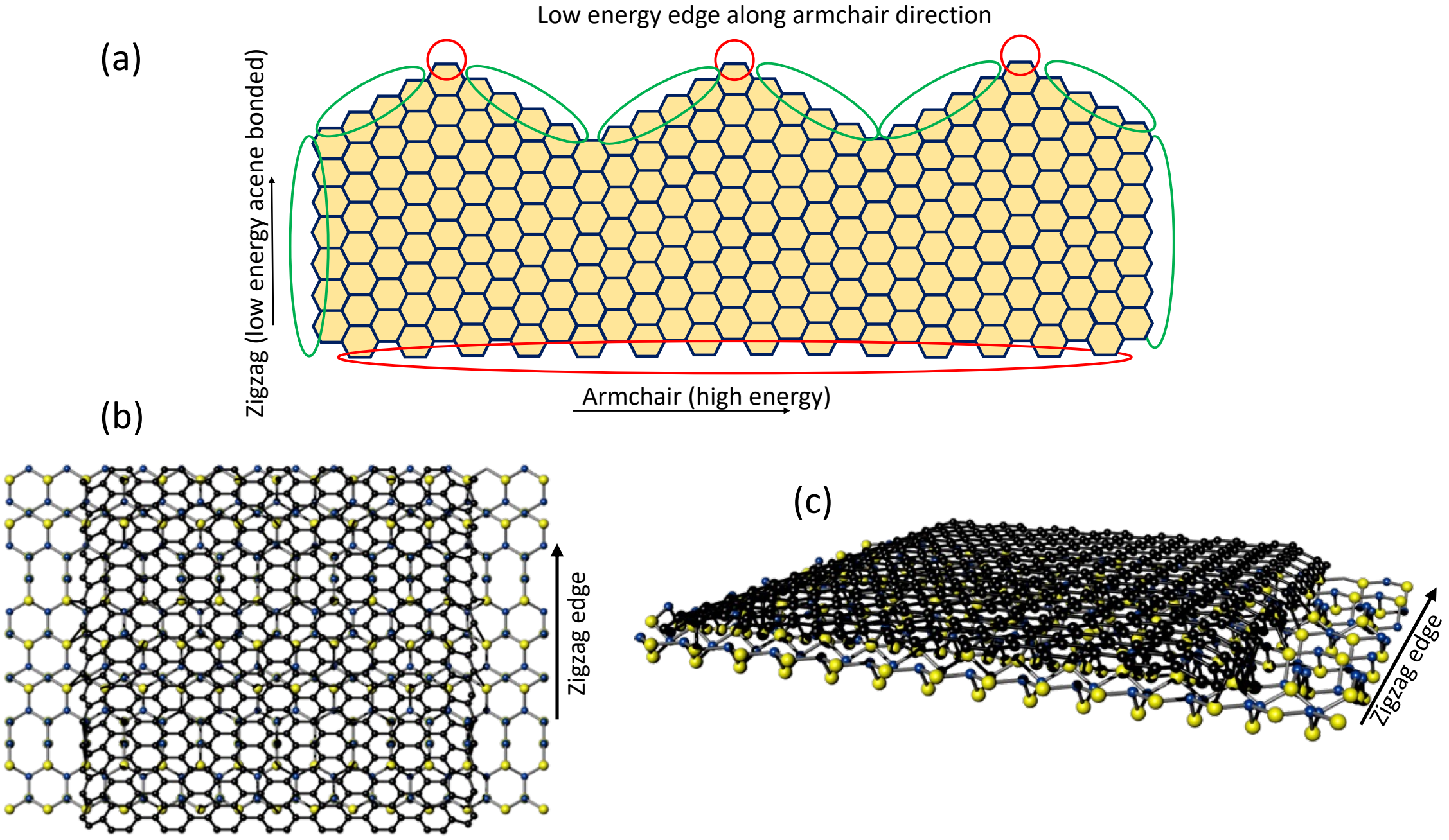

**Figure S21.** (a) *Schematic drawing of a nominally armchair ribbon with effective mixed edge chirality (here mostly zigzag). (b) top view and (c) perspective view of a ball and stick model of a zigzag ribbon on a bulk terminated 4H-SiC(1-105) facet (only top SiC bilayer shown). The graphene alignment corresponds to that of the Si(0001) face per continuity. The bonds between the graphene edge and SiC are hand drawn and for illustration only. A realistic structure should be fully calculated and account for surface reconstruction, which is far beyond the scope of this paper.*

Since the graphene edges are bonded to the substrate, these bond energies play an essential role in the morphology of the edge. Cross sectional TEM images of side wall ribbons show that the zigzag edges bond to the silicon and/or carbon atoms of the substrate.[36,37] The images indicate that the graphene bends down at the edge so that these bonds are perpendicular to the SiC surface and match up with the direction of the Si and C bonds of the SiC at the surface. These strong bonds passivate and mechanically stabilize the edge. In contrast, the dangling edge atoms of an armchair edge will not naturally line up with the substrate Si and C dangling bonds, so that these bonds will be strained and therefore of higher energy. If this is the case, then an annealed edge along the armchair direction will reconstruct to maximize the number of zigzag edge atoms (acene edge atoms), as schematically shown above. Hence, the nominal armchair edge can be seen as a string of zigzag edges, which is similar to a general chiral edge, which a 0-DoS peak and an edge state as shown by Beenakker et al. [38]

**Supplementary References**

1 Das Sarma, S., Adam, S., Hwang, E. H. & Rossi, E. Electronic Transport in Two-Dimensional Graphene. *Review of Modern Physics* **83**, 407-466 (2011).

2 Martin, J. *et al.* Observation of electron-hole puddles in graphene using a scanning single-electron transistor. *Nature Physics* **4**, 144-148 (2008).

3 Chen, J. H. *et al.* Charged-impurity scattering in graphene. *Nature Physics* **4**, 377-381 (2008).

4 Trushin, M. & Schliemann, J. Minimum Electrical and Thermal Conductivity of Graphene: A Quasiclassical Approach. *Physical Review Letters* **99**, 216602 (2007).

5 Novoselov, K. S. *et al.* Two-dimensional gas of massless Dirac fermions in graphene. *Nature* **438**, 197-200 (2005).

6 Cho, S. & Fuhrer, M. S. Charge transport and inhomogeneity near the minimum conductivity point in graphene. *Physical Review B* **77**, 081402 R (2008).

7 Tan, Y. W. *et al.* Measurement of scattering rate and minimum conductivity in graphene. *Physical Review Letters* **99**, 246803 (2007).

8 Stolyarova, E. *et al.* High-resolution scanning tunneling microscopy imaging of mesoscopic graphene sheets on an insulating surface. *Proceedings of the National Academy of Sciences* **104**, 9209 (2007).

9 Miller, D. L. *et al.* Observing the Quantization of Zero Mass Carriers in Graphene. *Science* **324**, 924-927 (2009).

10 Chen, J. H., Jang, C., Xiao, S. D., Ishigami, M. & Fuhrer, M. S. Intrinsic and Extrinsic Performance Limits of Graphene Devices on SiO2. *Nat Nanotechnol* **3**, 206-209 (2008).

11 Nguyen, T. T. N. *et al.* Topological Surface State in Epitaxial Zigzag Graphene Nanoribbons. *Nano Lett* **21**, 2876-2882 (2021).

12 Datta, S. *Electronic transport in mesoscopic systems*. (Cambridge University Press, 1995).

13 Hankinson, J. *Spin dependent current injection into epitaxial graphene nanoribbons* PhD thesis, PhD - Georgia Institute of Technology, (2015).

14 Huan, C. *et al.* Tunnel magnetoresistance of magnetic junctions based on side-wall epitaxial graphene nanoribbons. *APS March meeting*, B7.00001 (2013).

15 Wakabayashi, K. & Aoki, T. Electrical conductance of zigzag nanographite ribbons with locally applied gate voltage. *Int J Modern Physics B* **16**, 4897-4909 (2002).

16 Chen, A., Hutchby, J., Zhirnov, V. & Bourianoff, G. *Emerging Nanoelectronic Devices*. (Wiley publisher, 2014).

17 Geim, A. K. & Novoselov, K. S. The rise of graphene. *Nature Materials* **6**, 183 (2007).

18 Frank, S., Poncharal, P., Wang, Z. L. & de Heer, W. A. Carbon nanotube quantum resistors. *Science* **280**, 1744-1746 (1998).

19 Fujita, M., Wakabayashi, K., Nakada, K. & Kusakabe, K. Peculiar localized state at zigzag graphite edge. *J Phys Soc Jpn* **65**, 1920-1923 (1996).

20 Nakada, K., Fujita, M., Dresselhaus, G. & Dresselhaus, M. S. Edge state in graphene ribbons: Nanometer size effect and edge shape dependence. *Physical Review B* **54**, 17954-17961 (1996).

21 Wakabayashi, K., Takane, Y. & Sigrist, M. Perfectly conducting channel and universality crossover in disordered graphene nanoribbons. *Physical Review Letters* **99**, 036601 (2007).

22 de Heer, W. A., Berger, C. & First, P. N. Patterned thin films graphite devices and methods for making the same. *US patent US7015142B2 (Provisional filed Jun. 12, 2003). This experimentally supported patent provides the foundation of graphene nanoelectronics in general and expliclty describes seamless zero-mode epigraphene electronics.*

23 Berger, C. *et al.* Ultrathin Epitaxial Graphite: 2D Electron Gas Properties And a Route Toward Graphene-Based Nanoelectronics. *J Phys Chem B* **108**, 19912-19916 (2004).

24 Berger, C. *et al.* Electronic confinement and coherence in patterned epitaxial graphene. *Science* **312**, 1191-1196 (2006).

25 de Heer, W. A. Patterned Graphite Nanoelectronics. *NSF proposal 01-157* **12/19/2001**, https://graphene.gatech.edu/historicalpersp.html (2001).

26 Berger, C., Conrad, E. & de Heer, W. A. in *Physics of Solid Surfaces, Landolt Börstein encyclopedia* Vol. Subvolume B *Physics of Solid Surfaces* (ed P. Chiaradia G. Chiarotti) Ch. 164-171, 727-807. ArXiv:1704.00374 (Springer-Verlag, 2018).

27 Epping, A. *et al.* Insulating State in Low-Disorder Graphene Nanoribbons. *physica status solidi (b)* **256**, 1900269 (2019).

28 Stampfer, C. *et al.* Energy Gaps in Etched Graphene Nanoribbons. *Physical Review Letters* **102**, 056403 (2009).

29 Gallagher, P., Todd, K. & Goldhaber-Gordon, D. Disorder-induced gap behavior in graphene nanoribbons. *Physical Review B* **81**, 115409 (2010).

30 Sprinkle, M. *et al.* Scalable templated growth of graphene nanoribbons on SiC. *Nat Nanotechnol* **5**, 727-731 (2010).

31 Ruan, M. *Structured epitaxial graphene for electronics* PhD thesis, PhD - Georgia Institute of Technology, (2012).

32 Ruan, M. *et al.* Epitaxial graphene on silicon carbide: Introduction to structured graphene. *MRS Bulletin* **37**, 1138-1147 (2012).

33 Hicks, J. *et al.* A wide-bandgap metal-semiconductor-metal nanostructure made entirely from graphene. *Nature Physics* **9**, 49-54 (2013).

34 Baringhaus, J. *et al.* Exceptional ballistic transport in epitaxial graphene nanoribbons. *Nature* **506**, 349-354 (2014).

35 Hu, Y. *The edge states of epitaxial graphene on SiC, PhD dissertation* PhD dissertation thesis, Georgia Tech, (2021).

36 Palacio, I. *et al.* Atomic Structure of Epitaxial Graphene Sidewall Nanoribbons: Flat Graphene, Miniribbons, and the Confinement Gap. *Nano Lett* **15**, 182−189 (2015).

37 Norimatsu, W. & Kusunoki, M. Growth of graphene from SiC{0001} surfaces and its mechanisms. *Semicond Sci Tech* **29**, 064009 (2014).

38 Akhmerov, A. R. & Beenakker, C. W. J. Boundary conditions for Dirac fermions on a terminated honeycomb lattice. *Physical Review B* **77**, 085423 (2008).